\documentclass[letterpaper,notitlepage,12pt]{gvbarticle}
\usepackage{amsmath}
\usepackage{times}
\usepackage{mhequ}			 
\usepackage{caption2}
\usepackage{epsfig}
\usepackage{amssymb}
\usepackage{rotating}

\newtheorem{theorem}{Theorem}[section]
\newtheorem{corollary}[theorem]{Corollary}
\newtheorem{definition}[theorem]{Definition}
\newtheorem{lemma}[theorem]{Lemma}
\newtheorem{proposition}[theorem]{Proposition}
\newtheorem{remark}[theorem]{Remark}

\def\triplet{({\bf v},\omega)}
\def\dirac{1\kern-0.6mm\rule{0.28mm}{8pt}\rule{0.55mm}{0.18mm}}
\def\LA{{\cal L}_{1}}
\def\LB{{\cal L}_{2}}

\def\csupun{C\hspace{-2mm}\sup_{x_0\leq x\leq2x_0}}
\def\pu{q}
\def\pv{p}
\def\pdv{r}
\def\S{{\cal S}}
\def\M{{\cal M}}

\def\xzero{x_0}
\def\Rp{[x_0,\infty)}
\def\Lambdazero{\Lambda_0}
\def\Lambdam{\Lambda_{-}}
\def\Lambdap{\Lambda_{+}}
\def\strouhal{\tau
}
\def\reynolds{{\rm R\kern-1.5pt e}
}

\def\olddelta{\varphi}
\def\omegazero{w}

\def\uzero{\nu}
\def\huzero{\mu}
\def\qzero{P}
\def\qun{Q}
\def\qA{R}
\def\qB{S}
\def\falpha{F}
\def\galpha{G}
\def\Fomega{{\cal F}_{2,\omega}}
\def\Fu{{\cal F}_{2,u}}
\def\Fv{{\cal F}_{2,v}}
\def\Lu{{\cal L}_u}
\def\Lv{{\cal L}_v}
\def\I{{\cal I}}
\def\P{{\cal P}}
\def\Q{{\cal P}_0}
\def\au{\eta}
\def\av{\xi}
\def\olddeltam{\frac{1}{2}}
\def\W{{\cal W}}
\def\Cu{{\cal C}_u}
\def\Ci{{\cal C}_i}
\def\myast{}

\def\kc{c}
\def\kzero{0}
\def\kun{1}
\def\kdeux{2}
\def\kcinq{5}
\def\ksix{6}
\def\ksept{7}
\def\khuit{8}
\def\kdix{10}
\def\kdouze{12}
\def\ktreize{13}
\def\kr{r}
\def\ki{i}

\newenvironment{proof}[1][Proof]{\noindent\textit{#1.} }{\ \rule{0.5em}{0.5em}\par}

\renewcommand{\vector}[2]{
\Big(
\begin{matrix}   
#1\\ #2
\end{matrix}\Big)
}

\newcommand{\A}[2]{
A\Big[
\begin{array}{l}
{\scriptstyle #2}\\[-0.5mm]
{\scriptstyle #1}
\end{array}
\Big]
}

\newcommand{\B}[2]{
B\Big[
\begin{array}{l}
{\scriptstyle #2}\\[-0.5mm]
{\scriptstyle #1}
\end{array}
\Big]
}

\newcommand{\D}[2]{
D\Big[
\begin{array}{l}
{\scriptstyle #2}\\[-0.5mm]
{\scriptstyle #1}
\end{array}
\Big]
}

\newcommand{\K}[3]{#1,\{#2,#3\}}
\newcommand{\chinese}[1]{\langle#1\rangle}

\renewcommand{\epsilon}{\varepsilon}

\def\L{{\rm L}}
\def\H{{\cal H}}
\def\Ltx#1#2{#1,#2}
\def\Lx#1#2#3{#1,#2,#3}

\newcommand{\ed}{e}
\newcommand{\essup}{\displaystyle\mathop{\rm ess~sup}}
\renewcommand{\Re}{{\rm Re}}
\renewcommand{\Im}{{\rm Im}}

\makeatletter
\@addtoreset{equation}{section}
\def\theequation{\thesection.\the\c@equation}
\def\newappendix#1{%
        \let\@oldform\@seccntformat%
        \def\@seccntformat##1{Appendix~\csname the##1\endcsname:~}%
        \section{#1}%
        \let\@seccntformat\@oldform%
        }
\makeatother

\begin{document}

\date{\today}
\title{Downstream asymptotics in exterior domains: from stationary wakes to
time periodic flows}
\author{Guillaume van Baalen\thanks{Supported by the Fonds National Suisse.}}
\institute{
Institut Fourier\\
Universit\'e J. Fourier / Grenoble\\
France\\
\email{guillaume.vanbaalen@ujf-grenoble.fr}}

\maketitle
\tableofcontents
\vspace{10mm}
\abstract{
In this paper, we consider the time-dependent Navier-Stokes equations in
the half-space $[\xzero,\infty)\times{\bf R}\subset{\bf R}^2$, with
boundary data on the line $x=\xzero$ assumed to be time-periodic (or
stationary) with a fixed asymptotic velocity ${\bf u}_{\infty}=(1,0)$ at
infinity. We show that there exist (locally) unique solutions for all data
satisfying a center-stable manifold compatibility condition in a certain
class of fuctions. Furthermore, we prove that as $x\to\infty$, the
vorticity decompose itself in a dominant stationary part on the parabolic
scale $y\sim\sqrt{x}$ and corrections of order $x^{-\frac{3}{2}+\epsilon}$,
while the velocity field decompose itself in a dominant stationary part in
form of an explicit multiscale expansion on the scales $y\sim\sqrt{x}$ and
$y\sim x$ and corrections decaying at least like
$x^{-\frac{9}{8}+\epsilon}$. The asymptotic fields are made of linear
combinations of universal functions with coefficients depending mildly on the
boundary data. The asymptotic expansion for the component
parallel to ${\bf u}_{\infty}$ contains `non-trivial' terms in the
parabolic scale with amplitude $\ln(x)x^{-1}$ and $x^{-1}$. To first order,
our results also imply that time-periodic wakes behave like stationary ones
as $x\to\infty$.

The class of functions used to prove these results is `natural' in the
sense that the well known `Physically Reasonable' (in the sense of Finn \&
Smith) stationary solutions of the Navier-Stokes equations around an
obstacle fall into that class if the half-space extends in the downstream
direction and the boundary ($x=\xzero$) is sufficiently far downstream. In
that case, the coefficients appearing in the asymptotics can be linearly
related to the net force acting on the obstacle. In particular, the
asymptotic description holds for `Physically Reasonable' stationary
solutions in exterior domains, {\em without restrictions on the size of the
drag acting on the obstacle}. To our knowledge, it is the first time that
estimates uncovering the $\ln(x)x^{-1}$ correction are proved in this
setting.}

\section{Introduction}
In this paper, we consider the time-dependent Navier-Stokes
equations\footnote{Vectors are denoted by boldface letters, generic
positions in the physical space ${\bf R}^2$ are denoted either by
${\bf x}$ or by $(x,y)$.}
\begin{equa}[1][eqn:NavierStokes]
\partial_t{\bf u}+{\bf u}\cdot\nabla{\bf u}&=
\frac{1}{\reynolds}\triangle{\bf u}-\nabla p\\
\nabla\cdot{\bf u}&=0\\
{\bf u}({\bf x},t)|_{x=\xzero}&={\bf u}_b(y,t)\\
\lim_{|{\bf x}|\to\infty}{\bf u}({\bf x},t)&={\bf u}_{\infty}\equiv
\vector{u_{\infty}}{0}
\end{equa}
in the half-space $\Omega_{+}=\Rp\times{\bf R}$. We will restrict
ourselves to the time-periodic setting, i.e. ${\bf u}_b(y,t)=\sum_{n\in{\bf
Z}}\ed^{i n\strouhal t}{\bf u}_{b,n}(y)$, with basic frequency (Strouhal
number) $\strouhal$ and ${\bf u}_{b}\in l^1({\bf Z},{\cal B})$ for some
functional space ${\cal B}$ to be defined later on. Note that with
appropriate scalings, we can set without loss of generality $|{\bf
u}_{\infty}|=u_{\infty}=1$ and $\reynolds=1$. The scale of the Reynolds
number then translates to the scale of ${\bf u}_{b}$.

We consider this problem as a simplified version of the `usual' exterior
problem around an obstacle
\begin{equa}[1][eqn:NavierStokesUsual]
\partial_t{\bf u}+{\bf u}\cdot\nabla{\bf u}&=
\frac{1}{\reynolds}\triangle{\bf u}-\nabla p\\
\nabla\cdot{\bf u}&=0\\
{\bf u}({\bf x},t)|_{\partial\Omega}&=0\\
\lim_{|{\bf x}|\to\infty}{\bf u}({\bf x},t)&={\bf u}_{\infty}\equiv
\vector{u_{\infty}}{0}
\end{equa}
in ${\bf R}^2\backslash\Omega$, where $\partial\Omega$, the obstacle, is
compact and connected. Getting rid of the obstacle by considering the flow
only in the downstream region is a brutal simplification. We hope that by
capturing the main difficulty of (\ref{eqn:NavierStokesUsual}), (the
spatial asymetry introduced by (\ref{eqn:NavierStokesUsual}.4), as seen in
the slow decay of vorticity as $x\to\infty$ for instance), techniques used
in this paper could shed a new light on the theory on the Navier-Stokes
equations (\ref{eqn:NavierStokesUsual}) which began with J. Leray's
pioneering work in the 1930's (see also \cite{Gal94} and references
therein). 

The question we address here is to give a quantitative description of the flow
in the so-called `wake region' which extends downwards of the obstacle
(i.e.\ as $x\to\infty$). In previous papers \cite{Peter2D2003,wake} such a
description has been obtained in the stationary case by assuming that the
restriction of the solution of (\ref{eqn:NavierStokesUsual}) on a given
line $x=x_{0}\gg1$ was in a certain function class. Then it follows from
\cite{Peter2D2003,wake} that as $x\to\infty$, the velocity field ${\bf u}$
and the vorticity $\omega=\nabla\times{\bf u}$ satisfy
\begin{equa}[1][eqn:asympt]
{\bf u}(x,y)&=
{\textstyle
{\bf u}_{\infty}+
\vector{\tilde{u}_{\bf a}(x,y)}{\tilde{v}_{\bf a}(x,y)}
+{\cal O}(x^{-1+\olddelta_0})}
~,~~~~~
\omega(x,y)=\omega_{{\bf a}}(x,y)
+{\cal O}(x^{-\frac{3}{2}+\olddelta_0})~,
\end{equa}
for some $\olddelta_0>0$, where
\begin{equs}
\label{eqn:defuafirst}
\tilde{u}_{{\bf a}}(x,y)&=
{\textstyle\frac{a_1}{\sqrt{x}}}f_0({\textstyle\frac{y}{\sqrt{x}}})
+
{\textstyle\frac{1}{x}}\Big(
a_2g_0({\textstyle\frac{y}{x}})-
a_3g_1({\textstyle\frac{y}{x}})\Big)
\\
\label{eqn:defvafirst}
\tilde{v}_{{\bf a}}(x,y)&={\textstyle\frac{a_1}{2x}}
f_1({\textstyle\frac{y}{\sqrt{x}}})+
{\textstyle\frac{1}{x}}\Big(
a_2g_1({\textstyle\frac{y}{x}})+
a_3g_0({\textstyle\frac{y}{x}})\Big)\\
\omega_{{\bf a}}(x,y)&={\textstyle\frac{a_1}{2x}}
f_1({\textstyle\frac{y}{\sqrt{x}}})
\label{eqn:omegaadef}
\end{equs}
for some ${\bf a}=(a_1,a_2,a_3)$ and
\begin{equs}
f_m(z)&=
{\textstyle
\frac{z^m\ed^{-\frac{z^2}{4}}}{\sqrt{4\pi}}~,~~~~
g_m(z)=\frac{1}{\pi}\frac{z^m}{1+z^2}}~.
\end{equs}
This result was expected to hold for a long time, see e.g.
\cite{Batchelor}. It should be noted that the terms on the $y\sim
x$ scale are of smaller order than the stated correction order. It is
however argued in \cite{Batchelor,Peter2D2003,wake} that on the given
scales ($y\sim x$ or $y\sim\sqrt{x}$) the velocity field indeed converge to
its asymptotic form and furthermore that the upstream asymptotics
($x\to-\infty$ is given by (\ref{eqn:defuafirst}) and
(\ref{eqn:defvafirst}) with $a_1=0$ {\em and the same coefficients} $a_2$
and $a_3$ as in the downstream direction. Integration of the equations
(\ref{eqn:NavierStokesUsual}) in the domain comprised between the lines
$x=-\xzero\ll0$ and $x=\xzero\gg0$ then implies (see e.g.\ Appendix II in
\cite{Peter04}) in the limit $\xzero\to\infty$ that $a_1+2a_2=0$ (mass
conservation), and that the force ${\bf F}$ acting on the obstacle is given
by
\begin{equs}
{\bf F}=\vector{2a_2}{-2a_3}+
\int_{{\bf R}^2\setminus\Omega}
\hspace{-3mm}
{\rm d}{\bf x}~\partial_t{\bf u}({\bf x})
\equiv
\vector{\mbox{drag}}{\mbox{lift}}
~.
\label{eqn:force}
\end{equs}
In other words, for stationary flows, this shows that $a_2$ resp. $a_3$
are linearly related to the drag resp. lift acting on the obstacle, see also
\cite{Batchelor,Peter2D2003,wake} for more physical interpretations. 

For completeness, we note that (\ref{eqn:defuafirst}) and
(\ref{eqn:defvafirst}) can be easily derived heuristically in the two
following steps. We first note that the field $(\tilde{u}_{\bf
a},\tilde{v}_{\bf a})$ with $a_1=0$ would be a solution of the
Navier-Stokes equations (\ref{eqn:NavierStokes}) or
(\ref{eqn:NavierStokesUsual}) (for an appropriate pressure) but for the
boundary conditions. And then, as we may expect
$\partial^2_x\omega\ll\partial_x\omega$ as $x\to\infty$, the vorticity
satisfies (to first order) $\partial_x\omega=\partial_y^2\omega$ whose
solutions corresponding to decaying velocity fields behave asymptotically
like (\ref{eqn:omegaadef}). It is then easy to see using
$\omega\approx-\partial_y u$ and $\partial_y v=-\partial_x u$
that the corresponding velocity fields are as stated in 
(\ref{eqn:defuafirst}) and (\ref{eqn:defvafirst}).

Unfortunately, the function class used in a first attempt to give a
rigorous foundation to these heuristics in \cite{Peter2D2003,wake} is
rather unorthodox and the question wether the (restriction of) solutions of
(\ref{eqn:NavierStokesUsual}) were in this class was completely left open.

On the other hand, it is well known from experiences and numerical
simulations that stationary solutions of (\ref{eqn:NavierStokesUsual}) in
exterior domains are only stable at low Reynolds numbers, and it is commonly
believed (see e.g.\ \cite{Dusek94,Provansal87,Schumm94,Sreenivasan86}) that
at a (first) critical Reynolds number, the stationary flow loses its
stability through a Hopf bifurcation and becomes time-periodic before
eventually leading for larger Reynolds number to von Karman's vortex street
and then to turbulence.

In this paper, we will give a more detailed asymptotic description than
(\ref{eqn:asympt}), as we will prove that in both the stationary and
time-periodic case, the solutions of (\ref{eqn:NavierStokes}) satisfy
\begin{equs}
{\bf u}(x,y,t)={\bf u}_{\infty}+
\vector{u_{{\bf a}(t)}(x,y)}{v_{{\bf a}(t)}(x,y)}
+
{\cal O}\vector{x^{-\frac{9}{8}+\olddelta_0}}
{x^{-\frac{3}{2}+\olddelta_0}}~,~~~~
\omega(x,y)=
\omega_{{\bf a}(t)}(x,y)+{\cal O}(x^{-\frac{3}{2}+\olddelta_0})~,
\label{eqn:asymptfinal}
\end{equs}
uniformly in time, where $0<\olddelta_0<\frac{1}{8}$, ${\bf
a}(t)=(a_1,a_2(t),a_3(t),a_4,a_5,a_6)$ for some constants $a_1$,
$a_4$, $a_5$ and $a_6$ and {\em time periodic functions} $a_2$ and $a_3$,
$\omega_{{\bf a}}$ is as above and
\begin{equa}[1][eqn:defuaomegaa]
u_{{\bf a}(t)}(x,y)&=
{\textstyle\frac{a_1}{\sqrt{x}}}f_0({\textstyle\frac{y}{\sqrt{x}}})
-
{\textstyle\frac{1}{2x}}\Big(
a_5h({\textstyle\frac{y}{\sqrt{x}}})
+(a_6\ln(x)+a_4)f_1({\textstyle\frac{y}{\sqrt{x}}})
\Big)
+
{\textstyle\frac{1}{x}}\Big(
a_2(t)g_0({\textstyle\frac{y}{x}})-
a_3(t)g_1({\textstyle\frac{y}{x}})\Big)
\\
v_{{\bf a}(t)}(x,y)&={\textstyle\frac{a_1}{2x}}
f_1({\textstyle\frac{y}{\sqrt{x}}})+
{\textstyle\frac{1}{x}}\Big(
a_2(t)g_1({\textstyle\frac{y}{x}})+
a_3(t)g_0({\textstyle\frac{y}{x}})\Big)
\end{equa}
where 
\begin{equs}
f_m(z)&=
{\textstyle
\frac{z^m\ed^{-\frac{z^2}{4}}}{\sqrt{4\pi}}~,~~~~
g_m(z)=\frac{1}{\pi}\frac{z^m}{1+z^2}}~,~~~~
h(z)=
{\textstyle
f_0(z)^2
+\frac{1}{8\sqrt{\pi}}
\,z\,{\rm erf}\big({\textstyle\frac{z}{2}}\big)\ed^{-\frac{z^2}{4}}
}~.
\end{equs}
By the use of functional spaces more adapted than in
\cite{Peter2D2003,wake}, we will prove that existing results on
(\ref{eqn:NavierStokesUsual}) implies that (\ref{eqn:asymptfinal}) also
holds for (\ref{eqn:NavierStokesUsual}). This rigorous link between
(\ref{eqn:NavierStokes}) and (\ref{eqn:NavierStokesUsual}) will only be
estabished for the stationary case, as this case has been widely studied in
the litterature (see e.g.\ \cite{Gal94}). Though we believe it should also
hold just after the Hopf bifurcation, we are not aware of any rigorous
treatment of the exterior periodic problem in 2D (see \cite{Sazonov03} for
some rigorous work on the 3D case).

In analogy with the stationary case, we may also expect that for the
solution of (\ref{eqn:NavierStokesUsual}), the asymptotic velocity field
upstream ($x\to-\infty$) is given by (\ref{eqn:defuaomegaa}) with
$a_1=a_4=a_5=a_6=0$ and the same coefficients $a_2(t)$ and $a_3(t)$
than in the downstream direction. If this holds, then integrating 
$\nabla\cdot{\bf u}=0$ and $\omega=\nabla\times{\bf u}$ in the domain
comprised between $x=-\xzero\ll0$ and $x=\xzero\gg0$, we get (letting
$\xzero\to\infty$) $a_2(t)=-\frac{1}{2}a_1$ and $a_3(t)=\int_{{\bf
R}^2\setminus\Omega}\omega({\bf x})\,{\rm d}{\bf x}$. As this last quantity
(the total vorticity) is preserved by (\ref{eqn:NavierStokesUsual}), we see
that $a_2(t)$ and $a_3(t)$ are in fact constant in time\footnote{Note that
it would be wrong to conclude that the drag and lift are constant in time
from the fact that $a_2$ and $a_3$ are constant, as the volume integral of 
$\partial_t{\bf u}$ in (\ref{eqn:force}) will generically no longer be zero
for time-periodic flows. This is in agreement with results of numerical
simulations, see e.g.\ \cite{Homescu02,Posdziech01}.}. This implies that to
the order given by (\ref{eqn:asymptfinal}), time-periodic wakes cannot be
distinguished from stationary ones, though the actual value of the
coefficients will differ from the to case. Without rigorous proof
that the upstream asymptotics are as expected, we consider these physical
interpretations as conjectural ones.

We end this section by noting that asymptotical results like
(\ref{eqn:asympt}) have been successfully used in numerical simulations of 
the stationary Navier-Stokes equations (\ref{eqn:NavierStokesUsual}) in
exterior domains, see \cite{Peter03,Peter04}. In particular, in fixed
simulations domains, it allows to compute drag and lift coefficients with
better accuracy than usual methods, while for fixed accuracy, smaller
simulations domains can be used, thereby reducing significantly the CPU
time needed for these computations. It is our hope that
(\ref{eqn:asymptfinal}) could also be of such use in the time-periodic
setting.

\subsection{Reformulation of the problem}\label{sec:reformulation}

As in \cite{Peter2D2003,wake}, the starting point of the analysis is to
write (\ref{eqn:NavierStokes}) as a dynamical system where $x$ plays the
r\^ole of time. To do so, we write ${\bf u}={\bf u}_{\infty}+{\bf v}$ where
${\bf v}=(u,v)$ and introduce the
vorticity $\omega=\partial_x v-\partial_y u$ and its derivative w.r.t.\ $x$
as $\eta=\partial_x\omega$. Since the boundary data is assumed to be
time-periodic, it is natural to
assume that there is also a (discrete) Fourier decomposition of the various
fields (this corresponds to the so-called {\em global mode} behavior, see
also \cite{Protas02,Wesfreid96,Zielinska95}) given by
\begin{equa}[1][eqn:fourierdec]
u(x,y,t)= \sum_{n\in{\bf Z}}\ed^{i\,n\strouhal\,t}~u_{n}(x,y)~,~~~~
v(x,y,t)= \sum_{n\in{\bf Z}}\ed^{i\,n\strouhal\,t}~v_{n}(x,y)~,\\
\omega(x,y,t)= \sum_{n\in{\bf Z}}\ed^{i\,n\strouhal\,t}~\omega_{n}(x,y)~,~~~~
\eta(x,y,t)= \sum_{n\in{\bf Z}}\ed^{i\,n\strouhal\,t}~\eta_{n}(x,y)~.
\end{equa}
In terms of this decomposition, the $n$-th Fourier coefficient of
(\ref{eqn:NavierStokes}) read (see also \cite{Peter2D2003})
\begin{equa}[1][eqn:sysdynper]
\partial_x\omega&=\eta\\
\partial_x\eta&=\eta-\partial_y^2\omega+i n\strouhal\omega+q\\
\partial_x u&=-\partial_y v\\
\partial_x v&=\partial_y u+\omega~,
\end{equa}
with $q=u\partial_x\omega+v\partial_y\omega$, and where we dropped the $n$
indices on the fields for concision. Namely, the third equation
in (\ref{eqn:sysdynper}) is the incompressibility relation $\nabla\cdot{\bf
u}=0$, the last one is the definition of the vorticity, while substituting
the first one in the second, one recovers the `dynamical' part of
(\ref{eqn:NavierStokes}). We note here that using the incompressibility
relation $\partial_xu=-\partial_y v$ and the definition of the vorticity,
we may cast the nonlinearity $q$ in the following equivalent form
\begin{equs}
q&=\partial_x(u\omega)+\partial_y(v\omega)\equiv\partial_x(\qzero)+\partial_y(
\qun)~.
\end{equs}
We also note that using $\partial_xu=-\partial_y v$ and defining
$\qA=uv$, $\qB={\textstyle\frac{1}{2}}(v^2-u^2)$, we have the
decompositions
\begin{equs}
\qzero&=u\omega=\partial_x \qA+\partial_y \qB~,~~~~~
\qun=v\omega=-\partial_y \qA+\partial_x \qB~~~~~\mbox{ and }~~~~~
q=(\partial_x^2+\partial_y^2)\qA+2\qun~.
\end{equs}
We interpret (\ref{eqn:sysdynper}) as a new dynamical system where the
space variable $x$ plays the role of time (the $x$ derivatives on the
r.h.s.~of (\ref{eqn:sysdynper}) can be eliminated using $\eta-\qzero$
instead of $\eta$ as auxiliary field).

Using Duhamel's variation of constants formula, we can cast
(\ref{eqn:sysdynper}) in an integral form, whose structure (omitting also
the time argument for concision) is given by
\begin{equs}
u(x)&=K_{\kun}(x-\xzero)\myast\Lu\omegazero+
K_{\kzero}(x-\xzero)\uzero
+{\cal F}_{1,u}(x)+{\cal
F}_{2,u}(x)
+\LA \qB(x)-\LB \qA(x)
~,\label{eqn:foromega}\\
v(x)&=K_{\kun}(x-\xzero)\myast\Lv\omegazero+
K_{\kzero}(x-\xzero)\huzero
+{\cal F}_{1,v}(x)+{\cal
F}_{2,v}(x)
-\LA \qA(x)-\LB \qB(x)
~,\label{eqn:foru}\\
\omega(x)&=K_{\kun}(x-\xzero)\myast\omegazero
+{\cal F}_{1,\omega}(x)+{\cal
F}_{2,\omega}(x)
~.\label{eqn:forv}
\end{equs}
The derivation of this integral formulation for the solution of
(\ref{eqn:NavierStokes}) is given in Section \ref{sec:integral}, as well as
precise definitions of the different expressions. The terms involving the
kernels $K_{\kzero}$ and $K_{\kun}$ depend on $\omega$ and $\uzero$, which are
functions defined on the boundary $x=\xzero$, while $\huzero$ is given by
the Hilbert transform $\huzero=\H\uzero$ on the boundary (in particular,
$\huzero$ is {\em not} an independent quantity). The ${\cal F}$
terms depend on the values of $\qzero$ and $\qun$ on
$[\xzero,\infty)\times{\bf R}$ and {\em do not vanish} on the boundary
$x=\xzero$. Thus the integral formulation above does not satisfy the
boundary condition ${\bf u}(\xzero,y,t)={\bf u}_{b}(y,t)$, unless for
specific choices of $\uzero$ and $\omegazero$, but then the boundary
condition for the vorticity cannot be satisfied. This is related to the
fact that the equation (\ref{eqn:NavierStokes}) with a boundary condition
for the vorticity, the velocity and the pressure is ill-posed. The boundary
values on $x=\xzero$ thus have to be taken on the so-called central-stable
manifold. As we will explain in section \ref{sec:discussion}, the
parametrization of that manifold by the functions $\omegazero$ and $\uzero$
is a convenient one.

We will give precise statements of our main results in subsection
\ref{sec:mainresults} below after the definition of some functional spaces
and related norms. On an informal level, our results are twofold. We will
use the integral formulation (\ref{eqn:foromega})-(\ref{eqn:forv}) to
prove that if $\omega$ and $\uzero$ are in a certain class ${\cal C}_{i}$,
there exist a (locally) unique solution of (\ref{eqn:NavierStokes}) in the
Banach space $\W$ defined in the next section. We will then show that the
asymptotic structure of these solutions is indeed given by
(\ref{eqn:asymptfinal}) with $\olddelta_0>0$. On the other hand,
time-periodic solutions of (\ref{eqn:NavierStokesUsual}) must satisfy
(\ref{eqn:NavierStokes}) for all $\xzero$ sufficiently large. We will then
show that for solutions of (\ref{eqn:NavierStokesUsual}) in a certain class
${\cal C}_u$, the ${\cal F}$'s are well defined, and thus solutions of
(\ref{eqn:NavierStokesUsual}) in ${\cal C}_u$ must also
satisfy (\ref{eqn:foromega})-(\ref{eqn:forv}) for certain functions
$\omegazero$ and $\uzero$. As $K_{\kzero}(0)=K_{\kun}(0)=1$, the functions
$\omegazero$ and $\uzero$ can be determined by inverting any two of the
three (linear and local) relations (\ref{eqn:foromega})-(\ref{eqn:forv}) at
$x=\xzero$, the third relation, which correspond to the central-stable
manifold condition in the dynamical system formulation
(\ref{eqn:sysdynper}), is then automatically satisfied {\em since we know
that the solution exist}. We will then show that the functions
$\omegazero$ and $\uzero$ obtained in this way are in the class ${\cal
C}_i$, which finally implies that solutions of
(\ref{eqn:NavierStokesUsual}) in ${\cal C}_u$ also satisfy 
(\ref{eqn:asympt}) with $\olddelta_0>0$.

\subsection{Definitions}\label{sec:functionalspaces}
To state our main result, we need some definitions. We first give the
topology we will use to control the decompositions (\ref{eqn:fourierdec}).
Let $1\leq p<\infty$, and $f(x,y,t)=\sum_{n\in{\bf Z}}\ed^{i n\strouhal
t}f_{n}(x,y)$ for $(x,y,t)\in\Rp\times{\bf
R}\times[0,\frac{2\pi}{\strouhal}]$. Let $\chinese{x}=\sqrt{1+x^2}$, we
define
\begin{equs}
\|f\|_{\Ltx{p}{\sigma}}&=\sup_{x\geq\xzero}\|f\|_{\Lx{x}{p}{\sigma}}~,~~~
\|f\|_{\Lx{x}{p}{\sigma}}
=
\chinese{x}^{\sigma}\|f(x)\|_{p}
=\chinese{x}^{\sigma}\sum_{n\in{\bf Z}}
\Big(
\int_{{\bf R}}
\hspace{-1mm}{\rm d}y~|f_n(x,y)|^p
\Big)^{\frac{1}{p}}~,\\
\|f\|_{\Ltx{\infty}{\sigma}}&=
\sup_{x\geq\xzero}\|f\|_{\Lx{x}{\infty}{\sigma}}~,~~~
\|f\|_{\Lx{x}{\infty}{\sigma}}
=
\chinese{x}^{\sigma}
\|f(x)\|_{\infty}
=\chinese{x}^{\sigma}\sum_{n\in{\bf Z}}
\essup_{y\in{\bf R}}|f_n(x,y)|~,\\
\|f\|_{\Ltx{p}{\{\sigma_1,\sigma_2\}}}
&=\sup_{x\geq0}
\chinese{x}^{-\sigma_1}
x^{\sigma_2}
\|f(x)\|_{\L^p}~,~~~~
\|f(x)\|_{\L^p}=
\sup_{n\in{\bf Z}}
\Big(
\int_{{\bf R}}
\hspace{-1mm}{\rm d}y~|f_n(x,y)|^p
\Big)^{\frac{1}{p}}~,\\
\|f\|_{\Ltx{\infty}{\{\sigma_1,\sigma_2\}}}&=
\sup_{x\geq0}
\chinese{x}^{-\sigma_1}
x^{\sigma_2}
\|f(x)\|_{\L^{\infty}}~,~~~~
\|f(x)\|_{\L^{\infty}}=
\sup_{n\in{\bf Z}}
\essup_{y\in{\bf R}}|f_n(x,y)|~,
\end{equs}
where we use the notation $\|f(x)\|_{p}$ as a shorthand to
the more rigorous $\|f(x,\cdot)\|_{p}$. We will refer to $\P$ and $\Q$ as
the projection operators on Fourier series defined by
\begin{equs}
\P\Big(\sum_{n\in{\bf Z}}\ed^{i n\strouhal t}f_{n}\Big)
=\sum_{n\in{\bf Z},n\neq0}\ed^{i n\strouhal t}f_{n}~,~~~~
\Q\Big(\sum_{n\in{\bf Z}}\ed^{i n\strouhal t}f_{n}\Big)
=f_{0}~,
\end{equs}
as well as the operators $\I$ (the `primitive'), $\S$ (the symmetrization)
and $\M$ (the `mean value') defined by
\begin{equs}
(\I f)(y)&=
\int_{-\infty}^{y}
\hspace{-3mm}{\rm d}z~
\frac{f(z)}{2}-
\int_{y}^{\infty}
\hspace{-3mm}{\rm d}z~
\frac{f(z)}{2}~,
\label{eqn:defI}~~~~
(\S f)(y)=f(y)+f(-y)~,~~~~
\M(f)=\int_{{\bf R}}f(y){\rm d}y~.
\end{equs}
Note that $\I$ is the inverse of $\partial_y$ (when it is defined). We can
now specify our basic functional space
\begin{definition}\label{def:fctspaces}
Let ${\cal C}_0^{\infty}=\left\{\{(u_n,v_n,\omega_n)\}_{n\in{\bf Z}}~\mbox{s.t.}~
(u_n,v_n,\omega_n)\in
{\cal C}_0^{\infty}(\Rp\times{\bf R},{\bf R}^3)~\forall n\in{\bf
Z}\right\}$. We denote by $\W$ the Banach space obtained by closure of
${\cal C}_0^{\infty}$ under the norm
\begin{equs}
\|\triplet\|&=\sup_{x\geq\xzero}\|\triplet\|_{x}\\
\|\triplet\|_{x}&=
\|u\|_{\Lx{x}{\infty}{\frac{1}{2}}}+
\|u\|_{\Lx{x}{\pu}{\frac{1}{2}-\frac{1}{\pu}}}+
\|\partial_yu\|_{\Lx{x}{\pdv}{1-\frac{1}{2\pdv}-\au}}
\\
&\phantom{=~}+
\|v\|_{\Lx{x}{\infty}{1-\olddelta}}+
\|v\|_{\Lx{x}{\pv}{1-\olddelta-\frac{1}{\pv}}}+
\|\partial_yv\|_{\Lx{x}{\pdv}{\frac{3}{2}-\frac{1}{2\pdv}-\av}}
\\&\phantom{=~}+
\|\omega\|_{\Lx{x}{2}{\frac{3}{4}}}+
\||y|^{\beta}\omega\|_{\Lx{x}{2}{\frac{3}{4}-\frac{\beta}{2}}}+
\|\partial_y\omega\|_{\Lx{x}{\infty}{\frac{3}{2}}}
+\|\partial_y\omega\|_{\Lx{x}{1}{1}}
~.
\end{equs}
\end{definition}
This choice is discussed at the end of this section. Note at this point
that the `expected' asymptotic decomposition (\ref{eqn:asympt}) is in $\W$
if $p>1$. We now specify the class of solutions of
(\ref{eqn:NavierStokesUsual}) for which our results can be applied:
\begin{definition}\label{def:expected}
A solution $\triplet$ of (\ref{eqn:NavierStokesUsual}) is in the class
$\Cu$ if $\|\triplet\|\leq\rho$ for some finite constant
$\rho$ and 
\begin{equa}[1][eqn:restrictions]
{\textstyle
\frac{13}{7}}&{\textstyle\leq\beta\leq3~,~~~~
1-\frac{1}{p}<\olddelta<\olddeltam~,~~~~
1<\pv\leq\pu~,~~~~
\pdv>2~,~~~~
\frac{1}{2}\geq\av\geq\au\geq0~,}\\
{\textstyle
\av}&{\textstyle\geq\olddelta~,~~~
\frac{1}{4}-\frac{\olddelta}{2}-\au
>0~,~~~~
\frac{1}{2}-\big(1+\frac{1}{2\pdv}\big)\olddelta>0~,~~~~
\frac{1}{2}+\av-\au-2\olddelta>0~,}\\
{\textstyle\frac{\chinese{\strouhal}}{\strouhal}}
&\leq\chinese{\xzero}^{\olddelta}~,~~~~~
{\textstyle
\frac{1}{2}+\au-\av-\frac{\olddelta}{\pdv}>0
}~.
\end{equa}
\end{definition}
The condition
$\frac{\chinese{\strouhal}}{\strouhal}\leq\chinese{\xzero}^{\olddelta}$
will be a convenient way to
get bounds depending on $\xzero$ only and not on the Strouhal number
$\strouhal$. It should be noted that this condition is only restrictive in
the limit of vanishing Strouhal number. This is {\em not} expected to occur
for time-periodic solutions of (\ref{eqn:NavierStokesUsual}), if the Hopf
bifurcation picture of \cite{Dusek94,Provansal87,Schumm94,Sreenivasan86} is
correct.

We now define the class ${\cal C}_i$, consisting essentially of those
functions $\omegazero$, $\uzero$ and $\huzero$ for which the part of
r.h.s.\ of (\ref{eqn:foromega})--(\ref{eqn:forv}) depending on
$\omegazero$, $\uzero$ and $\huzero$ is in $\W$ (see Lemma
\ref{lem:evollmforomega}, \ref{lem:evollmforu} and \ref{lem:thescaryone}):
\begin{definition}\label{def:leszeros}
We say that $\uzero$ and $\omegazero$ are in the class $\Ci$ if
$\M(\Q\omegazero)=0$ and $\|(\uzero,\huzero,\omegazero)\|_{\xzero}\leq\rho$
for parameters satisfying (\ref{eqn:restrictions}).
\end{definition}
Note that $\M(\Q\omegazero)$ is always well defined if
$\|(0,0,\omegazero)\|_{\xzero}<\infty$ since
$\|\omegazero\|_{\L^1}\leq\chinese{\xzero}^{-\frac{1}{2}}\|(0,0,\omegazero)
\|_{\xzero}$ (see Lemma \ref{lem:Lun}). In the case of symmetric flows
(i.e.\ $u$ even in $y$ and $v$ odd in $y$), $\M(\Q\omegazero)=0$ is
a trivial consequence of the fact that $\omegazero$ is an odd function of
$y$ (it is also expected from (\ref{eqn:asympt}) or
(\ref{eqn:asymptfinal})). Our results will in particulat imply that the
vorticity decomposes itself in a first order part with zero mean value with
second order corrections with generically non-zero mean value (see
(\ref{eqn:foromega})).

We end this section by making some comments on Definition
\ref{def:fctspaces}. First, for the $v$ component, we will need $\olddelta>0$.
Namely, as we will see, the optimal decay rate for $v$ as $x\to\infty$ can only
be obtained if $\huzero\in\L^{1}({\bf R},{\rm d}y)$, since the integral
expression for $v$ contains a convolution of $\huzero$ with
$K_{\kzero}(x-\xzero,y)=\frac{1}{\pi}\frac{x-\xzero}{(x-\xzero)^2+y^2}$.
But (apart from symetric flows), $\huzero(y)\sim1/y$ as $y\to\infty$ (see
(\ref{eqn:asympt})), so in general $\huzero\not\in\L^{1}({\bf R},{\rm
d}y)$.

The second comment is on the need of $\au$ and $\av$.
Basically, the problem is that $\partial_yu$ and
$\partial_yv$ are naturally build of sum of functions on two length scales
($y\sim\sqrt{x}$ and $y\sim x$, see e.g.\ (\ref{eqn:defuaomegaa})). Dependence on $\pdv$ of the decay
exponents as $x\to\infty$ of $\L^{\pdv}$ norms of such functions either
vary like $1/(2\pdv)$ for functions on the shorter scale or like $1/\pdv$
for functions on the longer one. Our choice of exponents are thus `wrong'
on the scale $y\sim x$ and is `corrected' by introducing $\au$ and
$\av$. These additional parameters would not be needed if we choose
$\pdv=\infty$, but in that case we would lose the boundedness of the
Dirichlet-Neumann operator (or exchange operator) ${\bf v}\to\H{\bf v}$ in
$\W$, which is needed to compare solutions of (\ref{eqn:NavierStokes}) and
(\ref{eqn:NavierStokesUsual}) (see Section \ref{sec:discussion}).

The last comment is that if (\ref{eqn:restrictions}) holds, we control
the nonlinearities $\qA$, $\qB$, $\qzero$ and $\qun$ in terms of the
$\|\cdot\|$-norm by
\begin{equa}[1][eqn:nonlinbounds]
\|\qA\|_{m,\frac{3}{2}-\olddelta-\frac{1}{m}}+
\|\qB\|_{m,1-\frac{1}{m}}+
\|\partial_y\qA\|_{\pdv,\frac{3}{2}-\au-\frac{1}{2\pdv}}+
\|\partial_y\qB\|_{\pdv,\frac{3}{2}-\au-\frac{1}{2\pdv}}
&\leq
C\|\triplet\|^2~,
\\
\|\qzero\|_{\Ltx{m}{\frac{3}{2}-\frac{1}{2m}}}+
\|\qun\|_{\Ltx{m}{2-\olddelta-\frac{1}{2m}}}+
\|\partial_y\qzero\|_{\Ltx{n}{2-\frac{1}{2n}-\au}}+
\|\partial_y\qun\|_{\Ltx{n}{\frac{5}{2}-\frac{1}{2n}-\av}}
&\leq
C\|\triplet\|^2~,\\
\||y|^{\beta}\qzero\|_{\Ltx{2}{\frac{5}{4}-\frac{\beta}{2}}}+
\||y|^{\beta}\qun\|_{\Ltx{2}{\frac{7}{4}-\olddelta-\frac{\beta}{2}}}
&\leq
C\|\triplet\|^2~,
\end{equa}
for all $1\leq m\leq\infty$ and $1\leq n\leq\pdv$. To establish these
estimates, we used
\begin{equs}
\|\omega\|_{\Ltx{\infty}{1}}+
\|\omega\|_{\Ltx{1}{\frac{1}{2}}}&\leq\|\triplet\|~,
\label{eqn:simplificatrice}
\end{equs}
which follows (see Lemma \ref{lem:Lun}) from
\begin{equs}
\|\omega\|_{\L^{1}}\leq
C_{\beta}
\|\omega\|_{\L^{2}}^{1-\frac{1}{2\beta}}
\||y|^{\beta}\omega\|_{\L^{2}}^{\frac{1}{2\beta}}~~~~\mbox{ and }~~~~
\|\omega\|_{\L^{\infty}}\leq\sqrt{\|\omega\|_{\L^2}\|\partial_y\omega\|_{\L
^2}}~.
\end{equs}

\subsection{Main results}\label{sec:mainresults}

We are now in position to state our results in a precise manner.
The first one states that the topology of Definition
\ref{def:fctspaces} is well adapted to (\ref{eqn:NavierStokes}):
\begin{theorem}\label{thm:equivalence}
If $\xzero$ is sufficiently large, then the two following statements are
equivalent
\begin{enumerate}
\item\label{it:un} There exist a unique solution to
(\ref{eqn:NavierStokes}) in $\Cu$ with parameters satisfying
(\ref{eqn:restrictions}).
\item\label{it:deux} $\uzero$ and $\omegazero$ are in the class $\Ci$
with parameters satisfying (\ref{eqn:restrictions}).
\end{enumerate}
Furthermore if \ref{it:un}.\ holds and additionally
\begin{equs}
\||y|^{\frac{1}{2}}{\bf v}(\xzero)\|_{4}
+\||y|^{\frac{1}{2}-(1+\epsilon)\olddelta}\S{\bf v}(\xzero)\|_{1}\leq
C(\xzero,\|\triplet\|)~,
\end{equs}
then for all $\epsilon>0$, it holds
\begin{equs}
\||y|^{\frac{1}{2}-(1+\epsilon)\olddelta}\S\uzero\|_{1}+
\||y|^{\frac{1}{2}-(1+\epsilon)\olddelta}\S\huzero\|_{1}
&\leq C_1(\xzero,\triplet\|)~.
\label{eqn:additionalsymmetric}
\end{equs}
\end{theorem}
Our next result is that stationary solutions to the `usual' exterior
problem (\ref{eqn:NavierStokesUsual}) are in the class $\Cu$:
\begin{proposition}\label{prop:fromphysically}
For any stationary solution of (\ref{eqn:NavierStokesUsual}) "Physically
Reasonable" (PR) in the sense of Finn and Smith (see e.g.
\cite{PR1,Gal94,PRgaldi}),
the fields $u$, $v$ and $\omega$ satisfy $\|\triplet\|\leq C$ with
parameters satisfying (\ref{eqn:restrictions}) if $\xzero$ is sufficiently
large. Furthermore 
$\||y|^{\frac{1}{2}}{\bf v}(\xzero)\|_4
+\||y|^{\frac{1}{2}-(1+\epsilon)\olddelta}\S{\bf v}(\xzero)\|_1<\infty$ for
all $\epsilon>0$.
\end{proposition}
From this, we conclude that (PR) solutions satisfy the integral
formulation (\ref{eqn:foromega})-(\ref{eqn:forv}):
\begin{corollary}\label{cor:onthesymmetricpart}
For any (PR) stationary solution of (\ref{eqn:NavierStokesUsual}), $\uzero$
and $\omegazero$ are in the class $\Ci$ with parameters
satisfying (\ref{eqn:restrictions}) and
$\||y|^{\frac{1}{2}-(1+\epsilon)\olddelta}\S\uzero\|_1
+\||y|^{\frac{1}{2}-(1+\epsilon)\olddelta}\S\huzero\|_1<\infty$ for
all $\epsilon>0$.
\end{corollary}
The proof of this Corollary follows directly from Theorem
\ref{thm:equivalence}, Proposition \ref{prop:fromphysically} and
uniqueness. Once these results are proved, we will use the integral
formulation (\ref{eqn:foromega})-(\ref{eqn:forv}) to get the asymptotic
structure of the solutions to (\ref{eqn:NavierStokes}) or (PR) solutions to
(\ref{eqn:NavierStokesUsual}):
\begin{corollary}\label{cor:asymptotics}
Let ${\bf a}_1=(-\M(\I\Q\omegazero)-\int_{\Omega_{+}}\Q\qun(x,y)\,
{\rm d}x{\rm d}y,0,0)$, then all solutions to (\ref{eqn:NavierStokes}) in
$\Cu$ satisfy (\ref{eqn:asympt}) with
$\olddelta_0=(1+\epsilon)\olddelta>\olddelta$.
\end{corollary}
Note that since $\olddelta>0$, in (\ref{eqn:asympt}), the terms containing
$a_2$ and $a_3$ are of smaller order than the remainder, which explains why
these parameters are not specified in the Corollary. Once this Corollary is
proved, we will be able to get the more precise asymptotic form as shows
the
\begin{corollary}\label{cor:asymptoticsrefined}
Assume that $\||y|^{\frac{1}{2}}{\bf v}(\xzero)\|_4<\infty$ and
$\||y|^{\frac{1}{2}-(1+\epsilon)\olddelta}\S\uzero\|_1+
\||y|^{\frac{1}{2}-(1+\epsilon)\olddelta}\S\huzero\|_1<\infty$, and let 
$a_1=-\M(\I\Q\omegazero)-\int_{\Omega_{+}}\Q\qun(x,y)\,{\rm d}x{\rm d}y$, 
$a_2=\M(\S\uzero)-\int_{\Omega_{+}}\Q\qun(x,y)\,{\rm d}x{\rm d}y$, $a_3=\M(\S\huzero)$ and
$\olddelta<\olddelta_0=(1+\epsilon)\olddelta<\frac{1}{8}$. Then there exist
a constant $a_4$ such that all solutions to
(\ref{eqn:NavierStokes}) in $\Cu\subset\W$ satisfy (\ref{eqn:asymptfinal})
with ${\bf a}=(a_1,a_2,a_3,a_4,a_1^2,a_1\Q a_3)$, 
in $\|\cdot\|_{\infty}$ for ${\bf u}$, and in
$\|\cdot\|_{\infty}+\|\cdot\|_{1}+\||y|^{1-2(1+\epsilon)\olddelta}\cdot\|_{2}$ for $\omega$.
\end{corollary}
We conclude this section by noting that the constant $a_1$ can be expressed
in the following equivalent forms:
\begin{remark}\label{rem:equivalentaun}
The constant $a_1$ is also given by the value of the following
{\em constant} function
\begin{equs}
\tilde{a}_1(x)&=
\M\Big(
\I\Big(
\Q\omega(x)
+\int_x^{\infty}
\hspace{-3mm}
{\rm d}\tilde{x}~\ed^{x-\tilde{x}}\Q\qzero(\tilde{x})
\Big)
+
\int_x^{\infty}
\hspace{-3mm}
{\rm d}\tilde{x}~(\ed^{x-\tilde{x}}-1)\Q\qun(\tilde{x})
\Big)~,\\
&=
\M\Big(
\I\Big(
\Q\omega(x)+\int_x^{\infty}
\hspace{-3mm}
{\rm d}\tilde{x}~\ed^{x-\tilde{x}}\Q\qzero(\tilde{x})
\Big)+
\int_x^{\infty}
\hspace{-3mm}
{\rm d}\tilde{x}~\ed^{x-\tilde{x}}\Q\qun(\tilde{x})
+
\Q\qB(x)
\Big)~,
\end{equs}
which is `almost local' in $x$ due to the exponential factors.
\end{remark}

\subsection{Structure of the paper}\label{sec:struct}

Our first task in the remainder of this paper is to establish the integral
formulation (\ref{eqn:foromega})-(\ref{eqn:forv}). This is done in the next
section (Section \ref{sec:integral}). The proof of Theorem
\ref{thm:equivalence} is then split in two parts. The proof that
(\ref{eqn:additionalsymmetric}) holds is delayed until Section
\ref{sec:discussion} together with the proof that \ref{it:un}.\ implies
\ref{it:deux}. The converse is given in Section \ref{sec:existeunique}. The
proof of Proposition \ref{prop:fromphysically} is also delayed until Section
\ref{app:checkingwithgaldi}. Finally, the proof of Corollary
\ref{cor:asymptotics} is given in Section \ref{sec:asymptotics}, that of
Corollary \ref{cor:asymptoticsrefined} in Section
\ref{sec:asymptoticsrefined}, while the proof that Remark
\ref{rem:equivalentaun} holds is left to the reader
as it follows very easily from the integral formulation in Fourier space
given in the next section.

\section{Integral formulation}\label{sec:integral}

We now derive the integral formulation
(\ref{eqn:foromega})-(\ref{eqn:forv}) of the solution of
(\ref{eqn:sysdynper}) and (\ref{eqn:NavierStokes}). All the material of
this section is very similar to \cite{Peter2D2003,wake} where the case
$\strouhal=0$ was treated. For completeness, we now reproduce some of the
analysis here, encompassing the additional term proportional to the
Strouhal number $\strouhal$.

We first note that performing a (continuous) Fourier transform\footnote{we
distinguish functions and their Fourier transform only
from their arguments `$k$', resp.\ `$y$' in Fourier resp.\ direct
space.}
$f(k)=\int_{\bf R}\ed^{iky}f(y)$ leads to a system of the form
$\partial_x{\bf z}=L{\bf z}+{\bf q}$, with
${\bf z}=(\omega,\eta,u,v)$, ${\bf q}=(0,q,0,0)$ and
\begin{equs}
L(k)=
\left(
\begin{matrix}   
0 & 1 & 0 & 0\\
k^2+in\strouhal& 1 & 0 & 0\\
0&0&0&ik\\
1&0&-ik&0
\end{matrix}\right)~.
\end{equs}
As in \cite{Peter2D2003}, the matrix $L$ can be diagonalized. Namely,
define $\sigma(k)={\rm sign}(k)$, $\Lambdazero=\sqrt{1+4(k^2+in\strouhal)}$
and $\Lambda_{\pm}=\frac{1\pm\Lambdazero}{2}$, and set ${\bf z}=S{\bf y}$, with 
${\bf y}=(\omega_{+},\omega_{-},u_{+},u_{-})$ and
\begin{equs}					   
S(k)=
\left(
\begin{matrix}   
1 & 1 & 0 & 0\\
\Lambdap& \Lambdam & 0 & 0\\
\frac{ik}{\Lambdap+in\strouhal} & 
\frac{ik}{\Lambdam+in\strouhal} &
1&1\\
\frac{\Lambdap}{\Lambdap+in\strouhal} & 
\frac{\Lambdam}{\Lambdam+in\strouhal} &
-i\sigma&i\sigma
\end{matrix}\right)
~,
\end{equs}
\begin{figure}[t]
\unitlength=1mm
\begin{center}
\begin{picture}(0,0)(0,0)
\put(-65,-70){\psfig{file=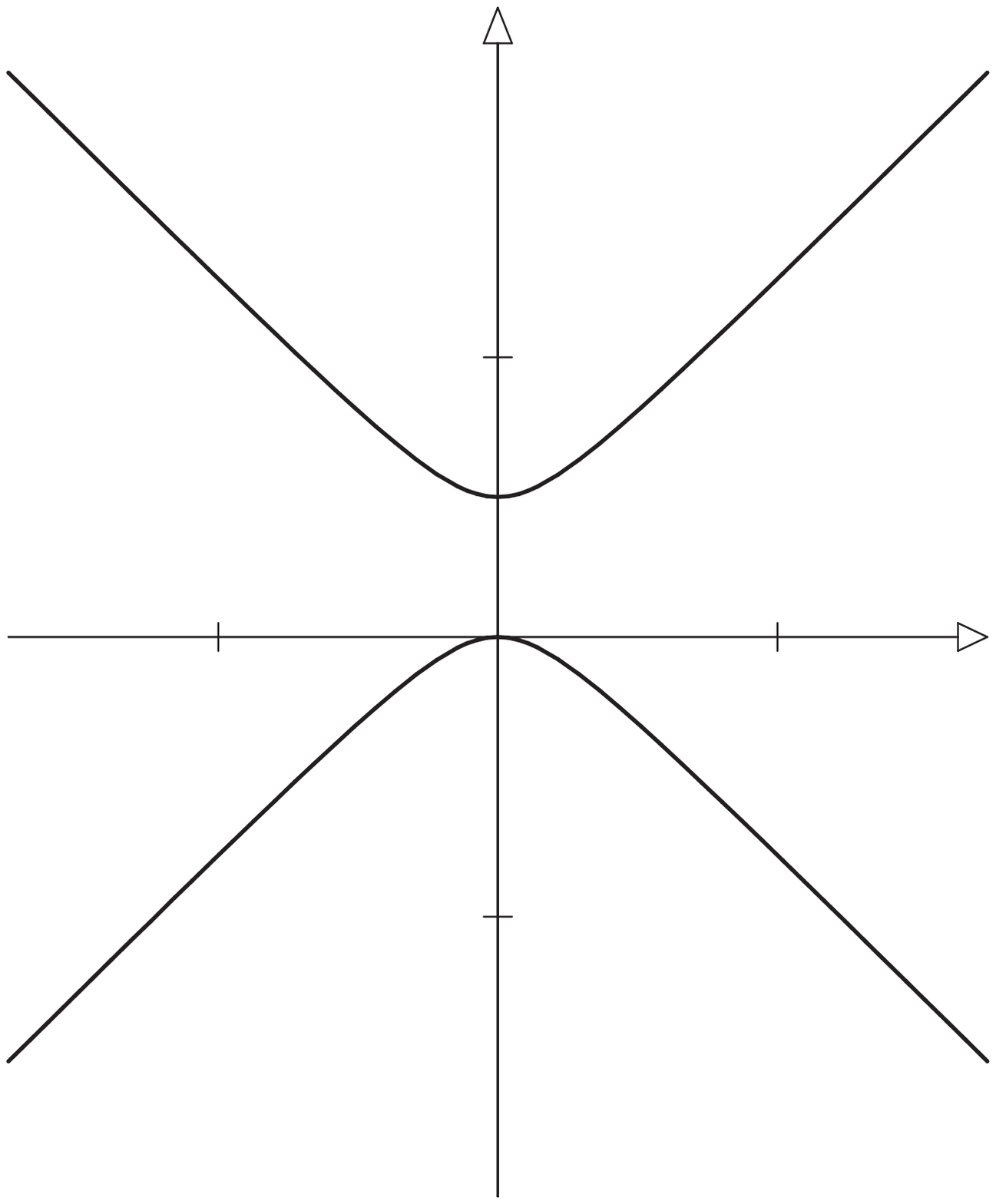,width=6cm}}
\put(5,-70){\psfig{file=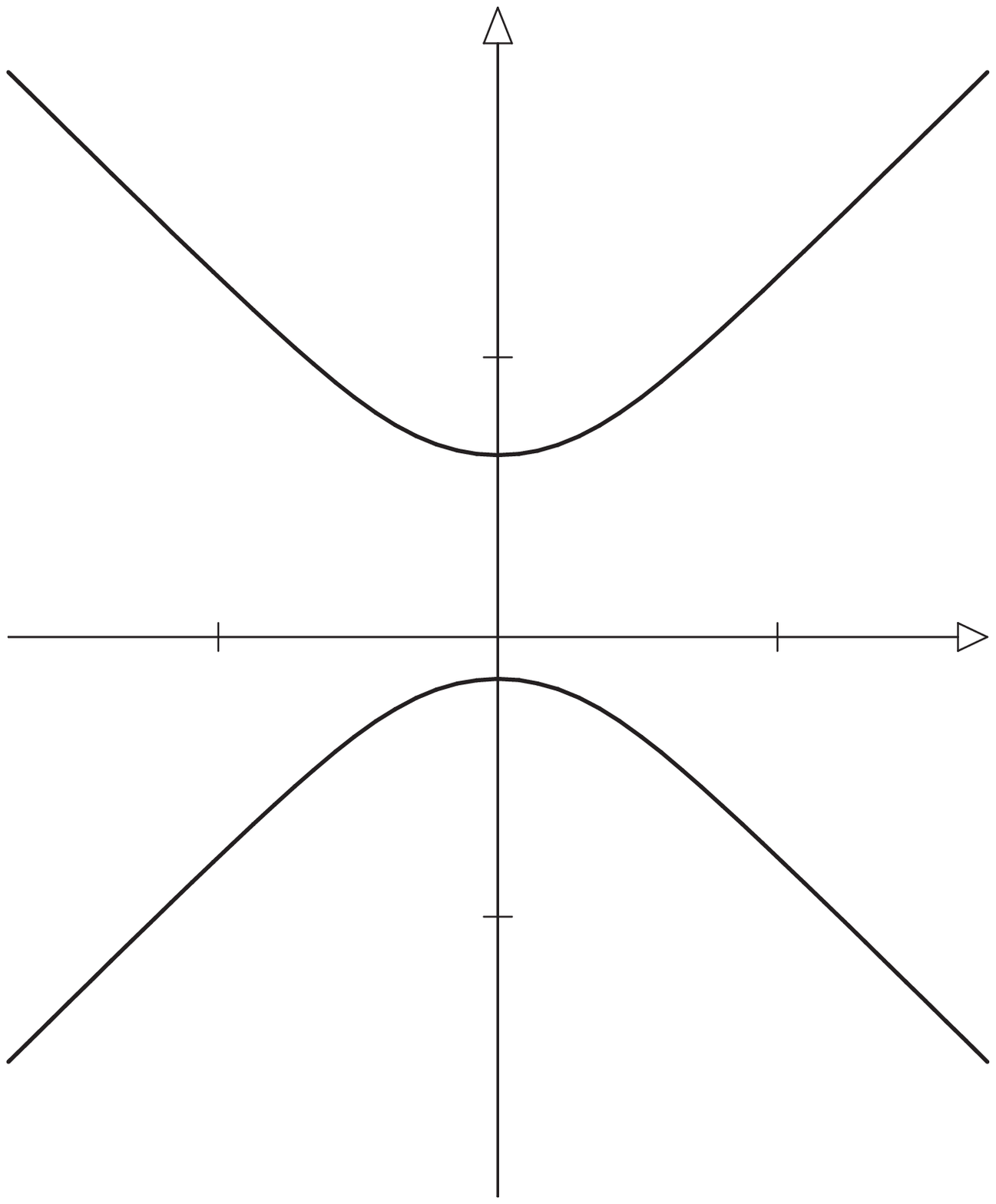,width=6cm}}
\end{picture}
\end{center}
\begin{center}
\begin{picture}(100,60)(0,0)
\put(-7,58){\small ${\rm Re}(\Lambdap)$}
\put(-15,17){\small ${\rm Re}(\Lambdam)$}
\put(63,58){\small ${\rm Re}(\Lambdap)$}
\put(55,17){\small ${\rm Re}(\Lambdam)$}
\put(44,25){\small $k$}
\put(114,25){\small $k$}
\put(31,25){\small $2$}
\put(-6,25){\small $-2$}
\put(101,25){\small $2$}
\put(64,25){\small $-2$}
\put(11,45){\small $2$}
\put(8,11){\small $-2$}
\put(81,45){\small $2$}
\put(78,11){\small $-2$}
\end{picture}
\end{center}
\label{fig:thefigure}
\setcaptionwidth{110mm}
\caption{Dispersion relations $\Lambda_{\pm}$ as a function of wavelength
$k$, with $n\strouhal=0$ in left panel, and $n\strouhal=1$ in right panel.}
\end{figure}%
then we get $S^{-1}LS={\rm diag}(\Lambdap,\Lambdam,|k|,-|k|)$
(see figure \ref{fig:thefigure} for a graphical display of the dispersion
relations $\Lambda_{\pm}$). The real part of $\Lambdap$ being positive, the
two equations corresponding to the `$+$' modes are linearly unstable. We
thus integrate these modes backwards from $x=\infty$, where we set them to
$0$ (see also \cite{wake}). We then get
\begin{equs}
\omega_{+}(x)&=
-
\int_{x}^{\infty}\hspace{-4mm}{\rm d}\tilde{x}
\frac{\ed^{\Lambdap(x-\tilde{x})}}{\Lambdazero}q(\tilde{x})\\
\omega_{-}(x)&=\ed^{\Lambdam(x-\xzero)}\tilde{\omega}_0
-
\int_{\xzero}^{x}\hspace{-3mm}{\rm d}\tilde{x}~
\frac{\ed^{\Lambdam(x-\tilde{x})}}{\Lambdazero}q(\tilde{x})\\
u_{+}(x)&=
-\frac{1}{2}
\int_{x}^{\infty}\hspace{-4mm}{\rm d}\tilde{x}~
\frac{\ed^{|k|(x-\tilde{x})}}{ik-n\strouhal\sigma}q(\tilde{x})\\
u_{-}(x,k)&=\ed^{-|k|(x-\xzero)}\tilde{u}_0
+\frac{1}{2}
\int_{\xzero}^{x}\hspace{-3mm}{\rm d}\tilde{x}~
\frac{\ed^{-|k|(x-\tilde{x})}}{ik+n\strouhal\sigma}q(\tilde{x})~,
\end{equs}
for some functions $\tilde{\omega}_0$ and $\tilde{u}_0$ to be specified.
Integrating by parts the integrals involving $\partial_x\qzero$ in
$\omega_{\pm}$, replacing $q=(\partial_x^2+\partial_y^2)\qA+2\partial_y\qun$
in $u_{\pm}$ and integrating twice by parts the term involving
$\partial_x^2\qA$, we find
\begin{equs}
\omega_{+}(x)&=
\frac{\qzero(x)}{\Lambdazero}
-\int_{x}^{\infty}\hspace{-4mm}{\rm d}\tilde{x}~
\frac{\ed^{\Lambdap(x-\tilde{x})}}{\Lambdazero}q_{+}(\tilde{x})\\
\omega_{-}(x)&=\ed^{\Lambdam(x-\xzero)}\omegazero
-\frac{\qzero(x)}{\Lambdazero}
-\int_{\xzero}^{x}\hspace{-3mm}{\rm d}\tilde{x}~
\frac{\ed^{\Lambdam(x-\tilde{x})}}{\Lambdazero}q_{-}(\tilde{x})\\
u_{+}(x)&=
\frac{P(x)+ik\qB(x)+|k|\qA(x)}{2(ik-n\strouhal\sigma)}+
\int_{x}^{\infty}\hspace{-4mm}{\rm d}\tilde{x}~
\frac{ik\ed^{|k|(x-\tilde{x})}}{ik-n\strouhal\sigma}\qun(\tilde{x})\\
u_{-}(x)&=\ed^{-|k|(x-\xzero)}\uzero
+\frac{P(x)+ik\qB(x)-|k|\qA(x)}{2(ik+n\strouhal\sigma)}
-\int_{\xzero}^{x}\hspace{-3mm}{\rm d}\tilde{x}~
\frac{ik\ed^{-|k|(x-\tilde{x})}}{ik+n\strouhal\sigma}\qun(\tilde{x})~,
\end{equs}
where $q_{\pm}=\Lambda_{\pm}\qzero-ik\qun$, 
$\uzero(k)=\tilde{u}_0(k)-\frac{P(\xzero)+ikB(\xzero)-|k|A(\xzero)}{2(ik+n\strouhal\sigma)}$ and
$\omegazero(k)=\tilde{\omega}_0(k)+\frac{\qzero(\xzero)}{\Lambdazero}$. For
convenience, we also introduce $\huzero(k)=i\sigma\uzero(k)$. Then, a
little algebra shows that when reconstructing $\omega$, $u$ and $v$, the
terms involving $\qzero(x)$ cancel out exactly, giving
\begin{equs}
\omega(x)&=\ed^{\Lambdam(x-\xzero)}\omegazero
-
\int_{\xzero}^{x}\hspace{-3mm}{\rm d}\tilde{x}~
\frac{\ed^{\Lambdam(x-\tilde{x})}}{\Lambdazero}q_{-}(\tilde{x})
-
\int_{x}^{\infty}\hspace{-4mm}{\rm d}\tilde{x}~
\frac{\ed^{\Lambdap(x-\tilde{x})}}{\Lambdazero}q_{+}(\tilde{x})
\label{eqn:foromegafourier}
\\
u(x)&=\frac{ik\ed^{\Lambdam(x-\xzero)}}{\Lambdam+in\strouhal}\omegazero
+\ed^{-|k|(x-\xzero)}\uzero+
\frac{k^2\qB(x)-|k|n\strouhal \qA(x)}{k^2+(n\strouhal)^2}
\\&\phantom{=}
-\frac{ik}{\Lambdazero}
\int_{\xzero}^{x}\hspace{-3mm}{\rm d}\tilde{x}~
\frac{\ed^{\Lambdam(x-\tilde{x})}}{\Lambdam+in\strouhal}q_{-}(\tilde{x})
-\frac{ik}{\Lambdazero}
\int_{x}^{\infty}\hspace{-4mm}{\rm d}\tilde{x}~
\frac{\ed^{\Lambdap(x-\tilde{x})}}{\Lambdap+in\strouhal}q_{+}(\tilde{x})
\\&\phantom{=}
-
\int_{\xzero}^{x}\hspace{-3mm}{\rm d}\tilde{x}~
\frac{ik\ed^{-|k|(x-\tilde{x})}}{ik+n\strouhal\sigma}\qun(\tilde{x})
+
\int_{x}^{\infty}\hspace{-4mm}{\rm d}\tilde{x}~
\frac{ik\ed^{ |k|(x-\tilde{x})}}{ik-n\strouhal\sigma}\qun(\tilde{x})\label{eqn:forubf}\\
v(x)&=\frac{\Lambdam\ed^{\Lambdam(x-\xzero)}}{\Lambdam+in\strouhal}\omegazero
+\ed^{-|k|(x-\xzero)}\huzero-
\frac{k^2\qA(x)+|k|n\strouhal \qB(x)}{k^2+(n\strouhal)^2}
\\&\phantom{=}
-\frac{\Lambdam}{\Lambdazero}
\int_{\xzero}^{x}\hspace{-3mm}{\rm d}\tilde{x}~
\frac{\ed^{\Lambdam(x-\tilde{x})}}{\Lambdam+in\strouhal}q_{-}(\tilde{x})
-\frac{\Lambdap}{\Lambdazero}
\int_{x}^{\infty}\hspace{-4mm}{\rm d}\tilde{x}~
\frac{\ed^{\Lambdap(x-\tilde{x})}}{\Lambdap+in\strouhal}q_{+}(\tilde{x})
\\&\phantom{=}
+
\int_{\xzero}^{x}\hspace{-3mm}{\rm d}\tilde{x}~
\frac{|k|\ed^{-|k|(x-\tilde{x})}}{ik+n\strouhal\sigma}\qun(\tilde{x})
+
\int_{x}^{\infty}\hspace{-4mm}{\rm d}\tilde{x}~
\frac{|k|\ed^{ |k|(x-\tilde{x})}}{ik-n\strouhal\sigma}\qun(\tilde{x})~.
\label{eqn:forvbf}
\end{equs}
Using inverse Fourier transform, we get
(\ref{eqn:foromega})-(\ref{eqn:forv}) where the operator $K_{\kun}(x)$ is
the convolution operator with the inverse Fourier transform of
$K_{\kun}(x,k)=\ed^{\Lambdam x}$, $K_{\kzero}(x)$ is the convolution
operator with $K_{\kzero}(x,y)=\frac{1}{\pi}\frac{x}{x^2+y^2}$, and, in
terms of their symbols, $\LA=\frac{k^2}{k^2+(n\strouhal)^2}$,
$\LB=\frac{|k|n\strouhal}{k^2+(n\strouhal)^2}$,
$\Lu=(\Lambdam+in\strouhal)^{-1}ik$,
$\Lv=(\Lambdam+in\strouhal)^{-1}\Lambdam$, and
\begin{equs}
{\cal F}_{1,\omega}(x)&=
\int_{\xzero}^{x}\hspace{-3mm}{\rm d}\tilde{x}~
K_{1,1,\omega}(x-\tilde{x})\qzero(\tilde{x})
+K_{1,2,\omega}(x-\tilde{x})\qun(\tilde{x})\label{eqn:f1omega}\\
{\cal F}_{2,\omega}(x)&=
\int_{x}^{\infty}\hspace{-4mm}{\rm d}\tilde{x}~
K_{2,1,\omega}(\tilde{x}-x)\qzero(\tilde{x})
+K_{2,2,\omega}(\tilde{x}-x)\qun(\tilde{x})
\label{eqn:f2omega}
\end{equs}
with similar definitions for ${\cal F}_{1,u}$, ${\cal F}_{2,u}$,
${\cal F}_{1,v}$ and ${\cal F}_{2,v}$ and
\begin{equs}
\begin{array}{ll}
K_{1,1,\omega}=-K_{\khuit}-K_{\kdix}
&K_{1,2,\omega}=-K_{\kdeux}\\
K_{2,1,\omega}=-\ed^{-x}(K_{\kun}+K_{\khuit}+K_{\kdix})
\hspace{10mm}
&K_{2,2,\omega}=-\ed^{-x}K_{\kdeux}
\\
K_{1,1,u}=
K_{\kdeux}-K_{\ktreize}
&K_{1,2,u}=
-\falpha-K_{\kdouze}
\\
K_{2,1,u}=
\ed^{-x}\big(K_{\kdeux}-K_{\ksix}\big)
&K_{2,2,u}=\falpha^{*}-\ed^{-x}K_{\kcinq}
\\
K_{1,1,v}=K_{1,1,\omega}+K_{\kr}+K_{\ki}~~~~~~~~~
&K_{1,2,v}=
K_{1,2,\omega}+\galpha+K_{\ktreize}\\
K_{2,1,v}=K_{2,1,\omega}-\ed^{-x}K_{\ksept}
&K_{2,2,v}=
K_{2,2,\omega}
-\galpha^{*}+\ed^{-x}K_{\ksix}
\end{array}
\end{equs}
with
\begin{equs}
\begin{array}{lll}
K_{\kun}(x,k)=\ed^{\Lambdam x} 
&K_{\kdeux}(x,k)=-\frac{ik\ed^{\Lambdam x}}{\Lambdazero}
&K_{\kcinq}(x,k)=\frac{k^2\ed^{\Lambdam x}}{\Lambdazero(\Lambdap+in\strouhal)}
\\[2mm]
K_{\ksix}(x,k)=\frac{kn\strouhal\ed^{\Lambdam x}}{\Lambdazero(\Lambdap+in\strouhal)}
&K_{\ksept}(x,k)=-\frac{in\strouhal\Lambdap\ed^{\Lambdam x}}{\Lambdazero(\Lambdap+in\strouhal)}~~~~~~~~
&K_{\khuit}(x,k)=\frac{\Re(\Lambdam)}{\Lambdazero}\ed^{\Lambdam x}
\\[2mm]
K_{\kdix}(x,k)=\frac{i\Im(\Lambdam)}{\Lambdazero}\ed^{\Lambdam x}
&K_{\kdouze}(x,k)=\frac{k^2\ed^{\Lambdam x}}{\Lambdazero(\Lambdam+in\strouhal)}
&K_{\ktreize}(x,k)=\frac{kn\strouhal\ed^{\Lambdam x}}{\Lambdazero(\Lambdam+in\strouhal)}
\\[2mm]
%
%
%
K_{\kr}(x,k)=\frac{in\strouhal\Re(\Lambdam)\ed^{\Lambdam x}}{\Lambdazero(\Lambdam+in\strouhal)}
&K_{\ki}(x,k)=\frac{-n\strouhal\Im(\Lambdam)\ed^{\Lambdam x}}{\Lambdazero(\Lambdam+in\strouhal)}
&K_{\kzero}(x,y)=\frac{1}{\pi}\frac{x}{x^2+y^2}~,
\end{array}
\end{equs}
and
\begin{equs}
\falpha(x,k)=\frac{ik\ed^{-|k|x}}{ik+n\strouhal\sigma}~,~~~~
\galpha(x,k)=\frac{|k|\ed^{-|k|x}}{ik+n\strouhal\sigma}~. 
\end{equs}
Various estimates of these kernels are given in Appendix
\ref{app:kernelestimates}. Intuitively, the two kernels $\falpha$ and
$\galpha$ behave like Poisson's kernels
$\frac{1}{\pi}\frac{x}{x^2+y^2}$ and $\frac{1}{\pi}\frac{y}{x^2+y^2}$,
while all the other kernels behave like $y$ derivatives or primitives of
$K_{\kun}$ according to the expansion of their prefactor as $|k|\to0$ or
$|k|\to\infty$. We thus need to understand the basic properties of
$\ed^{\Lambdam x}$. To do so, we define
\begin{equs}
\textstyle
b(\alpha)=\frac{1}{4}\Big(1-\sqrt{\frac{1+\sqrt{1+16\alpha^2}}{2}}\Big)
~~,~~
c(\alpha)=\frac{1}{2}\sqrt{\frac{1+\sqrt{1+16\alpha^2}}{2+32\alpha^2}}~,
\end{equs}
and note that (see also figure \ref{fig:thefigure} on page
\pageref{fig:thefigure})
\begin{equs}
{\rm Re}(\Lambdam)&\leq\Bigg\{
\begin{array}{ll}
b(n\strouhal)-c(n\strouhal)k^2&\forall|k|\leq1\\[1mm]
b(n\strouhal)-\frac{|k|}{2}&\forall|k|\geq1
\end{array}
~~~~~\mbox{and}~~~~
\left|\frac{1}{\Lambdazero}\right|\leq\Bigg\{
\begin{array}{l}
(1+(n\strouhal)^2)^{-1/4}\\[1mm]
(1+k^2)^{-1/2}
\end{array}~.
\end{equs}
For all practical purpose, the kernel $K_{\kun}$ corresponding to
$\ed^{\Lambdam x}$ thus behaves like a superposition of a kernel of
Poisson's type with a heat kernel (see also Lemma \ref{lem:alittlelemma}):
\begin{equs}
K_{\kun}(x,y)\approx
\ed^{b(n\strouhal)x}\left(
\frac{\ed^{-\frac{y^2}{4x}}}{\sqrt{4\pi\,x}}+
\frac{1}{\pi}
\frac{2x}{x^2+4y^2}
\right)~.
\end{equs}
Actually, most results of Appendix \ref{app:kernelestimates} can be easily
derived from this analogy. In particular since $b(0)=0$ and
$b(\strouhal)<0$, it is easy to see that $\L^p$ estimates on $K_{\kun}$
will decay {\em at most algebraically} as $x\to\infty$, while the same
estimates on $\P K_{\kun}$ will decay {\em exponentially faster}. We also
easily see that $\partial_y^mK_{\kun}\sim
x^{-m}\chinese{x}^{\frac{m}{2}}K_{\kun}$.

\section{`Evolution' estimates}\label{sec:estimates}

Our next task is to prove that for each boundary data in $\Ci$, there exist
in $\Cu$ a unique solution to (\ref{eqn:foromega})-(\ref{eqn:forv}). This
will be done in the next Section using a contraction mapping argument in
$\W$. We thus have to show that the r.h.s.\  of
(\ref{eqn:foromega})-(\ref{eqn:forv}) defines a Lipschitz map in (a ball
of) $\W$. Subsection \ref{sec:linearterms} is devoted to the terms
involving $\uzero$, $\huzero$ and $\omegazero$, Subsection
\ref{sec:localterms} to those involving $\qA$ and $\qB$, Subsection
\ref{sec:nonlineartermsun} to the terms ${\cal F}_{1,\cdot}$ and Subsection
\ref{sec:nonlineartermsdeux} to the terms ${\cal F}_{2,\cdot}$.

\subsection{Preliminaries}
In this whole section, we will encounter various convolution products like
\begin{equs}
K(x-z)f(z)\equiv
\int_{-\infty}^{\infty}
\hspace{-4mm}
{\rm d}\tilde{y}~
K(x-z,y-\tilde{y},n\strouhal)f_n(z,\tilde{y})
\end{equs}
on which we will use repeatedly the following inequalities
(see Subsection \ref{sec:functionalspaces} for the definitions of the
norms)
\begin{equs}
\||y|^{\beta}K(x-z)f(z)\|_{2}&\leq
\||y|^{\beta}K(x-z)\|_{\L^2}\|f(z)\|_{1}+
\|K(x-z)\|_{\L^1}\||y|^{\beta}f(z)\|_{2}~,
\label{eqn:onuseB}
\\
\|K(x-z)f(z)\|_{s}&\leq\min\Big(
\|K(x-z)\|_{\L^{p_1}}\|f(z)\|_{q_1}~,~
\|K(x-z)\|_{\L^{p_2}}\|f(z)\|_{q_2}
\Big)
\label{eqn:onuseI}
\\
\|K(x-z)f(z)\|_{s}&\leq\min\Big(
\|K(x-z)\|_{\L^{p_1}}\|f(z)\|_{q_1}~,~
\|\partial_yK(x-z)\|_{\L^{p_2}}\|\I f(z)\|_{q_2}
\Big)~,
\label{eqn:onuseII}
\\
\|\partial_y(K(x-z)f(z))\|_{s}&\leq\min\Big(
\|K(x-z)\|_{\L^{p_1}}\|\partial_yf(z)\|_{q_1}~,~
\|\partial_yK(x-z)\|_{\L^{p_2}}\|f(z)\|_{q_2}
\Big)\label{eqn:onuseD}
\end{equs}
where
$\frac{1}{p_1}+\frac{1}{q_1}=\frac{1}{p_2}+\frac{1}{q_2}=1+\frac{1}{s}$.
Note that (\ref{eqn:onuseII}) and (\ref{eqn:onuseD}) follow from
$K(x-z)f(z)=\partial_yK(x-z)\I f(z)$ and
$\partial_y(K(x-z)f(z))=(\partial_yK(x-z))f(z)=K(x-z)(\partial_yf(z))$.
In particular, (\ref{eqn:onuseI})-(\ref{eqn:onuseD}) give a
great freedom for the way we will actually do the estimates. Our main
concern and difficulty will be to get optimal decay rates as
$x\to\infty$. As a rule (particularly in Subsections
\ref{sec:nonlineartermsun} and \ref{sec:nonlineartermsdeux}), we will
choose the $p_1$ as small as possible to cover regions where $|x-z|$ is
small and $p_2$ as large as possible in regions where $|x-z|$ is large.
For concision, we will often omit the arguments in the $K$'s and $f$'s when
no confusion is possible. For the same reason, we will use
(\ref{eqn:onuseB})-(\ref{eqn:onuseD}) without reference or even sometimes
without explicit statement of the choice made for the parameters.

We also note for further reference that using
$\|f\|_{\infty}\leq(\|f\|_{2}\|\partial_yf\|_{2})^{\frac{1}{2}}$,
the interpolation inequality, $0<\olddelta<\olddeltam$ and
$\frac{1}{2}+\au-\av\geq0$, we have for some constant $C$ that
\begin{equs}
\|(f,0,0)\|\leq C\|(0,f,0)\|\leq C\|(0,0,f)\|~.
\label{eqn:interpolation}
\end{equs}

\subsection{The `linear' terms}\label{sec:linearterms}

In this subsection, we will prove the following inequalities,
\begin{equs}
\|(
\Lu K_{\kun}(x-\xzero)\omegazero(\xzero),
\Lv K_{\kun}(x-\xzero)\omegazero(\xzero),
K_{\kun}(x-\xzero)\omegazero(\xzero))\|
&\leq
C \|(0,0,\omegazero)\|_{\xzero}
\label{eqn:insteadun}
\\
\|
(
K_{\kzero}(x-\xzero)\uzero(\xzero),
K_{\kzero}(x-\xzero)\huzero(\xzero),
0
)
\|
&\leq C\|(\uzero,\huzero,0)\|_{\xzero}~,
\label{eqn:insteaddeux}
\end{equs}
which show that the $\|\cdot\|$ norm of the `linear' terms
in (\ref{eqn:foromega})--(\ref{eqn:forv}) is controlled by 
$\|(\uzero,\huzero,\omegazero)\|_{\xzero}$, provided $\uzero$,
$\huzero$ and $\omegazero$ are in the Class ${\cal C}_i$ of Definition 
\ref{def:leszeros}. By (\ref{eqn:interpolation}), it will be enough to
prove that
\begin{equs}
\|(
0,
0,
K_{\kun}(x-\xzero)\omegazero(\xzero))\|
&\leq
C \|(0,0,\omegazero)\|_{\xzero}
\label{eqn:insteaduna}
\\
\|(
\Lu K_{\kun}(x-\xzero)\omegazero(\xzero),
(\Lv-1) K_{\kun}(x-\xzero)\omegazero(\xzero),
0)\|
&\leq
C \|(0,0,\omegazero)\|_{\xzero}
\label{eqn:insteaddeuxa}
\\
\|
(
K_{\kzero}(x-\xzero)\uzero(\xzero),
K_{\kzero}(x-\xzero)\huzero(\xzero),
0
)
\|
&\leq C\|(\uzero,\huzero,0)\|_{\xzero}~.
\label{eqn:insteadtroisa}
\end{equs}
For convenience, these three inequalities are proved
in the three following Lemmas. The general idea of the proofs is to 
consider separately the regions $\xzero\leq x\leq2\xzero$ and 
$x\geq2\xzero$. In the first region, we will use the fact that
$\|K_{\kzero}(x-\xzero)\|_{\L^1}+\|K_{\kun}(x-\xzero)\|_{\L^1}$ is
uniformly bounded (thus $K_{\kzero}\cdot$ and $K_{\kun}\cdot$ are
$\L^p$-bounded operators for all $p\geq1$), whereas in the region $x\geq2\xzero$, we will
essentially use that
$\|K_{\kzero}(x-\xzero)\|_{\L^p}+\|K_{\kun}(x-\xzero)\|_{\L^p}$ decays as
$x\to\infty$ as soon as $p>1$.

\begin{lemma}\label{lem:evollmforomega}
Let $f=K_{\kun}(x-\xzero)\omegazero(\xzero)$. Assume that the parameters
satisfy (\ref{eqn:restrictions}), then there exist a constant $C$
such that
\begin{equs}
\|(0,0,f)\|
\leq
C \|(0,0,\omegazero)\|_{\xzero}~.
\end{equs}
\end{lemma}			  

\begin{proof}
We first note that
for $\xzero\leq x\leq2\xzero$, since $\|K_{\kun}\|_{1}\leq C$, we have
\begin{equs}
\|f\|_{x,2,\frac{3}{4}}+
\|\partial_y f\|_{x,\infty,\frac{3}{2}}+
\|\partial_y f\|_{x,1,1}\leq C\big(
\|\omegazero\|_{\xzero,2,\frac{3}{4}}+
\|\partial_y\omegazero\|_{\xzero,\infty,\frac{3}{2}}+
\|\partial_y\omegazero\|_{\xzero,1,1}\big)~,
\end{equs}
while for $x\geq2\xzero$, 
\begin{equs}
\|f\|_{x,2,\frac{3}{4}}&\leq\big(
\|\P K_{\kun}\|_{x,2,\frac{3}{4}}\|\omegazero\|_1+
\|\partial_y K_{\kun}\|_{x,2,\frac{3}{4}}\|\Q\I\omegazero\|_1\big)
~,\\
\|\partial_yf\|_{x,\infty,\frac{3}{2}}&\leq\big(
\|\P \partial_yK_{\kun}\|_{x,2,\frac{3}{4}}\|\omegazero\|_1+
\|\partial_y^2 K_{\kun}\|_{x,2,\frac{3}{4}}\|\Q\I\omegazero\|_1\big)
~,\\
\|\partial_yf\|_{x,1,1}&\leq\big(
\|\P \partial_yK_{\kun}\|_{x,1,1}\|\omegazero\|_1+
\|\partial_y^2 K_{\kun}\|_{x,1,1}\|\Q\I\omegazero\|_1\big)~.
\end{equs}					  
Using Lemma \ref{lem:kernelun}, that $x-x_0\geq\frac{x}{2}$ if $x\geq2x_0$,
and that
$\chinese{x}^{\frac{1}{2}}\ed^{b(\strouhal)x}\leq\frac{\chinese{\strouhal}
}{\strouhal}\leq\chinese{\xzero}^{\frac{1}{2}}$, we get
\begin{equs}
\|f\|_{x,2,\frac{3}{4}}+
\|\partial_yf\|_{x,\infty,\frac{3}{2}}+
\|\partial_yf\|_{x,1,1}&\leq
C(\chinese{\xzero}^{\frac{1}{2}}
\|\omegazero\|_1+
\|\Q\I\omegazero\|_1
)
~.
\end{equs}
Next, we note that for all
$z\in{\bf R}$, we can write $|y|^{\beta}=|y-z|^{\beta}+L(y,z)$ with
$|L(y,z)|\leq C(|z|^{\beta}+|z||y-z|^{\beta-1})$, so that
\begin{equs}
\| |y|^{\beta}\P f\|_{\Ltx{2}{\frac{3}{4}-\frac{\beta}{2}}}&\leq
C\chinese{x_0}^{\frac{3}{4}-\frac{\beta}{2}}
\||y|^{\beta}\omegazero\|_{2}+
C\sup_{x\geq\xzero}
\chinese{x}^{\frac{3}{4}-\frac{\beta}{2}}
\|\P |y|^{\beta}K_{\kun}\|_{\L^2}\|\omegazero\|_{1}
\\
&\leq
C\chinese{x_0}^{\frac{3}{4}-\frac{\beta}{2}}
\||y|^{\beta}\omegazero\|_{2}+
C\chinese{\xzero}^{\frac{1}{2}}\|\omegazero\|_{1}~,\\
\||y|^{\beta}\Q f\|_{\Ltx{2}{\frac{3}{4}-\frac{\beta}{2}}}&\leq
C\sup_{x\geq\xzero}
\chinese{x}^{\frac{3}{4}-\frac{\beta}{2}}
\Big(
\|(|y|^{\beta}K_{\kun})\Q\omegazero\|_{2}
+\||y|^{\beta}\omegazero\|_{2}
+\||y|^{\beta-1}K_{\kun}\|_{\L^2}
\|y\omegazero\|_{1}\Big)\\
&\leq
C\Big(
\sup_{x\geq\xzero}
\chinese{x}^{\frac{3}{4}-\frac{\beta}{2}}
\|\partial_y(|y|^{\beta}K_{\kun})\|_{\L^2}\|\I\Q\omegazero\|_{1}\Big)
+C\|(0,0,\omegazero)\|_{\xzero}~,
\end{equs}
where we used $\|(|y|^{\beta}K_{\kun})\Q\omegazero\|_{\L^2}=
\|(\partial_y|y|^{\beta}K_{\kun})\I\Q\omegazero\|_{\L^2}$ and that
since $\beta>\frac{3}{2}$, we have
\begin{equs}
\chinese{x}^{\frac{3}{4}-\frac{\beta}{2}}
\|y\omegazero\|_{1}\leq
\chinese{\xzero}^{\frac{3}{4}-\frac{\beta}{2}}\|\omegazero\|_{2}+
\chinese{\xzero}^{\frac{3}{4}-\frac{\beta}{2}}
\||y|^{\beta}\omegazero\|_{2}\leq \|(0,0,\omegazero)\|_{\xzero}~.
\end{equs}
The proof is completed using
$\chinese{\xzero}^{\frac{1}{2}}\|\omegazero\|_{1}\leq
\|(0,0,\omegazero)\|_{\xzero}$ (see (\ref{eqn:simplificatrice})) and
Lemma \ref{lem:key} below.
\end{proof}

\begin{lemma}\label{lem:key}
Let $\beta>\frac{3}{2}$ and $0\leq\gamma<\beta-\frac{3}{2}$.
, ${\cal Z}_{\beta}=\{\|(1+|y|^{\beta})f\|_{\L^2}<\infty$
and $\M(f)=\int_{{\bf R}}f(y){\rm d}y=0\}$. Then there exist
constants $C_{\beta},C_{\beta,\gamma}$ such that for all
$f\in{\cal Z}_{\beta}$,
\begin{equs}
\|\I f\|_{\L^{\infty}}&\leq C_{\beta}\|f\|_{\L^2}^{1-\frac{1}{2\beta}}
\||y|^{\beta}f\|_{\L^2}^{\frac{1}{2\beta}}~,\\
\||y|^{\gamma}\I f\|_{\L^1}&\leq 
C_{\beta,\gamma}\|f\|_{\L^2}^{1-\frac{3}{2\beta}-\frac{\gamma}{\beta}}
\||y|^{\beta}f\|_{\L^2}^{\frac{3}{2\beta}+\frac{\gamma}{\beta}}~.
\end{equs}
The first inequality is also valid if $\M(f)\neq0$.
\end{lemma}

\begin{proof}
Let $\beta>\frac{3}{2}$ and $a>0$. Since $\|\I f\|_{\L^{\infty}}\leq
\|f\|_{\L^1}$, the first inequality follows from Lemma \ref{lem:Lun}. Then,
since $\M(f)=0$, we have
\begin{equs}
\I f(y)=
-\int_y^{\infty}\hspace{-2mm}{\rm d}z~f(z)
=\int_{-\infty}^y\hspace{-2mm}{\rm d}z~f(z)~,
\end{equs}
from which we deduce
\begin{equs}
\||y|^{\gamma}\I f(y)\|_{\L^1}
&\leq2\Big(
a\|f\|_{\L^2}+\||y|^{\beta}f\|_{\L^2}\Big)
\int_{0}^{\infty}
\hspace{-3mm}{\rm d}y~|y|^{\gamma}\left(
\int_y^{\infty}\hspace{-3mm}{\rm d}z
~(a+|z|^{\beta})^{-2}\right)^{\frac{1}{2}}\\
&\leq\Big(
a^{\frac{3}{2\beta}+\frac{\gamma}{\beta}}\|f\|_{\L^2}
+a^{\frac{3}{2\beta}+\frac{\gamma}{\beta}-1}\||y|^{\beta}f\|_{\L^2}\Big)
\int_{0}^{\infty}
\hspace{-3mm}{\rm d}y~|y|^{\gamma}\left(
\int_y^{\infty}\hspace{-3mm}
\frac{{\rm d}z}{(1+|z|^{\beta})^{2}}~\right)^{\frac{1}{2}}~.
\end{equs}
Setting $a=\||y|^{\beta}f\|_{\L^2}/\|f\|_{\L^2}$ completes the
proof, since the last integral is bounded if $\gamma<\beta-\frac{3}{2}$.
\end{proof}

\begin{lemma}\label{lem:evollmforu}
Let $f_u=K_{\kun}(x-\xzero)\Lu\omegazero(\xzero)$, $\tilde{\Lv}=\Lv-1$ and
$f_v=K_{\kun}(x-\xzero)\tilde{\Lv}\omegazero(\xzero)$. If
(\ref{eqn:restrictions}) holds, then there exist a constant $C$ such that
\begin{equs}
\|(f_u,f_v,0)\|\leq
C \|(0,0,\omegazero)\|_{\xzero}~.
\end{equs}
\end{lemma}

\begin{proof}
By Lemma \ref{lem:kernelun}, for all $1\leq s\leq\infty$, we have
\begin{equs}
\|\partial_yf_u\|_{\Ltx{\pdv}{1-\frac{1}{2\pdv}-\au}}
&\leq
\csupun
\chinese{x}^{1-\frac{1}{2\pdv}-\au}
\|\partial_y\Lu\omegazero\|_{\pdv}
+\sup_{x\geq 2x_0}
\chinese{x}^{1-\frac{1}{2\pdv}-\au}
\|\partial_yK_{\kun}\|_{\L^{\pdv}}
\|\Lu\omegazero\|_{1}\\
&\leq C
\|\Lu\omegazero\|_{1}
+\chinese{x_0}^{1-\frac{1}{2\pdv}-\au}\|\partial_y\Lu\omegazero\|_{r}~,\\
\|f_u\|_{\Ltx{s}{\frac{1}{2}-\frac{1}{2s}}}
&\leq \csupun
\chinese{x}^{\frac{1}{2}-\frac{1}{2s}}
\|\Lu\omegazero\|_{s}
+
C\sup_{x\geq 2x_0}
\chinese{x}^{\frac{1}{2}-\frac{1}{2s}}
\|K_{\kun}\|_{\L^s}\|\Lu\omegazero\|_{1}\\
&\leq C
\|\Lu\omegazero\|_{1}
+\chinese{x_0}^{\frac{1}{2}}\|\Lu\omegazero\|_{\infty}~,
\end{equs}
since $\chinese{\xzero}^{\frac{1}{2}-\frac{1}{2p}}\|f\|_{\L^p}\leq
\|f\|_{\L^1}
+\chinese{\xzero}^{\frac{1}{2}}\|f\|_{\L^{\infty}}$ for all 
$1\leq p\leq\infty$. We then note that
$\tilde{\Lv}=\frac{-in\strouhal}{\Lambdam+in\strouhal}$,
in particular, $\P\tilde{\Lv}=\tilde{\Lv}$ and $\P\ed^{c
b(n\strouhal)x}\leq\ed^{c b(\strouhal)x}$ for all $c>0$.
As in Lemma \ref{lem:evollmforomega}, since $x-\xzero\geq\frac{x}{2}$ for
$x\geq2\xzero$ and $b(\strouhal)<0$, we have
\begin{equs}
\|f_v\|_{\Ltx{s}{1-\frac{1}{2s}-\olddelta}}
&\leq C
\chinese{\xzero}^{1-\frac{1}{2s}-\olddelta}
\|\tilde{\Lv}\omegazero\|_{s}
+
C\sup_{x\geq 2x_0}
\chinese{x}^{1-\frac{1}{2s}-\olddelta}
\|\P K_{\kun}\|_{\L^s}
\|\tilde{\Lv}\omegazero\|_{1}~,\\
&\leq
C\chinese{x_0}^{\frac{1}{2}-\olddelta}\|\tilde{\Lv}\omegazero\|_{1}
+C\chinese{x_0}^{1-\olddelta}\|\tilde{\Lv}\omegazero\|_{\infty}
+
\chinese{\xzero}^{\frac{1}{2}-\olddelta}
\|\tilde{\Lv}\omegazero\|_{1}~,\\
\|\partial_yf_v\|_{\Ltx{\pdv}{\frac{3}{2}-\frac{1}{2\pdv}-\av}}
&\leq \csupun
\chinese{x}^{\frac{3}{2}-\frac{1}{2\pdv}-\av}
\|\partial_y\tilde{\Lv}\omegazero\|_{\pdv}
+
C\sup_{x\geq 2x_0}
\chinese{x}^{\frac{3}{2}-\frac{1}{2\pdv}-\av}
\|\P\partial_yK_{\kun}\|_{\L^{\pdv}}\|\tilde{\Lv}\omegazero\|_{1}~,\\
&\leq C
\Big(
\chinese{\xzero}^{\frac{3}{2}-\frac{1}{2\pdv}-\av}
\|\partial_y\tilde{\Lv}\omega\|_{\pdv}+
\chinese{\xzero}^{\frac{1}{2}-\olddelta}\|\tilde{\Lv}\omega\|_{1}
\Big)~.
\end{equs}
Since
$\chinese{x}\ed^{\frac{b(\strouhal)x}{8}}\leq\frac{\chinese{\strouhal}}{\strouhal}
\leq\chinese{\xzero}^{\frac{1}{2}}$ and $\av\geq\olddelta$. The proof is
then completed using Lemma \ref{lem:hormander} and \ref{lem:lunforLu} below.
\end{proof}

\begin{lemma}[Mikhlin-H\"ormander]\label{lem:hormander}
Let $m:{\bf R}\to{\bf C}$, and define
$m_0=\sup_{k\in{\bf R}}|m(k)|+|k\partial_km(k)|$ and
$m_1=\sup_{k\in{\bf R}}|\partial_km(k)|$. Let ${\cal F}$ denotes the
(continuous) Fourier transform and $M:f\to{\cal F}^{-1}m(\cdot){\cal F}f$.
Then there exist constants $C_p$ such that
\begin{equs}
\|Mf\|_{\L^{\infty}}&\leq C_{\infty}m_{0}
\sqrt{\|f\|_{\L^2}\|\partial_yf\|_{\L^2}}~,~~~~~
\|Mf\|_{\L^p}\leq C_pm_0\|f\|_{\L^p}\\
\|Mf\|_{\L^{1}}&\leq C_{1}\Big(m_{0}
\sqrt{\|f\|_{\L^2}\|yf\|_{\L^2}}+
\sqrt{m_0m_1}\|f\|_{\L^2}
\Big)
\end{equs}
for all $1<p<\infty$.
\end{lemma}

\begin{proof}
The $\L^p$ estimate for $1<p<\infty$ is a consequence of the classical
Mikhlin-H\"ormander condition (see, e.g. \cite{Hormander}), the 
$\L^{\infty}$ and $\L^1$ estimates are immediate consequences of the Sobolev
and Plancherel inequalities.
\end{proof}

\begin{lemma}\label{lem:lunforLu}
Let $\tilde{\Lv}=\Lv-1$ and $\tilde{\Lu}=\Lu+\I\Q$ and assume that
(\ref{eqn:restrictions}) holds, then
\begin{equs}
\|\Lu\omegazero\|_{1}+
\chinese{\xzero}^{\frac{1}{2}}\|\Lu\omegazero\|_{\infty}
+\chinese{\xzero}^{1-\frac{1}{2\pdv}-\au}\|\partial_y\Lu\omegazero\|_{\pdv}
&\leq
C(\|\I\Q\omegazero\|_{1}+\|(0,0,\omegazero)\|_{\xzero})~,\\
\|\tilde{\Lu}\omegazero\|_{1}+
\chinese{\xzero}^{\frac{1}{2}}\|\tilde{\Lu}\omegazero\|_{\infty}
+\chinese{\xzero}^{1-\frac{1}{2\pdv}-\au}\|\partial_y\tilde{\Lu}\omegazero\|_{\pdv}
&\leq
C \|(0,0,\omegazero)\|_{\xzero}~,\\
\chinese{\xzero}^{\frac{1}{2}-\olddelta}\|\tilde{\Lv}\omegazero\|_{1}+
\chinese{\xzero}^{1-\olddelta}
\|\tilde{\Lv}\omegazero\|_{\infty}
+\chinese{\xzero}^{\frac{3}{2}-\frac{1}{2\pdv}-\av}
\|\partial_y\tilde{\Lv}\omegazero\|_{\pdv}
&\leq
C\|(0,0,\omegazero)\|_{\xzero}~,
\\
\chinese{\xzero}^{\frac{1}{2}-\olddelta}\|\Lv\omegazero\|_{1}+
\chinese{\xzero}^{1-\olddelta}
\|\Lv\omegazero\|_{\infty}
+\chinese{\xzero}^{\frac{3}{2}-\frac{1}{2\pdv}-\av}
\|\partial_y\Lv\omegazero\|_{\pdv}
&\leq
C\|(0,0,\omegazero)\|_{\xzero}~.
\end{equs}
\end{lemma}

\begin{proof}
We have $\Lu=-\I\Q+\tilde{\Lu}$. Then
the symbol $T(k,n)$ of $\tilde{\Lu}$ is given by
$T(k,n)=\frac{-ik}{\Lambdam+in\strouhal}$ if $n\neq0$ and
$T(k,0)=\frac{-ik}{\Lambdap}$, and it satisfies (uniformly in $n\in{\bf Z}$)
the hypothesis of Lemma \ref{lem:hormander} with
$m_0=C\frac{\chinese{\strouhal}}{\strouhal}\leq
C\chinese{\xzero}^{\frac{1}{2}}$
and $m_1=C\frac{\chinese{\strouhal}^2}{\strouhal^2}\leq C\chinese{\xzero}$.
Using $\partial_y\I f=f$ and Lemma \ref{lem:key} and \ref{lem:hormander},
we get that
\begin{equs}
\|\tilde{\Lu}\omegazero\|_{1}
&\leq
\big(\chinese{\xzero}^{\frac{3}{4}}\|\omegazero\|_{2}
\chinese{\xzero}^{\frac{1}{4}}
\|y\omegazero\|_{2}\big)^{\frac{1}{2}}
+\chinese{\xzero}^{\frac{3}{4}}
\|\omegazero\|_{2}\\
\chinese{\xzero}^{\frac{1}{2}}\|\tilde{\Lu}\omegazero\|_{\infty}
&\leq
\big(
\chinese{\xzero}^{\frac{3}{4}}\|\omegazero\|_{2}
\chinese{\xzero}^{\frac{5}{4}}\|\partial_y\omegazero\|_{2}
\big)^{\frac{1}{2}}~,\\
\chinese{\xzero}^{1-\frac{1}{2\pdv}-\au}
\|\partial_y\tilde{\Lu}\omegazero\|_{\pdv}
&\leq
\chinese{\xzero}^{\frac{3}{2}-\frac{1}{2\pdv}-\au}
\|\partial_y\omegazero\|_{\pdv}
\leq
\chinese{\xzero}
\|\partial_y\omegazero\|_{1}+
\chinese{\xzero}^{\frac{3}{2}}
\|\partial_y\omegazero\|_{\infty}
~,\\
\chinese{\xzero}^{\frac{1}{2}}\|\I\Q\omegazero\|_{\infty}
&\leq
\chinese{\xzero}^{\frac{1}{2}}\|\Q\omegazero\|_{1}
\leq C
\big(\chinese{\xzero}^{\frac{3}{4}}
\|\omegazero\|_{2}
\chinese{\xzero}^{\frac{1}{4}}
\|y\omegazero\|_{2}\big)^{\frac{1}{2}}~,\\
\chinese{\xzero}^{1-\frac{1}{2\pdv}-\au}
\|\partial_y\I\Q\omegazero\|_{\pdv}
&\leq
\chinese{\xzero}^{1-\frac{1}{2\pdv}-\au}
\|\omegazero\|_{\pdv}\leq
\chinese{\xzero}^{\frac{1}{2}}
\|\omegazero\|_{1}+
\chinese{\xzero}
\|\omegazero\|_{\infty}
~.
\end{equs}
Similarly, since $\P\tilde{\Lv}$ satisfies the hypothesis of Lemma
\ref{lem:hormander} with $m_0=2\frac{\chinese{\strouhal}}{\strouhal}\leq
2\chinese{\xzero}^{\olddelta}<
2\chinese{\xzero}^{\frac{1}{2}}$ and
$m_1=m_0^2\leq4\chinese{\xzero}^{2\olddelta}$, we
get
\begin{equs}
\chinese{x_0}^{1-\olddelta}\|\tilde{\Lv}\omegazero\|_{\infty}&\leq
\Big(
\chinese{\xzero}^{\frac{3}{4}}\|\omegazero\|_{2}
\chinese{\xzero}^{\frac{5}{4}}\|\partial_y\omegazero\|_{2}
\Big)^{\frac{1}{2}}\\
\chinese{x_0}^{\frac{1}{2}-\olddelta}\|\tilde{\Lv}\omegazero\|_{1}&
\leq C\Big(
\chinese{\xzero}^{\frac{3}{4}}\|\omegazero\|_{2}
\chinese{\xzero}^{\frac{1}{4}}\|y\omegazero\|_{2}
\Big)^{\frac{1}{2}}
+C\chinese{\xzero}^{\frac{3}{4}}\|\omegazero\|_{2}\\
\chinese{\xzero}^{\frac{3}{2}-\frac{1}{2\pdv}-\av}
\|\partial_y\tilde{\Lv}\omegazero\|_{\pdv}
&\leq
\chinese{\xzero}^{\frac{3}{2}-\frac{1}{2\pdv}}\|\partial_y\omega\|_{\pdv}
\leq
\chinese{\xzero}\|\partial_y\omega\|_{1}+
\chinese{\xzero}^{\frac{3}{2}}\|\partial_y\omega\|_{\infty}~.
\end{equs}
Since $\Lv=\tilde{\Lv}+1$, the proof is completed using
$\chinese{\xzero}^{\frac{1}{2}}\|\omegazero\|_{1}+
\chinese{\xzero}\|\omegazero\|_{\infty}\leq
C\|(0,0,\omegazero)\|_{\xzero}$ (see also (\ref{eqn:simplificatrice})) and
$\chinese{\xzero}^{\frac{1}{4}}\|y\omegazero\|_{2}\leq
\chinese{\xzero}^{\frac{3}{4}}\|\omegazero\|_{2}+
\chinese{\xzero}^{\frac{3}{4}-\frac{\beta}{2}}\||y|^{\beta}\omegazero\|_{2}
\leq C\|(0,0,\omegazero)\|$.
\end{proof}

\begin{lemma}\label{lem:thescaryone}
Let $g_u(x)=K_{\kzero}(x-\xzero)\uzero(\xzero)$ and
$g_v(x)=K_{\kzero}(x-\xzero)\huzero(\xzero)$, then if
(\ref{eqn:restrictions}) holds, 
\begin{equs}
\|(g_u,g_v,0)\|
&\leq C\|(\uzero,\huzero,0)\|_{\xzero}~,\\
\|g_u(x)\|_{\infty}+
\|g_v(x)\|_{\infty}&\leq
C\chinese{x}^{-1+\olddelta}\|(\uzero,\huzero,0)\|_{\xzero}
\end{equs}
for all $x\geq2\xzero$.
\end{lemma}

\begin{proof}
We first note that $\|K_{\kzero}(x)\|_{\L^s}\leq C x^{\frac{1}{s}-1}$ and
$\|\partial_yK_{\kzero}(x)\|_{\L^s}\leq C x^{\frac{1}{s}-2}$. Then let 
$\pu\leq p_0\leq\infty$ and $\pv\leq p_1\leq\infty$, since
$(x-\xzero)^{\frac{1}{s}-1}\leq C\chinese{x}^{\frac{1}{s}-1}$ if
$x\geq2\xzero$, we get
\begin{equs}
\|g_u\|_{\Ltx{p_0}{\frac{1}{2}-\frac{1}{p_0}}}&\leq
\csupun
\chinese{x}^{\frac{1}{2}-\frac{1}{p_0}}
\|\uzero\|_{p_0}+
C\sup_{x\geq2\xzero}
\chinese{x}^{\frac{1}{2}-\frac{1}{\pu}}
\|\uzero\|_{\pu}~,\\
\|g_v\|_{\Ltx{p_1}{1-\frac{1}{p_1}-\olddelta}}&\leq
\csupun
\chinese{x}^{1-\frac{1}{p_1}-\olddelta}
\|\huzero\|_{p_1}+
C\sup_{x\geq2\xzero}
\chinese{x}^{1-\frac{1}{\pv}-\olddelta}
\|\huzero\|_{\pv}~,\\
\|\partial_yg_u\|_{\Ltx{\pdv}{1-\frac{1}{2\pdv}-\au}}&\leq
\csupun
\chinese{x}^{1-\frac{1}{2\pdv}-\au}
\|\partial_y\uzero\|_{\pdv}+
C\sup_{x\geq2\xzero}
\chinese{x}^{\frac{1}{2}-\frac{1}{\pu}}
\|\uzero\|_{\pu}~,\\
\|\partial_yg_v\|_{\Ltx{\pdv}{\frac{3}{2}-\frac{1}{2\pdv}-\av}}&\leq
\csupun
\chinese{x}^{\frac{3}{2}-\frac{1}{2\pdv}-\av}
\|\partial_y\huzero\|_{\pdv}+
C\sup_{x\geq2\xzero}
\chinese{x}^{1-\frac{1}{p}-\olddelta}
\|\huzero\|_{p}~,
\end{equs}
while for $x\geq2\xzero$, we have
\begin{equs}
\|g_u(x)\|_{\infty}+
\|g_v(x)\|_{\infty}\leq
\chinese{x}^{-1+\olddelta}
\Big(
\chinese{x}^{1-\frac{1}{\pv}-\olddelta}
\big(
\|\huzero\|_{\pv}+
\|\H\huzero\|_{\pv}\big)\Big)
\leq
\chinese{x}^{-1+\olddelta}
\Big(
\chinese{x}^{1-\frac{1}{\pv}-\olddelta}
\|\huzero\|_{\pv}\Big)
\end{equs}
The proof is completed since $\av\geq\olddelta$, $1\leq\pu<2$ and
$1-\frac{1}{\pv}\leq\olddelta<\olddeltam$.
\end{proof}

\subsection{The `local' terms}\label{sec:localterms}

From now on, we begin the estimates of the contribution of the
nonlinear terms in (\ref{eqn:foromega})-(\ref{eqn:forv}). We first consider
the `local' terms first.
\begin{proposition}\label{prop:localterms}
Assume that (\ref{eqn:restrictions}) holds then
for $\kappa_0=\min(\frac{\olddelta}{2},\frac{1}{2}-\au+\av-\olddelta)$, we
have
\begin{equs}
\|(\LA \qB-\LB \qA,
-\LA \qA-\LB \qB,0)\|
&\leq C
\chinese{\xzero}^{-\kappa_0}\|\triplet\|^2
~,\\
\|\LA \qB(x)-\LB \qA(x)\|_{\infty}+
\|\LA \qA(x)+\LB \qB(x)\|_{\infty}&
\leq C\chinese{x}^{-1}\|\triplet\|^2~.
\label{eqn:firstforuniv}
\end{equs}
\end{proposition}
\begin{proof}
The proof follows at once from Lemma \ref{lem:onLALB} and
(\ref{eqn:nonlinbounds}).
\end{proof}
We already see at this point (see (\ref{eqn:firstforuniv})) that these
terms are of smaller order as $x\to\infty$ than most terms of the preceding
section.

\subsection{The nonlinear terms I}\label{sec:nonlineartermsun}

In this section, we prove the
\begin{theorem}\label{thm:onthefun}
Assume that $\qzero$ and $\qun$ satisfy the bounds
(\ref{eqn:nonlinbounds}), and that the parameters satisfy
(\ref{eqn:restrictions}), then there exist constants $C$ and
$\kappa_1>0$ such that
\begin{equs}
\|({\cal F}_{1,u},{\cal F}_{1,v},{\cal F}_{1,\omega})\|&\leq
C\chinese{\xzero}^{-\kappa_1}\|\triplet\|^2~.
\label{eqn:lasecondomegaevthm}
\end{equs}
\end{theorem}
This is incidentally the hardest part of the paper in that the parameters
in (\ref{eqn:onuseB})-(\ref{eqn:onuseD}) need to be chosen in the right
way to get a bound that {\em decays} as $\xzero\to\infty$. The proof
of (\ref{eqn:lasecondomegaevthm}) is split component-wise in the three
Propositions ending this section. During the course of these proofs, we
will encounter repeatedly the following functions
\begin{equs}
\A{p_1,q_1}{p_2,q_2,s}(x,\xzero)&=
\int_{\xzero}^{x}\hspace{-3mm}{\rm d}\tilde{x}~
\min\Big(
\frac{\chinese{\tilde{x}}^{-q_1}}{(x-\tilde{x})^{p_1}},
\frac{\chinese{x}^{s}\chinese{\tilde{x}}^{-q_2}}{
(x-\tilde{x})^{p_2}}\Big)~,\\
\B{p_1,q_1,s_1}{p_2,q_2,s_2}(x,\xzero)&=
\int_{\xzero}^{x}\hspace{-3mm}
{\rm d}\tilde{x}~
\ed^{\frac{b(\strouhal)(x-\tilde{x})}{4}}
\min\Big(
\frac{\chinese{\tilde{x}}^{-q_1}\chinese{x-\tilde{x}}^{s_1}
}{
(x-\tilde{x})^{p_1}
},
\frac{\chinese{\tilde{x}}^{-q_2}\chinese{x-\tilde{x}}^{s_2}
}{
(x-\tilde{x})^{p_2}
}
\Big)~,
\end{equs}
which occur naturally from (\ref{eqn:onuseB})-(\ref{eqn:onuseD}). For
further reference, we note that these functions satisfy the
\begin{lemma}\label{lem:blackbox}
Let $p_1<1$, $s\geq0$ and $p_2,q_1,q_2\in{\bf R}$, there exist a constant
$C$ such that for all $x\geq\xzero\geq1$, it holds
\begin{equs}
\A{p_1,q_1}{p_2,q_2,s}(x,\xzero)&\leq
C\left(\chinese{x}^{1-q_1-p_1}+
\chinese{x}^{s-p_2}
\max(
\chinese{x}^{1-q_2}
,
\chinese{\xzero}^{1-q_2}
)
\right)
~,
\label{eqn:onA}
\end{equs}
if $q_2\neq1$, while the same inequality holds with
$\max(\chinese{x}^{1-q_2},\chinese{\xzero}^{1-q_2})$ replaced by
$\ln(1+x)$ if $q_2=1$. If furthermore we have $s_1,s_2\geq0$ and
$\frac{\chinese{\strouhal}}{\strouhal}\leq\chinese{\xzero}^{\olddelta}$,
then for all $m\geq0$, there exist a constant $C$ such that for all
$x\geq\xzero\geq1$, it holds 
\begin{equs}
\B{p_1,q_1,s_1}{p_2,q_2,s_2}(x,\xzero)
&\leq 
C\Big(
\chinese{x}^{-q_1}
\chinese{\xzero}^{2(1+s_1-p_1)\olddelta}
+
\chinese{x}^{-p_2-m}
\chinese{\xzero}^{2(1+m+s_2)\olddelta}
\max(\chinese{x}^{-q_2},\chinese{\xzero}^{-q_2})
\Big)~.
\end{equs}
\end{lemma}

\begin{proof}
We first note that for all $p>-1$, there exist a constant $C$ such that
\begin{equs}
\int_{\xzero}^{x}\hspace{-3mm}
{\rm d}\tilde{x}~\ed^{\frac{b(\strouhal)(x-\tilde{x})}{4}}
(x-\tilde{x})^{p}&\leq
C
\int_{0}^{x-\xzero}\hspace{-5mm}
{\rm d}z~\ed^{\frac{-|b(\strouhal)|z}{4}}z^{p}
\leq C|b(\strouhal)|^{-1-p}
\leq C\strouhal^{-2(1+p)}
\leq C\chinese{\xzero}^{2(1+p)\olddelta}~,
\label{eqn:avecleb}
\end{equs}
since $|b(\strouhal)|\leq C\strouhal^{-2}\leq C\chinese{\xzero}^{2\olddelta}$. 
We then note that since $x\geq\xzero\geq1$, we have
$\frac{\chinese{x}}{\sqrt{2}}\leq x\leq\chinese{x}$.
We first consider the case of finite $x$, that is precisely, $\xzero\leq
x\leq2\xzero$, then
\begin{equs}
\A{p_1,q_1}{p_2,q_2,s}(x,\xzero)&\leq
C\chinese{\xzero}^{-q_1}(x-\xzero)^{1-p_1}
\leq C\chinese{\xzero}^{1-p_1-q_1}
~,\\
\B{p_1,q_1,s_1}{p_2,q_2,s_2}(x,\xzero)
&\leq 
\chinese{\xzero}^{-q_1}
\int_{\xzero}^{x}\hspace{-3mm}
{\rm d}\tilde{x}~\ed^{\frac{b(\strouhal)(x-\tilde{x})}{4}}
(x-\tilde{x})^{s_1-p_1}\leq
C\chinese{\xzero}^{-q_1+2(1+s_1-p_1)\olddelta}~.
\end{equs}
However, in the applications of the result of this Lemma, we will
generically have e.g.\ $1-q_1-p_1<0$, that is, the integrals we seek to
bound {\em decay} as $x\to\infty$. To get the
optimal decay rate, the idea is to consider $x\geq2\xzero$, and split the
integration domain $\xzero\leq\tilde{x}\leq x$ in two equal parts. Since
$x\geq2\xzero$ implies $\frac{x}{2}\leq(x-\xzero)\leq x$ and
$\xzero\leq\tilde{x}\leq\frac{x+\xzero}{2}$ implies
$\frac{x}{4}\leq\frac{x-\xzero}{2}\leq x-\tilde{x}\leq x-\xzero\leq x$, we
have
\begin{equs}
\A{p_1,q_1}{p_2,q_2,s}(x,\xzero)
&\leq
C\chinese{x}^{s-p_2}
\int_{\xzero}^{\frac{x+\xzero}{2}}\hspace{-3mm}{\rm d}\tilde{x}~
\chinese{\tilde{x}}^{-q_2}+
C\chinese{x}^{-q_1}
\int_{\frac{x+\xzero}{2}}^{x}\hspace{-3mm}{\rm d}\tilde{x}~
(x-\tilde{x})^{-p_1}~.
\end{equs}
The proof of (\ref{eqn:onA}) is completed using
$\int_{\xzero}^{\frac{x+\xzero}{2}}{\rm d}\tilde{x}~
\chinese{\tilde{x}}^{-q_2}\leq
\int_{\xzero}^{x}{\rm d}\tilde{x}~
\chinese{\tilde{x}}^{-q_2}$ and considering separately $q_2<1$, $q_2=1$ and
$q_2>1$. In the same way, we have
\begin{equs}
\B{p_1,q_1,s_1}{p_2,q_2,s_2}(x,\xzero)
&\leq 
\frac{C\max(\chinese{x}^{-q_2},\chinese{\xzero}^{-q_2})}{
\chinese{x}^{p_2+m}}
\int_{\xzero}^{\frac{x+\xzero}{2}}\hspace{-3mm}
{\rm d}\tilde{x}\ed^{\frac{b(\strouhal)(x-\tilde{x})}{4}}
(x-\tilde{x})^{s_2+m}
\\&\phantom{~\leq}
+
\frac{C}{\chinese{x}^{q_1}}
\int_{\frac{x+\xzero}{2}}^x\hspace{-3mm}
{\rm d}\tilde{x}\ed^{\frac{b(\strouhal)(x-\tilde{x})}{4}}
(x-\tilde{x})^{s_1-p_1}~,
\end{equs}
which completes the proof with the help of (\ref{eqn:avecleb}).
\end{proof}
We now turn to the proof of the part of Theorem \ref{thm:onthefun} that
involves ${\cal F}_{1,\omega}$. To prepare the ground for the asymptotic
results of Section \ref{sec:asymptotics}, we also show that
most terms in ${\cal F}_{1,\omega}$ have decay rates as $x\to\infty$
faster by (almost) $x^{-\frac{1}{2}+\olddelta}$ than those of $\omega$.
\begin{proposition}\label{prop:thecalfisomegaevolve}
Assume that $\qzero$ and $\qun$ satisfy the bounds
(\ref{eqn:nonlinbounds}), and that the parameters satisfy
(\ref{eqn:restrictions}), then there exist a constant $C$ such that
for
$\kappa_{1,1}=\min(\frac{1}{4}-\frac{\olddelta}{2}-\au,\frac{1}{2}-\av)$,
we have
\begin{equs}
\|(0,0,{\cal F}_{1,\omega})\|&\leq
C
\chinese{\xzero}^{-\kappa_{1,1}}\|\triplet\|^2~,
\label{eqn:lasecondomegaev}
\end{equs}
and defining ${\cal F}_{1,1,\omega}(x)=
-\int_{\xzero}^{x}\hspace{0mm}{\rm d}\tilde{x}~
K_{\kdeux}(x-\tilde{x})\qun(\tilde{x})$, we have
\begin{equa}[1][eqn:universalunforomegaun]
\|{\cal F}_{1,\omega}(x)-{\cal F}_{1,1,\omega}(x)\|_{\infty}
&\leq C\chinese{x}^{-\frac{3}{2}+\olddelta}\|\triplet\|^2\\
\|{\cal F}_{1,\omega}(x)-{\cal F}_{1,1,\omega}(x)\|_{1}
&\leq C\chinese{x}^{-1+\olddelta}\|\triplet\|^2\\
\||y|^{\beta}
\big({\cal F}_{1,\omega}(x)-{\cal F}_{1,1,\omega}(x)\big)\|_{2}
&\leq C\chinese{x}^{-\frac{5}{4}+\frac{\beta}{2}+\olddelta}\|\triplet\|^2
~.
\end{equa}
\end{proposition}

\begin{proof}
We write ${\cal F}_{1,\omega}(x)=
{\cal F}_{1,1,\omega}(x)-{\cal F}_{1,2,\omega}(x)-{\cal
F}_{1,3,\omega}(x)$, where ${\cal F}_{1,1,\omega}(x)$ is defined above and
\begin{equs}
{\cal F}_{1,2,\omega}(x)=
\int_{\xzero}^{x}\hspace{-3mm}{\rm d}\tilde{x}~
K_{\khuit}(x-\tilde{x})\qzero(\tilde{x})
~,~~~~~
{\cal F}_{1,3,\omega}(x)=
\int_{\xzero}^{x}\hspace{-3mm}{\rm d}\tilde{x}~
K_{\kdix}(x-\tilde{x})\qzero(\tilde{x})~.
\end{equs}
Then, from the results of Section \ref{app:kernelestimates} and
(\ref{eqn:onuseB})-(\ref{eqn:onuseD}), it follows
easily that
\begin{equs}
\|(0,0,{\cal F}_{1,1,\omega})\|_{x}&\leq
C\|\triplet\|^2
\Big(
\chinese{x}^{\frac{3}{4}}
\A
{\frac{3}{4},\frac{3}{2}-\olddelta}
{\frac{3}{4},\frac{3}{2}-\olddelta,0}
(x,\xzero)
+\chinese{x}^{\frac{3}{4}-\frac{\beta}{2}}
\A
{\frac{3}{4}-\frac{\beta}{2},\frac{3}{2}-\olddelta}
{\frac{3}{4}-\frac{\beta}{2},\frac{3}{2}-\olddelta,0}
(x,\xzero)
\Big)
\\
&\phantom{~\leq}+C\|\triplet\|^2
\Big(
\chinese{x}^{\frac{3}{4}-\frac{\beta}{2}}
\A
{\frac{1}{2},\frac{7}{4}-\frac{\beta}{2}-\olddelta}
{\frac{1}{2},\frac{7}{4}-\frac{\beta}{2}-\olddelta,0}
(x,\xzero)
\Big)\\
&\phantom{~\leq}+C\|\triplet\|^2
\Big(
\chinese{x}^{\frac{3}{2}}
\A
{\frac{3}{4},\frac{9}{4}-\av}
{2,\frac{3}{2}-\olddelta,\frac{1}{2}}
(x,\xzero)
+\chinese{x}
\A
{\frac{1}{2},2-\av}
{\frac{3}{2},\frac{3}{2}-\olddelta,\frac{1}{2}}
(x,\xzero)
\Big)~.
\end{equs}
Using Lemma \ref{lem:blackbox} and $\beta\geq\frac{3}{2}$, we get
\begin{equs}
\|(0,0,{\cal F}_{1,1,\omega})\|&\leq
C
\Big(
\chinese{\xzero}^{-\frac{1}{2}+\olddelta}+
\chinese{\xzero}^{-\frac{1}{2}+\av}
\Big)
\|\triplet\|^2~.
\label{eqn:lasecondomegaevunun}
\end{equs}
Similarly, from the results of Lemma \ref{lem:kerneldeux}, it follows
easily, choosing $\xi_2=1-\epsilon_1$ and $\xi_3=2-2\epsilon_2$ with
$\epsilon_i>0$, that
\begin{equs}
\|(0,0,{\cal F}_{1,2,\omega})\|_{x}&\leq
C\|\triplet\|^2
\Big(
\chinese{x}^{\frac{3}{4}}
\A
{1-\epsilon_1,\frac{5}{4}}
{1-\epsilon_1,\frac{5}{4},0}
(x,\xzero)
+\chinese{x}^{\frac{3}{4}-\frac{\beta}{2}}
\A
{\frac{5}{4}-\frac{\beta}{2},1}
{\frac{5}{4}-\frac{\beta}{2},1,0}
(x,\xzero)
\Big)
\\
&\phantom{~\leq}+C\|\triplet\|^2
\Big(
\chinese{x}^{\frac{3}{4}-\frac{\beta}{2}}
\A
{1-\epsilon_1,\frac{5}{4}-\frac{\beta}{2}}
{1-\epsilon_1,\frac{5}{4}-\frac{\beta}{2},0}
(x,\xzero)
\Big)\\
&\phantom{~\leq}+C\|\triplet\|^2
\Big(
\chinese{x}^{\frac{3}{2}}
\A
{1-\epsilon_2,\frac{7}{4}-\au}
{2,\frac{3}{2},\frac{1}{2}}
(x,\xzero)
+\chinese{x}
\A
{1-\epsilon_1,\frac{3}{2}-\au}
{2,1,\frac{1}{2}}
(x,\xzero)
\Big)~,\\
\|{\cal F}_{1,2,\omega}(x)\|_{\infty}&\leq
C\|\triplet\|^2
\A
{1-\epsilon_1,\frac{3}{2}}
{2,1,\frac{1}{2}}
(x,\xzero)\leq
C\|\triplet\|^2\big(
\chinese{x}^{-\frac{3}{2}+\epsilon_1}+
\chinese{x}^{-\frac{3}{2}}\ln(x)
\big)~,\\
\|{\cal F}_{1,2,\omega}(x)\|_{1}&\leq
C\|\triplet\|^2
\A
{1-\epsilon_1,1}
{1,1,0}
(x,\xzero)\leq
C\|\triplet\|^2\big(
\chinese{x}^{-1+\epsilon_1}+
\chinese{x}^{-1}\ln(x)
\big)~,\\
\||y|^{\beta}{\cal F}_{1,2,\omega}(x)\|_{2}&\leq
C\|\triplet\|^2
\Big(
\A
{\frac{5}{4}-\frac{\beta}{2},1}
{\frac{5}{4}-\frac{\beta}{2},1,0}
(x,\xzero)
+\A
{1-\epsilon_1,\frac{5}{4}-\frac{\beta}{2}}
{1-\epsilon_1,\frac{5}{4}-\frac{\beta}{2},0}
(x,\xzero)
\Big)~.
\end{equs}
Using Lemma \ref{lem:blackbox}, $\ln(1+x)\leq C\chinese{x}^{\olddelta}$ and
$\epsilon_i>0$, we get
\begin{equs}
\|(0,0,{\cal F}_{1,2,\omega})\|&\leq
C
\Big(
\chinese{\xzero}^{-\frac{1}{2}+\olddelta}+
\chinese{\xzero}^{-\frac{1}{2}+\epsilon_1+\au}+
\chinese{\xzero}^{-\frac{1}{4}+\epsilon_2+\au}
\Big)
\|\triplet\|^2~,
\label{eqn:lasecondomegaevundeux}\\
\|{\cal F}_{1,2,\omega}(x)\|_{\infty}&\leq
C\|\triplet\|^2\chinese{x}^{-\frac{3}{2}+\epsilon_1}~,
~~~~
\|{\cal F}_{1,2,\omega}(x)\|_{1}\leq
C\|\triplet\|^2\chinese{x}^{-1+\epsilon_1}~,
\label{eqn:forequivun}
\\
\||y|^{\beta}{\cal F}_{1,2,\omega}(x)\|_{2}&\leq
C\|\triplet\|^2\chinese{x}^{-\frac{5}{4}+\frac{\beta}{2}+\epsilon_1}~.
\label{eqn:forequivdeux}
\end{equs}
Finally, from the results of Lemma \ref{lem:kerneldeuxx}, it follows
easily that
\begin{equs}
\|(0,0,{\cal F}_{1,3,\omega})\|_{x}&\leq
C\|\triplet\|^2
\Big(
\chinese{x}^{\frac{3}{4}}
\B
{\frac{3}{4},1,0}
{\frac{3}{4},1,0}
(x,\xzero)
+\chinese{x}^{\frac{3}{4}-\frac{\beta}{2}}
\B
{\frac{9}{8}-\frac{3\beta}{8},1,\frac{3}{8}+\frac{\beta}{8}}
{\frac{9}{8}-\frac{3\beta}{8},1,\frac{3}{8}+\frac{\beta}{8}}
(x,\xzero)
\Big)
\\
&\phantom{~\leq}+C\|\triplet\|^2
\Big(
\chinese{x}^{\frac{3}{4}-\frac{\beta}{2}}
\B
{\frac{5}{8},\frac{5}{4}-\frac{\beta}{2},\frac{1}{8}}
{\frac{5}{8},\frac{5}{4}-\frac{\beta}{2},\frac{1}{8}}
(x,\xzero)
\Big)\\
&\phantom{~\leq}+C\|\triplet\|^2
\Big(
\chinese{x}^{\frac{3}{2}}
\B
{\frac{3}{4},\frac{7}{4}-\au,0}
{2,1,\frac{1}{2}}
(x,\xzero)
+\chinese{x}
\B
{\frac{5}{8},\frac{3}{2}-\au,\frac{1}{8}}
{\frac{13}{8},1,\frac{5}{8}}
(x,\xzero)
\Big)~,\\
\|{\cal F}_{1,3,\omega}(x)\|_{\infty}&\leq
C\|\triplet\|^2
\B
{\frac{5}{8},\frac{3}{2},\frac{1}{8}}
{1,1,0}
(x,\xzero)\leq
C\|\triplet\|^2
\chinese{x}^{-\frac{3}{2}+\olddelta}~,\\
\|{\cal F}_{1,3,\omega}(x)\|_{1}&\leq
C\|\triplet\|^2
\B
{\frac{5}{8},1,\frac{1}{8}}
{\frac{5}{8},1,\frac{1}{8}}
(x,\xzero)\leq
C\|\triplet\|^2
\chinese{x}^{-1+\olddelta}~,\\
%
%
\||y|^{\beta}{\cal F}_{1,3,\omega}(x)\|_{2}&\leq
C\|\triplet\|^2
\Big(
\B
{\frac{9}{8}-\frac{3\beta}{8},1,\frac{3}{8}+\frac{\beta}{8}}
{\frac{9}{8}-\frac{3\beta}{8},1,\frac{3}{8}+\frac{\beta}{8}}
(x,\xzero)
+
\B
{\frac{5}{8},\frac{5}{4}-\frac{\beta}{2},\frac{1}{8}}
{\frac{5}{8},\frac{5}{4}-\frac{\beta}{2},\frac{1}{8}}
(x,\xzero)
\Big)~,\\
&\leq
C\|\triplet\|^2
\chinese{x}^{-\frac{5}{4}+\frac{\beta}{2}+\olddelta}~,
\end{equs}
where in the last inequality, we used $\beta\geq\frac{1}{2}$.
Using Lemma \ref{lem:blackbox} and $\beta\geq1$, we get
\begin{equs}
\|(0,0,{\cal F}_{1,3,\omega})\|&\leq
C
\chinese{\xzero}^{-\frac{1}{4}+\frac{\olddelta}{2}+\au}
\|\triplet\|^2~.
\label{eqn:lasecondomegaevuntrois}
\end{equs}
The proof of (\ref{eqn:lasecondomegaev}) and
(\ref{eqn:universalunforomegaun}) is completed choosing
$\epsilon_1=\olddelta$ and $\epsilon_2=\frac{\olddelta}{2}$ in
(\ref{eqn:lasecondomegaevundeux})-(\ref{eqn:forequivdeux}).
\end{proof}
We now turn to ${\cal F}_{1,v}$. For further reference, we also show that
substracting some terms to ${\cal F}_{1,v}$ gives improved decay rates
compared to those of $v$.
\begin{proposition}\label{prop:thecalfisvev}
Assume that $\qzero$ and $\qun$ satisfy the bounds
(\ref{eqn:nonlinbounds}), and that the parameters satisfy
(\ref{eqn:restrictions}), 
then there exist a constant $C$ such that for
$\kappa_{1,2}=\min(\kappa_{1,1},\frac{\olddelta}{2},\frac{1}{2}-\au+\av
-2\olddelta)$, we have
\begin{equs}
\|(0,{\cal F}_{1,v},0)\|&\leq
C
\chinese{\xzero}^{-\kappa_{1,2}}
\|\triplet\|^2~,
\label{eqn:lasecondvev}\\
\|{\cal F}_{1,v}(x)-{\cal F}_{1,\omega}(x)-{\cal F}_{1,3,v}(x)\|_{\infty}
&\leq
C\chinese{x}^{-\frac{3}{2}+\olddelta}\chinese{\xzero}^{\olddelta}
\|\triplet\|^2~,
\end{equs}
where ${\cal F}_{1,3,v}(x)=\int_{\xzero}^{x}\hspace{0mm}{\rm d}\tilde{x}~
\galpha(x-\tilde{x})\qun(\tilde{x})$~.
\end{proposition}

\begin{proof}
We first note that we can write ${\cal F}_{1,v}(x)=
{\cal F}_{1,\omega}(x)+{\cal F}_{1,1,v}(x)
+{\cal F}_{1,2,v}(x)+{\cal F}_{1,3,v}(x)$ with ${\cal F}_{1,3}$ as above
and
\begin{equs}
{\cal F}_{1,1,v}(x)&=
\int_{\xzero}^{x}\hspace{-3mm}{\rm d}\tilde{x}~
\big(
 K_{\kr}(x-\tilde{x})
+K_{\ki}(x-\tilde{x})
\big)\qzero(\tilde{x})~,~~~~
{\cal F}_{1,2,v}(x)=
\int_{\xzero}^{x}\hspace{-3mm}{\rm d}\tilde{x}~
K_{\ktreize}(x-\tilde{x})\qun(\tilde{x})~.
\label{eqn:f23vev}
\end{equs}
Using (\ref{eqn:interpolation}), we see that the contribution of ${\cal
F}_{1,\omega}$ to (\ref{eqn:lasecondvev}) is already proved in
Proposition \ref{prop:thecalfisomegaevolve}. Then, from the results of
Lemma \ref{lem:withkr} and \ref{lem:withki}, it follows easily that
\begin{equs}
\|K_{\kr}(x)+K_{\ki}(x)\|_{\L^1}
&\leq
C\ed^{\frac{b(\strouhal)(x-\tilde{x})}{4}}
{\textstyle
\left(
\frac{1}{x^{\frac{1}{2}}}+
\frac{\chinese{x}^{\frac{1}{8}}}{x^{\frac{1}{8}}}+
\frac{
\chinese{x}^{\frac{1}{8}}
\chinese{\xzero}^{\olddelta}}{x^{\frac{1}{4}}}
\right)}
\equiv CB_1(x)
~,\\
\|K_{\ktreize}(x)\|_{\L^1}
&\leq
C\ed^{\frac{b(\strouhal)(x-\tilde{x})}{4}}
{\textstyle
\left(
1+
\frac{
\chinese{\xzero}^{\olddelta}}{x^{\frac{1}{4}}}
\right)}
\equiv C D_1(x)
~.
\end{equs}
We then have
\begin{equs}
\|(0,{\cal F}_{1,1,v},0)\|_{x}&\leq
C\|\triplet\|^2
\int_{\xzero}^{x}\hspace{-3mm}{\rm d}\tilde{x}
\chinese{x}^{1-\olddelta}
B_{1}(x-\tilde{x})\chinese{\tilde{x}}^{-\frac{3}{2}}
\\&\phantom{\leq~}+
C\|\triplet\|^2
\int_{\xzero}^{x}\hspace{-3mm}{\rm d}\tilde{x}
\chinese{x}^{1-\olddelta-\frac{1}{\pv}}
B_{1}(x-\tilde{x})\chinese{\tilde{x}}^{-\frac{3}{2}+\frac{1}{2\pv}}
\\&\phantom{\leq~}+
C\|\triplet\|^2
\int_{\xzero}^{x}\hspace{-3mm}{\rm d}\tilde{x}
\chinese{x}^{\frac{3}{2}-\frac{1}{2\pdv}-\av}
B_{1}(x-\tilde{x})\chinese{\tilde{x}}^{-2+\frac{1}{2\pdv}+\au}
~,\\
\|(0,{\cal F}_{1,2,v},0)\|_{x}&\leq
C\|\triplet\|^2
\int_{\xzero}^{x}\hspace{-3mm}{\rm d}\tilde{x}
\chinese{x}^{1-\olddelta}
D_{1}(x-\tilde{x})\chinese{\tilde{x}}^{-2+\olddelta}
\\&\phantom{\leq~}+
C\|\triplet\|^2
\int_{\xzero}^{x}\hspace{-3mm}{\rm d}\tilde{x}
\chinese{x}^{1-\olddelta-\frac{1}{\pv}}
D_{1}(x-\tilde{x})\chinese{\tilde{x}}^{-2+\olddelta+\frac{1}{2\pv}}
\\&\phantom{\leq~}+
C\|\triplet\|^2
\int_{\xzero}^{x}\hspace{-3mm}{\rm d}\tilde{x}
\chinese{x}^{\frac{3}{2}-\frac{1}{2\pdv}-\av}
D_{1}(x-\tilde{x})\chinese{\tilde{x}}^{-\frac{5}{2}+\frac{1}{2\pdv}+\av}
~,\\
\|{\cal F}_{1,1,v}(x)\|_{\infty}&\leq
C\|\triplet\|^2
\int_{\xzero}^{x}\hspace{-3mm}{\rm d}\tilde{x}
B_{1}(x-\tilde{x})\chinese{\tilde{x}}^{-\frac{3}{2}}~,\\
\|{\cal F}_{1,2,v}(x)\|_{\infty}&\leq
C\|\triplet\|^2
\int_{\xzero}^{x}\hspace{-3mm}{\rm d}\tilde{x}
D_{1}(x-\tilde{x})\chinese{\tilde{x}}^{-2+\olddelta}~.
\end{equs}
Using Lemma \ref{lem:blackbox}, we get 
\begin{equs}
\|(0,{\cal F}_{1,1,v},0)\|_{x}+\|(0,{\cal F}_{1,2,v},0)\|_{x}&\leq
C\Big(
\chinese{\xzero}^{-\frac{1}{2}+\olddelta}
+
\chinese{\xzero}^{-\frac{1}{2}+\au-\av+2\olddelta}
\Big)
\|\triplet\|^2~,\\
\|{\cal F}_{1,1,v}(x)\|_{\infty}+\|{\cal F}_{1,2,v}(x)\|_{\infty}&\leq
C\chinese{x}^{-\frac{3}{2}+\olddelta}
\chinese{\xzero}^{\olddelta}
\|\triplet\|^2~.
\end{equs}
In the same way, from the results of Lemma \ref{lem:sourcelikeesti}, it
follows easily that for all $q>1$ and $s\geq1$, we have
\begin{equs}
\|\partial_y\galpha(x)\|_{\L^2}
&\leq
Cx^{-\frac{3}{2}}~,~~~~~~
\|\galpha(x)\|_{\L^q}
\leq
Cx^{-1+\frac{1}{q}}\Big(
1+\frac{\chinese{\xzero}^{\frac{1}{4q}}}
{x^{\frac{1}{4q}}}
\Big)\equiv
C E_{q}(x)~.
\end{equs}
We then have
\begin{equs}
\|{\cal F}_{1,3,v}\|_{\Lx{x}{\infty}{1-\olddelta}}
&\leq
C\|\triplet\|^2
\chinese{x}^{1-\olddelta}
\int_{\xzero}^{x}\hspace{-3mm}{\rm d}\tilde{x}~
\chinese{\tilde{x}}^{-\frac{3}{2}+\frac{\olddelta}{2}}
E_{\frac{1}{\olddelta}}(x-\tilde{x})
\\
\|{\cal F}_{1,3,v}\|_{\Lx{x}{\pv}{1-\olddelta-\frac{1}{\pv}}}
&\leq
C\|\triplet\|^2
\chinese{x}^{1-\olddelta-\frac{1}{\pv}}
\int_{\xzero}^{x}\hspace{-3mm}{\rm d}\tilde{x}~
\chinese{\tilde{x}}^{-\frac{3}{2}+\olddelta}E_{\pv}(x-\tilde{x})
\\
\|\partial_y{\cal F}_{1,3,v}
\|_{\Lx{x}{\pdv}{\frac{3}{2}-\frac{1}{2\pdv}-\av}}
&\leq
C\|\triplet\|^2
\chinese{x}^{\frac{3}{2}-\frac{1}{2\pdv}-\av}
\int_{\xzero}^{x}
\hspace{-3mm}{\rm d}\tilde{x}
~\min\Big(
\chinese{\tilde{x}}^{-\frac{9}{4}+\frac{1}{2\pdv}+\av}
E_{2}(x-\tilde{x})
,
\frac{\tilde{x}^{-\frac{7}{4}+\frac{1}{2\pdv}}}{
(x-\tilde{x})^{\frac{3}{2}}}
\Big)~.
\end{equs}
Using Lemma \ref{lem:blackbox}, and
$\pdv>2$, we get
\begin{equs}
\|(0,{\cal F}_{1,3,v},0)\|_{x}\leq
C\chinese{\xzero}^{-\frac{\olddelta}{2}}\|\triplet\|^2~.
\end{equs}
The proof is completed.
\end{proof}
We conclude this section by estimating ${\cal F}_{1,u}$. In the spirit of 
Proposition \ref{prop:thecalfisomegaevolve}, we will also show that
substracting the `right' term to ${\cal F}_{1,u}$ improves its decay rate
as $x\to\infty$.
\begin{proposition}\label{lem:thecalfisuevolve}
Assume that $\qzero$ and $\qun$ satisfy the bounds
(\ref{eqn:nonlinbounds}), and that the parameters satisfy
(\ref{eqn:restrictions}),
then there exist a constant $C$ such that for 
$\kappa_{1,3}=\min(\kappa_{1,2},\frac{1}{2}-(1+\frac{1}{2\pdv})\olddelta,
\frac{1}{2}-\av+\au-\frac{\olddelta}{\pdv})$, we have
\begin{equs}
\|({\cal F}_{1,u},0,0)\|&\leq 
\chinese{\xzero}^{-\kappa_{1,3}}
\|\triplet\|^2
~,
\label{eqn:laseconduev}
\end{equs}
for all $x\geq\xzero$. Furthermore, let
\begin{equs}
{\cal F}_{1,2,u}(x)&=
-\int_{\xzero}^{x}\hspace{-3mm}{\rm d}\tilde{x}~
K_{\kdouze}(x-\tilde{x})\qun(\tilde{x})\\
\\
{\cal F}_{1,4,u}(x)&=
\int_{\xzero}^{x}\hspace{-3mm}{\rm d}\tilde{x}~\Big(\big(
\P K_{\kdeux}(x-\tilde{x})
-K_{\ktreize}(x-\tilde{x})\big)\qzero(\tilde{x})-
\P K_{\kdouze}(x-\tilde{x})\qun(\tilde{x})\Big)~.
\label{eqn:f24uev}
\end{equs}
then for all
$\epsilon>0$, there exists a constant $C$ such that
\begin{equs}
\|{\cal F}_{1,u}(x)-{\cal F}_{1,2,u}(x)\|_{\infty}
\leq C\chinese{x}^{-1+\olddelta}\|\triplet\|^2~,~~~~
\|{\cal F}_{1,4,u}(x)\|_{\infty}
\leq C\chinese{x}^{-\frac{3}{2}}
\chinese{\xzero}^{\frac{5\olddelta}{2}}\|\triplet\|^2~.
\end{equs}
\end{proposition}

\begin{proof}
We first note that with ${\cal F}_{1,2,u}$ as above, we can write
${\cal F}_{1,u}(x)={\cal F}_{1,1,u}(x)
+{\cal F}_{1,2,u}(x)+{\cal F}_{1,3,u}(x)$ with
\begin{equs}
{\cal F}_{1,1,u}(x)&=
\int_{\xzero}^{x}\hspace{-3mm}{\rm d}\tilde{x}~
\big(K_{\kdeux}(\tilde{x}-x)-K_{\ktreize}(\tilde{x}-x)\big)\qzero(\tilde{x})
~,~~~~
{\cal F}_{1,3,u}(x)=
-\int_{\xzero}^{x}\hspace{-3mm}{\rm d}\tilde{x}~
\falpha(\tilde{x}-x)\qun(\tilde{x})~.
\label{eqn:f21uev}
\end{equs}
Then we note that 
$\|({\cal F}_{1,3,u},0,0)\|\leq\|(0,{\cal F}_{1,3,u},0)\|$ (see
(\ref{eqn:interpolation})), and that
${\cal F}_{1,3,u}$ and ${\cal F}_{1,3,v}$ differ only
by signs and the exchange of the Kernels $\falpha$ and $\galpha$. The bound
on ${\cal F}_{1,3,v}$ in the proof of Proposition \ref{prop:thecalfisvev}
being insensitive to these details then apply mutatis mutandis, in
particular, we have $\|{\cal F}_{1,3,u}(x)\|_{\infty}\leq
C\chinese{x}^{-1+\olddelta}\|\triplet\|^2$. Then, by Lemma
\ref{lem:withkr}, we have
\begin{equs}
\|K_{\kdouze}(x)\|_{\L^p}&\leq
C\frac{
\chinese{x}^{\frac{1}{2}-\frac{1}{2p}}
}{x^{1-\frac{1}{p}}}
\Big(
1+
\frac{\chinese{\xzero}^{\frac{\olddelta}{p}}
}{x^{\frac{1}{4p}}}
\Big)
\equiv C E_{p}(x)~,\\
\|K_{\kdeux}(x)\|_{\L^p}+
\|K_{\ktreize}(x)\|_{\L^p}&\leq
C\left(
\frac{1}{x^{1-\frac{1}{2p}}}
+\frac{
\ed^{\frac{b(\strouhal)x}{4}}
\chinese{x}^{\frac{1}{2}-\frac{1}{2p}}
}{x^{1-\frac{1}{p}}}
\Big(
1+
\frac{\chinese{\xzero}^{\frac{\olddelta}{p}}
}{x^{\frac{1}{4p}}}
\Big)
\right)
\equiv C H_p(x)~,\\
\|\partial_yK_{\kdeux}(x)\|_{\L^2}+
\|\partial_yK_{\ktreize}(x)\|_{\L^2}&\leq
C\left(
\frac{\chinese{x}^{\frac{1}{2}}}{x^{\frac{7}{4}}}
+\frac{
\ed^{\frac{b(\strouhal)x}{4}}
\chinese{x}^{\frac{3}{4}}
}{x^{\frac{3}{2}}}
\right)
\equiv C J(x)~,
\end{equs}
so that for all $p_0\in[\pu,\infty)$, we have
\begin{equs}
\|{\cal F}_{1,2,u}\|_{\Lx{x}{p_0}{\frac{1}{2}-\frac{1}{p_0}}}
&\leq
C\|\triplet\|^2
\chinese{x}^{\frac{1}{2}-\frac{1}{p_0}}
\int_{\xzero}^{x}\hspace{-3mm}{\rm d}\tilde{x}
~\min\Big(
\chinese{\tilde{x}}^{-2+\olddelta+\frac{1}{2p_0}}
E_1(x-\tilde{x}),
\chinese{\tilde{x}}^{-\frac{3}{2}+\olddelta}
E_{p_0}(x-\tilde{x})
\Big)
\\
\|\partial_y{\cal F}_{1,2,u}
\|_{\Lx{x}{\pdv}{1-\frac{1}{2\pdv}-\au}}
&\leq
C\|\triplet\|^2
\chinese{x}^{1-\frac{1}{2\pdv}-\au}
\int_{\xzero}^{x}
\hspace{-3mm}{\rm d}\tilde{x}~\min\Big(
\chinese{\tilde{x}}^{-2+\xi}
E_{\pdv}(x-\tilde{x})
,
\frac{\chinese{x}^{1-\frac{1}{2\pdv}}\chinese{\tilde{x}}^{-\frac{3}{2}+\olddelta}
}{(x-\tilde{x})^{2-\frac{1}{\pdv}}}
\Big)~,\\
\|{\cal F}_{1,1,u}\|_{p_0}
&\leq
C\|\triplet\|^2
\int_{\xzero}^{x}\hspace{-3mm}{\rm d}\tilde{x}
~\min\Big(
\chinese{\tilde{x}}^{-\frac{3}{2}+\frac{1}{2p_0}}
H_1(x-\tilde{x}),
\chinese{\tilde{x}}^{-1}
H_{p_0}(x-\tilde{x})
\Big)
\\
\|\partial_y{\cal F}_{1,1,u}
\|_{\Lx{x}{\pdv}{1-\frac{1}{2\pdv}-\au}}
&\leq
C\|\triplet\|^2
\chinese{x}^{1-\frac{1}{2\pdv}-\au}
\int_{\xzero}^{x}
\hspace{-3mm}{\rm d}\tilde{x}~\min\Big(
\chinese{\tilde{x}}^{-\frac{3}{2}+\au}
H_{\pdv}(x-\tilde{x})
,
\chinese{\tilde{x}}^{-\frac{5}{4}+\frac{1}{2\pdv}}
J(x-\tilde{x})
\Big)~.
\end{equs}
By Lemma \ref{lem:blackbox}, using these bounds with 
$p_0=\pu$ and $p_0=\infty$ and $\ln(1+x)\leq C\chinese{x}^{\olddelta}$, we
get
\begin{equs}
\|({\cal F}_{1,2,u},0,0)\|_{x}&\leq
C\Big(
\chinese{\xzero}^{-\frac{1}{2}+\olddelta}+
\chinese{\xzero}^{-\frac{1}{2}+\av-\au+\frac{\olddelta}{\pdv}-\frac{1}{4\pdv}}
\Big)
\|\triplet\|^2~,\\
\|({\cal F}_{1,1,u},0,0)\|_{x}&\leq
C\Big(
\chinese{\xzero}^{-\frac{1}{2}+\olddelta}+
\chinese{\xzero}^{-\frac{1}{2}-\frac{1}{2\pdv}+(1+\frac{3}{2\pdv})\olddelta}
\Big)
\|\triplet\|^2~,\\
\|{\cal F}_{1,1,u}(x)\|_{\infty}&\leq
C
\chinese{x}^{-1+\olddelta}
\|\triplet\|^2~.
\end{equs}
We finally note that
\begin{equs}
\|{\cal F}_{1,4,u}(x)\|_{\infty}
&\leq
C\|\triplet\|^2
\int_{\xzero}^{x}
\hspace{-3mm}{\rm d}\tilde{x}~
\ed^{\frac{b(\strouhal)(x-\tilde{x}}{4}}
\Big(
1+\frac{1}{x^{\frac{1}{2}}}+
\frac{\chinese{\xzero}^{\olddelta}}{(x-\tilde{x})^{\frac{1}{4}}}
\Big)
\chinese{\tilde{x}}^{-\frac{3}{2}}\\
&\leq
C\chinese{x}^{-\frac{3}{2}}
\chinese{\xzero}^{\frac{5\olddelta}{2}}
\|\triplet\|^2~,
\end{equs}
which completes the proof.
\end{proof}

\subsection{The nonlinear terms II}\label{sec:nonlineartermsdeux}

In this section, we prove the
\begin{theorem}\label{thm:onthefdeux}
Assume that $\qzero$ and $\qun$ satisfy the bounds
(\ref{eqn:nonlinbounds}), and that the parameters satisfy
(\ref{eqn:restrictions}), then there exist constants $C$ and 
$\kappa_2>0$ such that
\begin{equs}
\|({\cal F}_{2,u},{\cal F}_{2,v},{\cal F}_{2,\omega})\|&\leq
C\chinese{\xzero}^{-\kappa_2}\|\triplet\|^2~.
\label{eqn:lasecondomeganonevthm}
\end{equs}
\end{theorem}
For convenience, the proof is split component-wise in the next three
Propositions. For further reference, we will also point out that most
decay rates on ${\cal F}_{2,\cdot}$ are in fact better than those of the
related fields.
\begin{proposition}\label{prop:thecalfisomega}
Assume that $\qzero$ and $\qun$ satisfy the bounds
(\ref{eqn:nonlinbounds}), and that the parameters satisfy
(\ref{eqn:restrictions}), then there exist a constant $C$ 
such that for $\kappa_{2,1}=\frac{1}{4}-\au$, we have
\begin{equs}
\|(0,0,{\cal F}_{2,\omega})\|&\leq
C\chinese{\xzero}^{-\kappa_{2,1}}\|\triplet\|^2~,~~~~~
\|{\cal F}_{2,\omega}(x)\|_{\infty}
\leq C\chinese{x}^{-\frac{3}{2}}\|\triplet\|^2
\\
\|{\cal F}_{2,\omega}(x)\|_{1}
&\leq C\chinese{x}^{-1}\|\triplet\|^2~,~~~~~
\||y|^{\beta}{\cal F}_{2,\omega}(x)\|_{2}
\leq C\chinese{x}^{-\frac{5}{4}+\frac{\beta}{2}}\|\triplet\|^2~,
\end{equs}
for all $x\geq\xzero$.
\end{proposition}

\begin{proof}
From the results of section \ref{app:kernelestimates}, it follows easily
that there are exponents $p\geq0$ and $q<1$ such that
\begin{equs}
\|\ed^xK_{2,1,\omega}\|_{\K{1}{p}{q}}+
\|\ed^xK_{2,2,\omega}\|_{\K{1}{p}{q}}+
\|\ed^xK_{2,1,\omega}\|_{\K{2}{p}{q}}+
\|\ed^xK_{2,2,\omega}\|_{\K{2}{p}{q}}&\leq C~,\\
\|\ed^x|y|^{\beta}K_{2,1,\omega}\|_{\K{1}{p}{q}}+
\|\ed^x|y|^{\beta}K_{2,2,\omega}\|_{\K{1}{p}{q}}&\leq C~,
\end{equs}
while for all $\xzero\leq x\leq\tilde{x}$, we have
\begin{equs}
\chinese{x}^{\frac{3}{4}}
\big(\|\qzero(\tilde{x})\|_{2}+\|\qun(\tilde{x})\|_{2}\big)
&\leq 
\chinese{x}^{-\frac{1}{2}}
\big(
 \|\qzero\|_{\Lx{\tilde{x}}{2}{\frac{5}{4}}}
+\|\qun\|_{\Lx{\tilde{x}}{2}{\frac{7}{4}-\olddelta}}\big)~,\\
\chinese{x}^{\frac{3}{2}}
\big(\|\partial_y\qzero(\tilde{x})\|_{2}
+\|\partial_y\qun(\tilde{x})\|_{2}\big)
&\leq 
\chinese{x}^{\au-\frac{1}{4}}
\big(
 \|\partial_y\qzero\|_{\Lx{\tilde{x}}{2}{\frac{7}{4}-\au}}
+\|\partial_y\qun\|_{\Lx{\tilde{x}}{2}{\frac{9}{4}-\av}}\big)~,\\
\chinese{x}
\big(\|\partial_y\qzero(\tilde{x})\|_{1}
+\|\partial_y\qun(\tilde{x})\|_{1}\big)
&\leq 
\chinese{x}^{\au-\frac{1}{2}}
\big(
 \|\partial_y\qzero\|_{\Lx{\tilde{x}}{1}{\frac{3}{2}-\au}}
+\|\partial_y\qun\|_{\Lx{\tilde{x}}{1}{2-\av}}\big)~,\\
\chinese{x}^{\frac{3}{4}-\frac{\beta}{2}}
\big(\||y|^{\beta}\qzero(\tilde{x})\|_{2}
+\||y|^{\beta}\qun(\tilde{x})\|_{2}\big)&\leq
\chinese{x}^{-\frac{1}{2}}
\big(
\||y|^{\beta}\qzero\|_{\Lx{\tilde{x}}{2}{\frac{5}{4}-\frac{\beta}{2}}}
+\||y|^{\beta}\qun\|_{\Lx{\tilde{x}}{2}{\frac{7}{4}-\olddelta-\frac{\beta}{2}}}
\big)~,\\
\chinese{x}^{\frac{3}{4}-\frac{\beta}{2}}
\big(\|\qzero(\tilde{x})\|_{1}
+\|\qun(\tilde{x})\|_{1}\big)
&\leq
\chinese{x}^{-\frac{1}{2}}\big(
\|\qzero\|_{\Lx{\tilde{x}}{1}{1}}
+\|\qun\|_{\Lx{\tilde{x}}{1}{\frac{3}{2}-\olddelta}}
\big)~,
\end{equs}
since $\beta>\frac{3}{2}$, $\olddelta\leq\av<\frac{1}{2}$ and
$\frac{1}{2}-\av+\au\geq0$. By
(\ref{eqn:nonlinbounds}), the above quantities are all bounded by 
$C\chinese{x}^{\au-\frac{1}{4}}\|\triplet\|^2$, while
$\chinese{x}^{\frac{3}{2}}\big(\|\qzero(\tilde{x})\|_{\infty}
+\|\qun(\tilde{x})\|_{\infty}\big)+
\chinese{x}\big(\|\qzero(\tilde{x})\|_{1}+\|\qun(\tilde{x})\|_{1}\big)\leq
C\|\triplet\|^2$. Easy estimates applied to 
(\ref{eqn:f2omega}) thus lead to
\begin{equs}
\|(0,0,{\cal F}_{2,\omega})\|_x&\leq
C\chinese{x}^{\au-\frac{1}{4}}\|\triplet\|^2
\int_{x}^{\infty}\hspace{-4mm}{\rm d}\tilde{x}~
\frac{\ed^{x-\tilde{x}}\chinese{\tilde{x}-x}^p}{(\tilde{x}-x)^{q}}~,\\
\|{\cal F}_{2,\omega}(x)\|_{\infty}&\leq
C\chinese{x}^{-\frac{3}{2}}\|\triplet\|^2
\int_{x}^{\infty}\hspace{-4mm}{\rm d}\tilde{x}~
\frac{\ed^{x-\tilde{x}}\chinese{\tilde{x}-x}^p}{(\tilde{x}-x)^{q}}~,\\
\|{\cal F}_{2,\omega}(x)\|_{1}&\leq
C\chinese{x}^{-1}\|\triplet\|^2
\int_{x}^{\infty}\hspace{-4mm}{\rm d}\tilde{x}~
\frac{\ed^{x-\tilde{x}}\chinese{\tilde{x}-x}^p}{(\tilde{x}-x)^{q}}~,\\
\||y|^{\beta}{\cal F}_{2,\omega}(x)\|_{2}&\leq
C\chinese{x}^{-\frac{5}{4}+\frac{\beta}{2}}\|\triplet\|^2
\int_{x}^{\infty}\hspace{-4mm}{\rm d}\tilde{x}~
\frac{\ed^{x-\tilde{x}}\chinese{\tilde{x}-x}^p}{(\tilde{x}-x)^{q}}~.
\end{equs}
This completes the proof.
\end{proof}

\begin{proposition}\label{prop:thecalfisv}
Assume that $\qzero$ and $\qun$ satisfy the bounds
(\ref{eqn:nonlinbounds}), and that the parameters satisfy
(\ref{eqn:restrictions}), then there exist a constant $C$ such that
for $\kappa_{2,2}=\min(\kappa_{2,1},\frac{\olddelta}{2})$, we have
\begin{equs}
\|(0,{\cal F}_{2,v},0)\|&\leq
C\chinese{\xzero}^{-\kappa_{2,2}}\|\triplet\|^2~,
\label{eqn:lasecondv}\\
\|{\cal F}_{2,v}(x)\|_{\infty}&\leq
C\chinese{\xzero}^{-\frac{3}{2}+(1+\epsilon)\olddelta}\|\triplet\|^2~.
\label{eqn:lasecondvuniv}
\end{equs}
for all $x\geq\xzero$ and $0<\epsilon\leq1$.
\end{proposition}

\begin{proof}
We first note that we can write ${\cal F}_{2,v}(x)={\cal F}_{2,\omega}(x)+
{\cal F}_{2,1,v}(x)+{\cal F}_{2,2,v}(x)$ where
\begin{equs}
{\cal F}_{2,1,v}(x)&=
-\int_{x}^{\infty}\hspace{-4mm}{\rm d}\tilde{x}~
\ed^{-(\tilde{x}-x)}K_{\ksept}(\tilde{x}-x)\qzero(\tilde{x})
-\ed^{-(\tilde{x}-x)}K_{\ksix}(\tilde{x}-x)\qun(\tilde{x})~,
\label{eqn:f21v}
\\
{\cal F}_{2,2,v}(x)&=-
\int_{x}^{\infty}\hspace{-4mm}{\rm d}\tilde{x}~
\galpha^{*}(\tilde{x}-x)\qun(\tilde{x})~.
\label{eqn:f22v}
\end{equs}
Using (\ref{eqn:interpolation}), we see again that the contribution of
${\cal F}_{2,\omega}$ to (\ref{eqn:lasecondv}) is already proved in
Proposition \ref{prop:thecalfisomega}. For the contribution of ${\cal
F}_{2,1,v}$ to (\ref{eqn:lasecondv}), we proceed as in Proposition
\ref{prop:thecalfisomega}. There are exponents $p\geq0$ and $q<1$ such that
\begin{equs}
\|K_{\ksix}\|_{\K{1}{p}{q}}+
\|K_{\ksept}\|_{\K{1}{p}{q}}\leq C~,
\end{equs}
while for all $\xzero\leq x\leq\tilde{x}$, we have
\begin{equs}
\chinese{x}^{1-\olddelta}
\big(
\|\qzero(\tilde{x})\|_{\infty}
+\|\qun(\tilde{x})\|_{\infty}\big)
&\leq 
\chinese{x}^{-\frac{1}{2}-\olddelta}\big(
\|\qzero\|_{\Lx{\tilde{x}}{\infty}{\frac{3}{2}}}
+
\|\qun\|_{\Lx{\tilde{x}}{\infty}{2-\olddelta}}\big)~,\\
\chinese{x}^{1-\olddelta-\frac{1}{p}}
\big(\|\qzero(\tilde{x})\|_{p}
+\|\qun(\tilde{x})\|_{p}&\leq
\chinese{x}^{-\frac{1}{2}-\olddelta-\frac{1}{2p}}\big(
\|\qzero\|_{\Lx{\tilde{x}}{p}{\frac{3}{2}-\frac{1}{2p}}}
+
\|\qun\|_{\Lx{\tilde{x}}{p}{2-\olddelta-\frac{1}{2p}}}\big)~,\\
\chinese{x}^{\frac{3}{2}-\frac{1}{2\pdv}-\av}
\big(\|\partial_y\qzero(\tilde{x})\|_{\pdv}
+\|\partial_y\qun(\tilde{x})\|_{\pdv}\big)
&\leq
\chinese{x}^{-\frac{1}{2}}\big(
\|\partial_y\qzero\|_{\Lx{\tilde{x}}{\pdv}{2-\frac{1}{2\pdv}-\au}}
+
\|\partial_y\qun\|_{\Lx{\tilde{x}}{\pdv}{\frac{5}{2}-\frac{1}{2r}-\av}}
\big)
~,
\end{equs}
since $\olddelta\leq\av<\frac{1}{2}$ and $\au\leq\av$. By
(\ref{eqn:nonlinbounds}), the above quantities are bounded by 
$C\chinese{x}^{-\frac{1}{2}}\|\triplet\|^2$, and we get
\begin{equs}
\|(0,{\cal F}_{2,1,v},0)\|_x&\leq
C\chinese{x}^{-\frac{1}{2}}\|\triplet\|^2
\int_{x}^{\infty}\hspace{-4mm}{\rm d}\tilde{x}~
\frac{\ed^{x-\tilde{x}}\chinese{\tilde{x}-x}^p}{(\tilde{x}-x)^{q}}~,\\
\|{\cal F}_{2,1,v}(x)\|_{\infty}&\leq
C\chinese{x}^{-\frac{3}{2}+\olddelta}\|\triplet\|^2~.
\end{equs}
Next, we note that for $x\leq\tilde{x}$ and $q>1$, we have
\begin{equs}
\|\Q\galpha\|_{\K{q}{0}{1-\frac{1}{q}}}
&\leq C~,~~~~~
\|\P\galpha\|_{\K{q}{0}{1-\frac{3}{4q}}}
\leq C\chinese{\xzero}^{\frac{1}{4q}}
\leq C\chinese{\tilde{x}}^{\frac{1}{4q}}~,
\end{equs}
where in the last inequality, we used that
$\strouhal^{-\frac{1}{4q}}\leq\chinese{\strouhal}^{-\frac{1}{4q}}
\chinese{\xzero}^{\frac{\olddelta}{4q}}\leq\chinese{\xzero}^{\frac{1}{4q}}$.
Then, for all $0<\epsilon\leq1$, after the change of variables
$\tilde{x}=xz$, we get
\begin{equs}
\|{\cal F}_{2,1,v}\|_{\Lx{x}{\infty}{1-\olddelta}}
&\leq
C\|\triplet\|^2\chinese{x}^{-\frac{1}{2}+\epsilon\olddelta}
\int_1^{\infty}
\hspace{-4mm}{\rm d}z
~
\frac{z^{-\frac{3}{2}+(1-\epsilon)\olddelta}}{
(z-1)^{1-2\epsilon\olddelta}}
\big(1+(z-1)^{-\frac{\olddelta}{4}}\big)~,
\\
\|{\cal F}_{2,1,v}\|_{\Lx{x}{p}{1-\olddelta-\frac{1}{p}}}
&\leq
C\|\triplet\|^2\chinese{x}^{-\frac{1}{2}}
\int_1^{\infty}
\hspace{-4mm}{\rm d}z
~\frac{z^{-\frac{3}{2}+\olddelta}}{
(z-1)^{1-\frac{1}{p}}}
\big(1+(z-1)^{-\frac{1}{4p}}\big)~,
\\
\|\partial_y{\cal F}_{2,1,v}
\|_{\Lx{x}{\pdv}{\frac{3}{2}-\frac{1}{2\pdv}-\av}}
&\leq
C\|\triplet\|^2
\chinese{x}^{-\frac{1}{4}}
\int_1^{\infty}
\hspace{-4mm}{\rm d}z
~z^{-\frac{9}{4}+\frac{1}{2\pdv}+\au}
\big((z-1)^{-\frac{1}{2}}+
(z-1)^{-\frac{5}{8}}\big)~.
\end{equs}
This completes the proof.
\end{proof}

\begin{proposition}\label{prop:thecalfisu}
Assume that $\qzero$ and $\qun$ satisfy the bounds
(\ref{eqn:nonlinbounds}), and that the parameters satisfy
(\ref{eqn:restrictions}),
then there exist a constant $C$ such that for 
$\kappa_{2,3}=\kappa_{2,2}$
\begin{equs}
\|({\cal F}_{2,u},0,0)\|&\leq
C\chinese{\xzero}^{-\kappa_{2,3}}\|\triplet\|^2~,
\label{eqn:lasecondu}\\
\|{\cal F}_{2,u}(x)\|_{\infty}&\leq
C\chinese{x}^{-\frac{3}{2}+(1+\epsilon)\olddelta}\|\triplet\|^2~,
\label{eqn:universalfdeuxu}
\end{equs}
for all $x\geq\xzero$ and $0<\epsilon\leq1$.
\end{proposition}

\begin{proof}
We first note that we can write
\begin{equs}
{\cal F}_{2,u}(x)&=
{\cal F}_{2,1,u}(x)
+{\cal F}_{2,2,u}(x)~,\\
{\cal F}_{2,1,u}(x)&=
\int_{x}^{\infty}\hspace{-4mm}{\rm d}\tilde{x}~
\ed^{-(\tilde{x}-x)}
\big(K_{\kdeux}(\tilde{x}-x)-K_{\ksix}(\tilde{x}-x)\big)\qzero(\tilde{x})
+\ed^{-(\tilde{x}-x)}K_{\kcinq}(\tilde{x}-x)\qun(\tilde{x})~,
\label{eqn:f21u}
\\
{\cal F}_{2,2,u}(x)&=
\int_{x}^{\infty}\hspace{-4mm}{\rm d}\tilde{x}~
\falpha^{*}(\tilde{x}-x)\qun(\tilde{x})~.
\label{eqn:f22u}
\end{equs}
We then note that
\begin{equs}
\|K_{\kdeux}\|_{\K{1}{0}{\frac{1}{2}}}+
\|K_{\kcinq}\|_{\K{1}{0}{\frac{1}{2}}}+
\|K_{\ksix}\|_{\K{1}{0}{\frac{1}{2}}}\leq C~,
\end{equs}
while for all $\xzero\leq x\leq\tilde{x}$, we have
\begin{equs}
\chinese{x}^{\frac{1}{2}}
\big(
\|\qzero(\tilde{x})\|_{\infty}
+\|\qun(\tilde{x})\|_{\infty}\big)
&\leq 
\chinese{x}^{-1}\big(
\|\qzero\|_{\Lx{\tilde{x}}{\infty}{\frac{3}{2}}}+
\|\qun\|_{\Lx{\tilde{x}}{\infty}{2-\olddelta}}\big)~,\\
\chinese{x}^{\frac{1}{2}-\frac{1}{p}}
\big(\|\qzero(\tilde{x})\|_{p}
+\|\qun(\tilde{x})\|_{p}&\leq
\chinese{x}^{-1-\frac{1}{2p}}\big(
\|\qzero\|_{\Lx{\tilde{x}}{p}{\frac{3}{2}-\frac{1}{2p}}}+
\|\qun\|_{\Lx{\tilde{x}}{p}{2-\olddelta-\frac{1}{2p}}}\big)~,\\
\chinese{x}^{1-\frac{1}{2\pdv}-\au}
\big(\|\partial_y\qzero(\tilde{x})\|_{\pdv}
+\|\partial_y\qun(\tilde{x})\|_{\pdv}\big)
&\leq
\chinese{x}^{-1}\big(
\|\partial_y\qzero\|_{\Lx{\tilde{x}}{\pdv}{2-\frac{1}{2\pdv}-\au}}+
\|\partial_y\qun\|_{\Lx{\tilde{x}}{\pdv}{\frac{5}{2}-\frac{1}{2\pdv}-\av}}
\big)
~,
\end{equs}
since $\olddelta\leq\av<\frac{1}{2}$ and
$\frac{1}{2}-\av+\au\geq0$. By (\ref{eqn:nonlinbounds}), the
above quantities are all bounded by
$C\chinese{x}^{-1}\|\triplet\|^2$, thus we get
\begin{equs}
\|({\cal F}_{2,1,u},0,0)\|_x&\leq
C\chinese{x}^{-1}\|\triplet\|^2
\int_{x}^{\infty}\hspace{-4mm}{\rm d}\tilde{x}~
\ed^{x-\tilde{x}}(\tilde{x}-x)^{-\frac{1}{2}}~.
\label{eqn:ici}
\end{equs}
Next, we use (\ref{eqn:interpolation}) and note that ${\cal F}_{2,2,u}$ and
${\cal F}_{2,2,v}$ differ only by signs and the exchange of the Kernels
$\falpha$ and $\galpha$ (see (\ref{eqn:f22v}) and (\ref{eqn:f22u})). The
bounds on ${\cal F}_{2,2,v}$ in the proof of Proposition
\ref{prop:thecalfisv} being insensitive to these details then apply mutatis
mutandis. Finally, the proof of (\ref{eqn:universalfdeuxu}) follows at once
from (\ref{eqn:ici}).
\end{proof}

\section{Existence and uniqueness results}\label{sec:existeunique}

Our next task is now to prove existence and (local) uniqueness result in
${\cal C}_u$ for solutions of (\ref{eqn:NavierStokes}). This was stated as
"\ref{it:deux}.\ implies \ref{it:un}." in
Theorem \ref{thm:equivalence}, or, to rephrase it, that
\begin{theorem}\label{thm:equivalencefirstpart}
If $\uzero$ and $\omegazero$ are in the class $\Ci$ with parameters
satisfying (\ref{eqn:restrictions}) and $\xzero$ is sufficiently large,
then there exist a (locally) unique solution to (\ref{eqn:NavierStokes}) in
$\Cu$ with parameters satisfying (\ref{eqn:restrictions}).
\end{theorem}
\begin{proof}
The proof follows from the contraction mapping principle. For fixed 
$\uzero$ and $\omegazero$ in $\Ci$, we define the map ${\cal F}:\W\to\W$ by
\begin{equs}
{\cal F}({\bf v},\omega)=\mbox{r.h.s.
of }(\ref{eqn:foromega})-(\ref{eqn:forv})~.
\end{equs}
By the results of Section \ref{sec:estimates}, it follows that if the
parameters satisfy (\ref{eqn:restrictions}), then for
$\kappa=\min(\kappa_0,\kappa_1,\kappa_2)$, we have
\begin{equs}
\|{\cal F}({\bf v},\omega)\|&\leq 
C_1\|(\uzero,\huzero,\omegazero)\|_{\xzero}+
C_2\chinese{\xzero}^{-\kappa}\|\triplet\|^2~,\\
\|{\cal F}({\bf v}_1,\omega_1)-{\cal F}({\bf v}_2,\omega_2)\|&\leq 
C_2\chinese{\xzero}^{-\kappa}
\big(\|({\bf v}_1-{\bf v}_2,\omega_1-\omega_2)\|\big)
\big(\|({\bf v}_1+{\bf v}_2,\omega_1+\omega_2)\|\big)~.
\end{equs}
Let $\rho>0$ and $0<\epsilon<\frac{1}{2}$. We easily see that if
$\|(\uzero,\huzero,\omegazero)\|_{\xzero}\leq\rho$, the map ${\cal F}$ is a
contraction in ${\cal B}_{0}((1+\epsilon)C_1\rho)\subset\W$ if 
$\chinese{\xzero}>\big(C_1C_2\rho\epsilon^{-1}\big)^{\frac{1}{\kappa}}$. By
classical arguments, the approximating sequence
$({\bf v}_{n+1},\omega_{n+1})={\cal F}({\bf v}_{n},\omega_{n})$ for $n>1$
and $({\bf v}_{1},\omega_{1})={\cal F}(0,0)$ converges to the unique
solution of (\ref{eqn:NavierStokes}) in ${\cal
B}_{0}((1+\epsilon)C_1\rho)\subset\W$. This completes the proof.
\end{proof}

\section{Asymptotics}\label{sec:asymptotics}

Now that we know that there exist (locally) unique solutions of
(\ref{eqn:NavierStokes}) in ${\cal C}_u$, we can turn to their asymptotic
description. As explained in Section \ref{sec:struct}, we will first prove
the partial description of Corollary \ref{cor:asymptotics}, and more
precisely that
\begin{theorem}\label{thm:asymptoticsrap}
Let ${\bf a}_1=(-\M(\I\Q\omegazero)-\int_{\Omega_{+}}\Q\qun(x,y)\,
{\rm d}x{\rm d}y,0,0,0,0,0)$ and $u_{{\bf a}_1}$,
$\omega_{{\bf a}_1}$ as in (\ref{eqn:defuaomegaa}),
then for all $\epsilon>0$, solutions to (\ref{eqn:NavierStokes}) in
$\Cu$ satisfy 
\begin{equa}[1][eqn:asymptrap]
\|u(x)-u_{{\bf a}_1}(x)\|_{\infty}&\leq C(\xzero,\|\triplet\|)
\;\chinese{x}^{-1+(1+\epsilon)\olddelta}
~,\\
\|\omega(x)-\omega_{{\bf a}_1}(x)\|_{\infty}
&\leq C(\xzero,\|\triplet\|)\;\chinese{x}^{-\frac{3}{2}+(1+\epsilon)\olddelta}\\
\|\omega(x)-\omega_{{\bf a}_1}(x)\|_{1}
&\leq C(\xzero,\|\triplet\|)\;\chinese{x}^{-1+(1+\epsilon)\olddelta}\\
\||y|^{\beta_0}\big(\omega(x)-\omega_{{\bf a}_1}(x)\big)\|_{2}
&\leq C(\xzero,\|\triplet\|)\;
\chinese{x}^{-\frac{5}{4}+\frac{\beta_0}{2}+(1+\epsilon)\olddelta}
\end{equa}
for all $\frac{1}{2}\leq\beta_0\leq1-2(1+\epsilon)\olddelta$ and
$x\geq\xzero$.
\end{theorem}
Note that here, in contrast with (\ref{eqn:asympt}) or the statement of
Corollary \ref{cor:asymptotics}, we did not include the
terms in the $y\sim x$ scale for $u$ nor the $v$ component, as they are of
order $x^{-1}$, resp. $x^{-1+\olddelta}$, which are smaller than the ${\cal
O}(x^{-1+(1+\epsilon)\olddelta})$ correction. These terms will appear later
in Section \ref{sec:asymptoticsrefined}. Note that we need only prove
(\ref{eqn:asymptrap}) for $x\to\infty$ (we will in fact 
prove them for $x\geq2\xzero$), as they are trivially satisfied for finite
$x$. Furthermore, for $x\geq2\xzero$, we can either compare, $u(x)$ to 
$u_{{\bf a}_1}(x)$ or $u_{{\bf a}_1}(x-\xzero)$ and $\omega(x)$ to 
$\omega_{{\bf a}_1}(x)$ or $\omega_{{\bf a}_1}(x-\xzero)$, as is proved in
the
\begin{lemma}\label{lem:equalasympt}
Let $K_{\kc}(x)=\frac{\ed^{-\frac{y^2}{4x}}}{\sqrt{4\pi x}}$,
$K_{\kzero}(x)=\frac{1}{\pi}\frac{x}{x^2+y^2}$,  
$f(x)=K_{\kc}(x-\xzero)-K_{\kc}(x)$ and
$g(x)=K_{\kzero}(x-\xzero)-K_{\kzero}(x)$, then for all $m\in{\bf N}$, there
exists a constant $C_m$ such that
\begin{equs}
\|\partial_y^m g(x)\|_{\infty}&
\leq C_m\chinese{x}^{-m-2}\chinese{\xzero}~,~~~~
\|\partial_y^m \H g(x)\|_{\infty}
\leq C_m\chinese{x}^{-m-2}\chinese{\xzero}\\
\|\partial_y^m f(x)\|_{\infty}&
\leq C_m\chinese{x}^{-\frac{3+m}{2}}\chinese{\xzero}~,~~~~
\|\partial_y^m f(x)\|_{1}
\leq C_m\chinese{x}^{-\frac{2+m}{2}}\chinese{\xzero}~,\\
\|y\partial_y^m f(x)\|_{2}&
\leq C_m\chinese{x}^{-\frac{3+2m}{4}}\chinese{\xzero}~,
\end{equs}
for all $x\geq2\xzero\geq2$.
\end{lemma}

\begin{proof}
Since $x-\xzero\geq\frac{x}{2}$ for $x\geq2\xzero$, we have
\begin{equs}
\|\partial_y^m g(x)\|_{\infty}+
\|\partial_y^m \H g(x)\|_{\infty}&\leq
\int_{-\infty}^{\infty}
\hspace{-3mm}{\rm d}k
|k|^m
\big|
\ed^{-|k|(x-\xzero)}-\ed^{-|k|x}
\big|
\leq
\xzero
\int_{-\infty}^{\infty}
\hspace{-3mm}{\rm d}k
|k|^{m+1}\ed^{-\frac{|k|x}{2}}\\
\|\partial_y^m f(x)\|_{\infty}&\leq
\int_{-\infty}^{\infty}
\hspace{-3mm}{\rm d}k
|k|^m
\big|
\ed^{-k^2(x-\xzero)}-\ed^{-k^2x}
\big|
\leq
\xzero
\int_{-\infty}^{\infty}
\hspace{-3mm}{\rm d}k
|k|^{m+2}\ed^{-\frac{k^2x}{2}}
\end{equs}
and similarly
\begin{equs}
\|\partial_y^m f(x)\|_{2}^2&\leq
\xzero^2
\int_{-\infty}^{\infty}
\hspace{-3mm}{\rm d}k
|k|^{2(m+2)}\ed^{-k^2x}
\leq C_m\xzero^2 x^{-\frac{5}{2}-m}~,\\
\|y\partial_y^m f(x)\|_{2}^2&\leq
\xzero^2\int_{-\infty}^{\infty}
\hspace{-3mm}{\rm d}k
|k|^{2(m+1)}(1+m^2+m^2k^2x^2)
\ed^{-k^2x}\leq C_m\xzero^2x^{-\frac{3}{2}-m}~.
\end{equs}
The proof is completed with the use of Lemma \ref{lem:Lun}.
\end{proof}
For convenience, the proof of Theorem \ref{thm:asymptoticsrap} is split
in the next two subsections. The terms coming from $\omegazero$ and
$\uzero$ in (\ref{eqn:foromega})-(\ref{eqn:forv}) will be studied in the
next subsection, the remainder in the second one. The basis of the proof of
Theorem \ref{thm:asymptoticsrap} is that the large time asymptotics of
$K_{\kun}(x)f$ is captured by\footnote{by abuse of notation, 
$K_{\kun}$ is here considered as a function and not as a convolution
operator.} $\M(f)K_{\kun}(x)$ if $f$ decays sufficiently fast, which is the
content of the next Lemma.
\begin{lemma}\label{lem:univ}
Let $0\leq\gamma\leq1$, $0\leq\gamma_2\leq2$ and $f$ satisfying
$\|\chinese{y}^{\gamma}f\|_{1}<\infty$. Then for all $m\geq0$, there exist
a constant $C_{\gamma}$
such that
\begin{equs}
\|\partial_y^mK_{\kun}(x)(f-\M(f))\|_{{\infty}}
&\leq
C_{\gamma}
\frac{\chinese{x}^{\frac{1+m+\gamma}{2}}
}{
x^{1+m+\gamma}}~
\||y|^{\gamma}f\|_{1}~,\\
\|\partial_y^mK_{\kun}(x)(f-\M(f))\|_{{2}}
&\leq
C_{\gamma}
\frac{\chinese{x}^{\frac{1}{4}+\frac{m+\gamma}{2}}
}{
x^{\frac{1}{2}+m+\gamma}}~
\||y|^{\gamma}f\|_{1}~,\\
\|y\partial_y^mK_{\kun}(x)(f-\M(f))\|_{{2}}
&\leq
C_{\gamma}
\frac{\chinese{x}^{\frac{5}{4}+\frac{m}{2}}
}{
x^{\frac{3}{2}+m}}~
\||y|f\|_{1}~,\\
\|\partial_y^mK_{\kun}(x)(f-\M(f))\|_{{1}}
&\leq
C_{\gamma}
\frac{\chinese{x}^{\frac{3+\gamma}{4}+\frac{m}{2}}
}{
x^{1+m+\frac{\gamma}{2}}}~
\sqrt{\||y|f\|_{1}\||y|^{\gamma}f\|_{1}}~,\\
\|K_{\kun}(x)(f-\M(f))-\partial_yK_{\kun}(x)\M(yf)\|_{{\infty}}
&\leq
C_{\gamma_2}
\frac{\chinese{x}^{\frac{1+\gamma_2}{2}}
}{
x^{1+\gamma_2}}~
\||y|^{\gamma_2}f\|_{1}~,\\
\|K_{\kdouze}(x)(f-\M(f))-\partial_yK_{\kdouze}(x)\M(yf)\|_{{\infty}}
&\leq
C_{\gamma_2}
\frac{\chinese{x}^{\frac{1+\gamma_2}{2}}
}{
x^{1+\gamma_2}}~
\||y|^{\gamma_2}f\|_{1}~,
\end{equs}
where $\M(f)=\int_{{\bf R}}f(y){\rm d}y$.
\end{lemma}

\begin{proof}
Let 
\begin{equs}
R_1(x)&=K_{\kdouze}(x)(f-\M(f))-\partial_yK_{\kdouze}(x)\M(yf)\\
R_2(x)&=K_{\kun}(x)(f-\M(f))-\partial_yK_{\kun}(x)\M(yf)
\end{equs}
 and $R_3(x)=K_{\kun}(x)(f-\M(f))$.
Using twice the Fourier Transform, we get
\begin{equs}
\|R_1(x)\|_{{\infty}}+\|R_2(x)\|_{{\infty}}
&\leq
\sum_{n\in{\bf Z}}
\int_{-\infty}^{\infty}
\hspace{-3mm}{\rm d}k~
\left||k|^{\gamma_2}\ed^{\Lambdam x}\right|
\int_{-\infty}^{\infty}
\hspace{-3mm}{\rm d}y~
\left|\frac{\ed^{iky}-1-iky}{|ky|^{\gamma_2}}\right|~
|y|^{\gamma_2}|f_n(y)|\\
\|\partial_y^mR_3(x)\|_{{\infty}}
&\leq
\sum_{n\in{\bf Z}}
\int_{-\infty}^{\infty}
\hspace{-3mm}{\rm d}k~
\left||k|^{m+\gamma}\ed^{\Lambdam x}\right|
\int_{-\infty}^{\infty}
\hspace{-3mm}{\rm d}y~
\left|\frac{\ed^{iky}-1}{|ky|^{\gamma}}\right|~
|y|^{\gamma}|f_n(y)|\\
\|\partial_y^mR_3(x)\|_{{2}}
&\leq
\sum_{n\in{\bf Z}}
\Big(\int_{-\infty}^{\infty}
\hspace{-3mm}{\rm d}k~
\left||k|^{2m+2\gamma}\ed^{2\Lambdam x}\right|
\Big|
\int_{-\infty}^{\infty}
\hspace{-3mm}{\rm d}y~
\Big|
\frac{\ed^{iky}-1}{|ky|^{\gamma}}
\Big|~
|y|^{\gamma}|f_n(y)|
\Big|^2
\Big)^{1/2}~,\\
\|y\partial_y^mR_3(x)\|_{{2}}
&\leq
x
\sum_{n\in{\bf Z}}
\Big(\int_{-\infty}^{\infty}
\hspace{-3mm}{\rm d}k~
\left||k|^{2(m+2)}\ed^{2\Lambdam x}\right|~
\|yf\|^2
\Big)^{1/2}+
\|\partial_y^m K_{\kun}(x)\|_{\L^2}\|\chinese{y}f\|_{1}~.
\end{equs}
The proof is completed using Lemma \ref{lem:L2} and $\|\partial_y^mR_3\|_{1}
\leq\big(\|\partial_y^mR_3\|_{2}\|y\partial_y^mR_3\|_{2}\big)^{\frac{1}{2}}$.
\end{proof}

\subsection{The `linear' terms}\label{sub:linear}
In this subsection, we consider the asymptotics of
\begin{equs}
U(x)&=K_{\kun}(x-\xzero)\myast\Lu\omegazero+
K_{\kzero}(x-\xzero)\uzero
~,~~~~
W(x)=K_{\kun}(x-\xzero)\myast\omegazero
~,\label{eqn:forvwzero}
\end{equs}
as $x\to\infty$. We first note that by Lemma \ref{lem:lunforLu},
\ref{lem:thescaryone} and \ref{lem:kernelun}, we have
\begin{equs}
\|K_{\kun}(x-\xzero)\myast(\Lu+\I\Q)\omegazero+
K_{\kzero}(x-\xzero)\uzero\|_{\infty}&\leq
C(\xzero,\|(\uzero,\huzero,\omegazero)\|_{\xzero})\;
\chinese{x}^{-1+\olddelta}~,\\
\|\P W(x)\|_{\infty}
&\leq C(\xzero,\|(\uzero,\huzero,\omegazero)\|_{\xzero})\;
\chinese{x}^{-\frac{3}{2}+\olddelta}~,\\
\|\P W(x)\|_{1}
&\leq C(\xzero,\|(\uzero,\huzero,\omegazero)\|_{\xzero})\;
\chinese{x}^{-1+\olddelta}~,\\
\||y|^{\beta_0}\P W(x)\|_{2}
&\leq C(\xzero,\|(\uzero,\huzero,\omegazero)\|_{\xzero})\;
\chinese{x}^{-\frac{5}{4}+\frac{\beta_0}{2}+\olddelta}~,
\end{equs}
for all $x\geq2\xzero$ and $0\leq\beta_0\leq\beta$. This means that the
first order contribution of $U$ and $W$ to (\ref{eqn:asymptrap}) is given
by $U_1(x)=K_{\kun}(x-\xzero) f$ and $W_1(x)=-\partial_y K_{\kun}(x-\xzero)
f$, where $f=-\I\Q\omegazero$. Using Lemma \ref{lem:univ} and
that by Lemma \ref{lem:key} we have $\|yf\|_{1}\leq
C\|(0,0,\omegazero)\|_{\xzero}$, we conclude that for
${\bf a}_{1,1}=(-\M(\I\Q\omegazero),0,0,0,0,0)$, we have
\begin{equa}[1][eqn:asymptrapfrap]
\|U(x)-u_{{\bf a}_{1,1}}(x-\xzero)\|_{\infty}&\leq C(\xzero,\|\triplet\|)
\;\chinese{x}^{-1+(1+\epsilon)\olddelta}
~,\\
\|W(x)-\omega_{{\bf a}_{1,1}}(x-\xzero)\|_{\infty}
&\leq C(\xzero,\|\triplet\|)\;\chinese{x}^{-\frac{3}{2}+(1+\epsilon)\olddelta}\\
\|W(x)-\omega_{{\bf a}_{1,1}}(x-\xzero)\|_{1}
&\leq C(\xzero,\|\triplet\|)\;\chinese{x}^{-1+(1+\epsilon)\olddelta}\\
\||y|^{\beta_0}\big(W(x)-\omega_{{\bf a}_{1,1}}(x-\xzero)\big)\|_{2}
&\leq C(\xzero,\|\triplet\|)\;
\chinese{x}^{-\frac{5}{4}+\frac{\beta_0}{2}+(1+\epsilon)\olddelta}
\end{equa}
for all $\frac{1}{2}\leq\beta_0\leq1$ and $x\geq2\xzero$.

\subsection{The nonlinear terms}
We now turn to the asymptotics of
\begin{equs}
U_1(x)&={\cal F}_{1,u}(x)+{\cal F}_{2,u}(x)
+\LA \qB(x)-\LB \qA(x)
~,~~~~~
W_1(x)={\cal F}_{1,\omega}(x)+{\cal F}_{2,\omega}(x)~.
\end{equs}
It follows from Propositions \ref{prop:localterms} to \ref{prop:thecalfisu}
that for all $\frac{1}{2}\leq\beta_0\leq\beta$ and $x\geq\xzero$, we have
\begin{equa}[1][eqn:asymptrapfrapcrap]
\|U_1(x)-{\cal F}_{1,2,u}(x)\|_{\infty}&\leq C(\xzero,\|\triplet\|)
\;\chinese{x}^{-1+(1+\epsilon)\olddelta}
~,\\
\|W_1(x)-{\cal F}_{1,1,\omega}(x)\|_{\infty}
&\leq C(\xzero,\|\triplet\|)\;\chinese{x}^{-\frac{3}{2}+(1+\epsilon)\olddelta}\\
\|W_1(x)-{\cal F}_{1,1,\omega}(x)\|_{1}
&\leq C(\xzero,\|\triplet\|)\;\chinese{x}^{-1+(1+\epsilon)\olddelta}\\
\||y|^{\beta_0}\big(W_1(x)-{\cal F}_{1,1,\omega}(x)\big)\|_{2}
&\leq C(\xzero,\|\triplet\|)\;
\chinese{x}^{-\frac{5}{4}+\frac{\beta_0}{2}+(1+\epsilon)\olddelta}~,
\end{equa}
where 
\begin{equs}
{\cal F}_{1,2,u}(x)=-\int_{\xzero}^x{\rm d}\tilde{x}~
K_{\kdouze}(x-\tilde{x})\qun(\tilde{x})~~~\mbox{ and }~~~
{\cal F}_{1,1,\omega}(x)=
-\int_{\xzero}^x{\rm d}\tilde{x}~K_{\kdeux}(x-\tilde{x})\qun(\tilde{x})~.
\end{equs}
The proof of Theorem \ref{thm:asymptoticsrap} is then an immediate
consequence of the preceding section, Lemma \ref{lem:equalasympt} and the
\begin{proposition}\label{prop:forkdouze}
Assume that $\qun$ satisfies (\ref{eqn:nonlinbounds}), and
define
${\bf a}_{1,2}=(-\int_{\Omega_{+}}\Q\qun(x,y)\,{\rm d}x{\rm d}y,0,0,0,0,0)$
and
\begin{equs}
D_1(x)&=
{\cal F}_{1,2,u}(x)
-u_{{\bf a}_{1,2}}(x-\xzero)~,~~~~
D_2(x)=
{\cal F}_{1,1,\omega}(x)
-\omega_{{\bf a}_{1,2}}(x-\xzero)~,
\end{equs}
then for all $\epsilon>0$, there exist a constant $C$ such that 
\begin{equs}
\|D_1(x)\|_{{\infty}}&\leq
C\chinese{x}^{-1+(1+\epsilon)\olddelta}\|\triplet\|^2~,~~~~~
\|D_2(x)\|_{{\infty}}\leq
C\chinese{x}^{-\frac{3}{2}+(1+\epsilon)\olddelta}\|\triplet\|^2\\
\|D_2(x)\|_{{1}}&\leq
C\chinese{x}^{-1+(1+\epsilon)\olddelta}\|\triplet\|^2~,~~~~~
\||y|^{\beta_0}D_2(x)\|_{2}\leq
C\chinese{x}^{-\frac{5}{4}+\frac{\beta_0}{2}+
(1+\epsilon)\olddelta}\|\triplet\|^2
\end{equs}
for all $x\geq2\xzero$ and $\frac{1}{2}\leq\beta_0\leq1-2(1+\epsilon)\olddelta$.
\end{proposition}

\begin{remark}\label{rem:heat}
This result is expected in view of the corresponding classical theory on
the nonlinear heat equation (see e.g.\ \cite{bricmont}). However in our
case, we can prove that in fact ${\bf a}_{1,2}$ {\em does not} depend on
$u,v$ and $\omega$ on the whole domain $\Omega_{+}$, but only on $u$ and
$v$ on the boundary $x=\xzero$. Namely, since
$\qun=-\partial_y\qA+\partial_x\qB$, we have
\begin{equs}
{\bf a}_{1,2}&=
\Q\int_{\Omega_{+}}\qun(x,y)\,{\rm d}x{\rm d}y=
\Q\int_{\Omega_{+}}\big(\partial_x \qB(x,y)-\partial_y \qA(x,y)\big)
\,{\rm d}x{\rm d}y=
-\M(\Q\qB(\xzero))
~.
\end{equs}
\end{remark}

\begin{proof}
Let $D_{1,1}(x)=-\P\int_{\xzero}^{x}{\rm d}\tilde{x}~
K_{\kdouze}(x-\tilde{x})\qun(\tilde{x})$ and
$D_{2,1}(x)=-\P\int_{\xzero}^{x}{\rm d}\tilde{x}~
K_{\kdeux}(x-\tilde{x})\qun(\tilde{x})$. Using Lemma \ref{lem:blackbox},
we have
\begin{equs}
\left\|D_{1,1}(x)\right\|_{\infty}
&\leq
\B{\frac{1}{2},\frac{7}{4}+\olddelta,\frac{1}{4}}{1,\frac{3}{2}-\olddelta,\frac{1
}{2}}(x,\xzero)\|\triplet\|^2\leq
C\chinese{x}^{-1+\olddelta}\|\triplet\|^2~,\\
\left\|D_{2,1}(x)\right\|_{\infty}
&\leq
\B{\frac{1}{2},2-\olddelta,0}{1,\frac{3}{2}-\olddelta,0}(x,\xzero)
\|\triplet\|^2
\leq C\chinese{x}^{-\frac{3}{2}+\olddelta}\|\triplet\|^2~,\\
\left\|D_{2,1}(x)\right\|_{1}
&\leq
\B{\frac{1}{2},1,0}{\frac{1}{2},1,0}(x,\xzero)
\|\triplet\|^2
\leq C\chinese{x}^{-1+\olddelta}\|\triplet\|^2~,\\
\left\||y|^{\beta}D_{2,1}(x)\right\|_{2}
&\leq
\Big(
\B{\frac{3}{4}-\frac{\beta}{2},\frac{3}{2}-\olddelta,0}
{\frac{3}{4}-\frac{\beta}{2},\frac{3}{2}-\olddelta,0}(x,\xzero)+
\B{\frac{1}{2},\frac{7}{4}-\frac{\beta}{2}-\olddelta,0}
{\frac{1}{2},\frac{7}{4}-\frac{\beta}{2}-\olddelta,0}(x,\xzero)
\Big)\|\triplet\|^2
\\&
\leq C\chinese{x}^{-\frac{5}{4}+\frac{\beta}{2}+\olddelta}\|\triplet\|^2~.
\end{equs}
Now, let $D_{1,2}(x)=-\Q\int_{\xzero}^{x}{\rm d}\tilde{x}~
K_{\kdouze}(x-\tilde{x})\qun(\tilde{x})$ and
$D_{2,2}(x)=-\Q\int_{\xzero}^{x}{\rm d}\tilde{x}~
K_{\kdeux}(x-\tilde{x})\qun(\tilde{x})$. Since
$\Q K_{\kdix}\equiv0$, $\partial_x\Q K_{\kdouze}=-\partial_y K_{\kdeux}$,
$\partial_x\Q K_{\kdeux}=\partial_y K_{\khuit}$ and
$\Q K_{\kdeux}=2\partial_y \Q K_{\khuit}+\partial_y\Q K_{\kun}$ and
$\Q K_{\kdouze}=-\Q K_{\kun}-\Q K_{\khuit}$,
integrating by parts in $\tilde{x}$, we get
\begin{equs}
D_{1,2}(x)&=
\Q\left(K_{\kun}(x-\xzero)+K_{\khuit}(x-\xzero)\right)
\int_{\xzero}^x\hspace{-3mm}{\rm d}z\,\qun(z)+
\int_{\xzero}^x\hspace{-3mm}{\rm d}\tilde{x}\,
\partial_yK_{\kdeux}(x-\tilde{x})
\int_{\tilde{x}}^x\hspace{-3mm}{\rm d}z\,\qun(z)~,\\
D_{2,2}(x)&=
-\Q\left(\partial_yK_{\kun}(x-\xzero)+2\partial_yK_{\khuit}(x-\xzero)\right)
\int_{\xzero}^x\hspace{-3mm}{\rm d}z\,\qun(z)-
\int_{\xzero}^x\hspace{-3mm}{\rm d}\tilde{x}\,
\partial_yK_{\khuit}(x-\tilde{x})
\int_{\tilde{x}}^x\hspace{-3mm}{\rm d}z\,\qun(z)~.
\end{equs}
We then have
\begin{equs}
\left\|K_{\kun}(x-\xzero)
\int_{x}^{\infty}\hspace{-3mm}{\rm d}z\,\qun(z)\right\|_{\infty}
&\leq C\|\triplet\|^2
\int_x^{\infty}\hspace{-3mm}{\rm d}z\,\chinese{z}^{-2+\olddelta}
\leq C\|\triplet\|^2\chinese{x}^{-1+\olddelta}~,
\\
\left\|K_{\khuit}(x-\xzero)
\int_{\xzero}^x\hspace{-3mm}{\rm d}z\,\qun(z)\right\|_{\infty}
&\leq C\|\triplet\|^2(x-\xzero)^{-1}
\int_{\xzero}^x\hspace{-3mm}{\rm d}z\,\chinese{z}^{-\frac{3}{2}+\olddelta}
\leq C\|\triplet\|^2\chinese{x}^{-1}~,\\
\left\|\partial_yK_{\kun}(x-\xzero)
\int_{x}^{\infty}\hspace{-3mm}{\rm d}z\,\qun(z)\right\|_{\infty}
&\leq C\|\triplet\|^2
\frac{\chinese{x}^{\frac{1}{2}}}{x-\xzero}
\int_x^{\infty}\hspace{-3mm}{\rm d}z\,\chinese{z}^{-2+\olddelta}
\leq C\|\triplet\|^2
\chinese{x}^{-\frac{3}{2}+\olddelta}~,
\\
\left\|\partial_yK_{\khuit}(x-\xzero)
\int_{\xzero}^x\hspace{-3mm}{\rm d}z\,\qun(z)\right\|_{\infty}
&\leq C\|\triplet\|^2
\frac{\chinese{x}^{\frac{1}{4}}}{(x-\xzero)^{\frac{7}{4}}}
\int_{\xzero}^x\hspace{-3mm}{\rm d}z\,\chinese{z}^{-2+\olddelta}
\leq C\|\triplet\|^2\chinese{x}^{-\frac{3}{2}}~,\\
\left\|\partial_yK_{\kun}(x-\xzero)
\int_{x}^{\infty}\hspace{-3mm}{\rm d}z\,\qun(z)\right\|_{1}
&\leq C\|\triplet\|^2
\frac{\chinese{x}^{\frac{1}{2}}}{x-\xzero}
\int_x^{\infty}\hspace{-3mm}{\rm d}z\,\chinese{z}^{-\frac{3}{2}+\olddelta}
\leq C\|\triplet\|^2
\chinese{x}^{-1+\olddelta}~,
\\
\left\|\partial_yK_{\khuit}(x-\xzero)
\int_{\xzero}^x\hspace{-3mm}{\rm d}z\,\qun(z)\right\|_{1}
&\leq C\|\triplet\|^2
\frac{\chinese{x}^{\frac{1}{4}}}{(x-\xzero)^{\frac{7}{4}}}
\int_{\xzero}^x\hspace{-3mm}{\rm d}z\,\chinese{z}^{-\frac{3}{2}+\olddelta}
\leq C\|\triplet\|^2\chinese{x}^{-\frac{3}{2}}~,
\end{equs}
where in the last six inequalities we used $x\geq2\xzero$. Similarly, for 
$x\geq2\xzero$, we have
\begin{equs}
\left\||y|^{\beta_0}\partial_yK_{\kun}(x-\xzero)
\int_{x}^{\infty}\hspace{-3mm}{\rm d}z\,\qun(z)\right\|_{2}
&\leq C\|\triplet\|^2\chinese{x}^{-\frac{5}{4}+\frac{\beta_0}{2}+\olddelta}~,
\\
\left\||y|^{\beta}\partial_yK_{\khuit}(x-\xzero)
\int_{\xzero}^x\hspace{-3mm}{\rm d}z\,\qun(z)\right\|_{2}
&\leq C\|\triplet\|^2
\chinese{x}^{-\frac{5}{4}+\frac{\beta}{2}+\olddelta}~.
\end{equs}
Note that the first of these two estimates is only valid if
$\beta_0<\frac{3}{2}-2\olddelta$. Next, for $D_{1,3}(x)=\int_{\xzero}^x{\rm
d}\tilde{x}\,\partial_yK_{\kdeux}(x-\tilde{x})
\int_{\tilde{x}}^x{\rm d}z\,\qun(z)$ and
$D_{2,3}(x)=\int_{\xzero}^x{\rm d}\tilde{x}\,
\partial_yK_{\khuit}(x-\tilde{x})
\int_{\tilde{x}}^x{\rm d}z\,\qun(z)$, we have, (see Lemma \ref{lem:onD}
below for the definition of $D\big[\cdot\big](x,\xzero)$ and related
estimates), that
\begin{equs}
\left\|D_{1,3}(x)\right\|_{\infty}
&\leq
\chinese{x}^{\frac{1}{2}}\D{\frac{3}{2},2-\olddelta}{2,\frac{3}{2}-\olddelta}
(x,\xzero)
\|\triplet\|^2\leq C\|\triplet\|^2\chinese{x}^{-1+\olddelta}~,\\
\left\|D_{2,3}(x)\right\|_{\infty}
&\leq
\chinese{x}^{\frac{1}{4}}
\D{\frac{7}{4},2-\olddelta}{\frac{7}{4},2-\olddelta}(x,\xzero)
\|\triplet\|^2\leq
C\|\triplet\|^2\chinese{x}^{-\frac{3}{2}+\olddelta}
~,\\
\left\|D_{2,3}(x)\right\|_{1}
&\leq
\chinese{x}^{\frac{1}{4}}
\D{\frac{7}{4},\frac{3}{2}-\olddelta}{\frac{7}{4},\frac{3}{2}-\olddelta}(x,\xzero)
\|\triplet\|^2\leq
C\|\triplet\|^2\chinese{x}^{-1+\olddelta}
~.
\end{equs}
Along the same lines, we find
$
\left\||y|^{\beta}D_{2,3}(x)\right\|_{2}
\leq
C\|\triplet\|^2\chinese{x}^{-\frac{5}{4}+\frac{\beta}{2}+\olddelta}
$.
We finally define
\begin{equs}
T(x,y)=
u_{{\bf a}_{1,2}}(x)-
\Q K_{\kun}(x-\xzero)
\int_{\xzero}^{\infty}\hspace{-3mm}{\rm d}z\,\qun(z)~.
\end{equs}
Since $\int_{\xzero}^x{\rm d}z\,\qun(z)=\int_{\xzero}^{\infty}{\rm
d}z\,\qun(z)-\int_x^{\infty}{\rm d}z\,\qun(z)$, we get, using Lemma
\ref{lem:univ} and $x\geq2\xzero$ that
\begin{equs}
\|D_1(x)\|_{\infty}&\leq
\left\|
T(x)
\right\|_{\infty}+C\chinese{x}^{-1+\olddelta}\|\triplet\|^2~,\\
&\leq
C\chinese{x}^{-\frac{1+\gamma}{2}}
\int_{\xzero}^{\infty}\hspace{-3mm}{\rm d}z\,\||y|^{\gamma}\qun(z)\|_{1}
+C\chinese{x}^{-1+\olddelta}\|\triplet\|^2~,\\
\|D_2(x)\|_{\infty}&\leq
\left\|
\partial_y T(x)\right\|_{\infty}+C\chinese{x}^{-\frac{3}{2}+\olddelta}\|\triplet\|^2~,\\
&\leq
C\chinese{x}^{-1-\frac{\gamma}{2}}
\int_{\xzero}^{\infty}\hspace{-3mm}{\rm d}z\,\||y|^{\gamma}\qun(z)\|_{1}
+C\chinese{x}^{-\frac{3}{2}+\olddelta}\|\triplet\|^2~,\\
\|D_2(x)\|_{1}&\leq
\left\|
\partial_y T(x)
\right\|_{1}+C\chinese{x}^{-1+\olddelta}\|\triplet\|^2~,\\
&\leq
C\chinese{x}^{-\frac{3+\gamma}{4}}
\int_{\xzero}^{\infty}\hspace{-3mm}{\rm d}z\,
\sqrt{\||y|\qun(z)\|_{1}\||y|^{\gamma}\qun(z)\|_{1}}
+C\chinese{x}^{-1+\olddelta}\|\triplet\|^2~,\\
\||y|^{\beta_0}D_2(x)\|_{2}&\leq
\left\|
|y|^{\beta_0}\partial_yT(x)\right\|_{2}
+C\chinese{x}^{-\frac{5}{4}+\frac{\beta_0}{2}+\olddelta}\|\triplet\|^2~,\\
&\leq
C\chinese{x}^{-\frac{3}{4}+\frac{\gamma}{2}(1-\beta_0)}
\int_{\xzero}^{\infty}\hspace{-3mm}{\rm d}z\,
\||y|^{\gamma}\qun\|_{1}^{1-\beta_0}
\||y|\qun\|_{1}^{\beta_0}
+C\chinese{x}^{-\frac{5}{4}+\frac{\beta_0}{2}+\olddelta}\|\triplet\|^2~,
\end{equs}
for any $0\leq\gamma\leq1$ (we used
$\||y|^{\beta_0}f\|_p\leq\|f\|_p^{1-\beta_0}\||y|f\|_p^{\beta_0}$ to
establish the last estimate). Then, for any $\gamma_1\leq1$ and
$\sigma>\frac{1}{2}$, we have
\begin{equs}
\int_{\xzero}^{\infty}\hspace{-3mm}{\rm d}z\,
\||y|^{\gamma_1}\qun(z)\|_{1}\leq
C\int_{\xzero}^{\infty}\hspace{-3mm}{\rm d}z\,
\|(1+|y|)^{\sigma}|y|^{\gamma_1}\qun(z)\|_{2}
\leq 
C\int_{\xzero}^{\infty}\hspace{-3mm}{\rm d}z\,
\chinese{z}^{\olddelta+\frac{\gamma_1+\sigma}{2}-\frac{7}{4}}\|\triplet\|^2~
,
\end{equs}
while with similar arguments we have for any $\gamma_2\leq1$ and
$\gamma_3\leq1$,
\begin{equs}
\int_{\xzero}^{\infty}\hspace{-3mm}{\rm d}z\,
\sqrt{\||y|\qun(z)\|_{1}\||y|^{\gamma_2}\qun(z)\|_{1}}
&\leq 
C\int_{\xzero}^{\infty}\hspace{-3mm}{\rm d}z\,
\chinese{z}^{\olddelta+\frac{\gamma}{4}+\frac{\sigma}{2}-\frac{3}{2}}
\|\triplet\|^2~,\\
\int_{\xzero}^{\infty}\hspace{-3mm}{\rm d}z\,
\||y|^{\gamma}\qun\|_{1}^{1-\beta_0}
\||y|\qun\|_{1}^{\beta_0}&\leq
C\int_{\xzero}^{\infty}\hspace{-3mm}{\rm d}z\,
\chinese{z}^{\olddelta+\frac{\beta_0}{2}+\frac{\gamma(1-\beta_0)}{2}+\frac{\sigma}{2}-\frac{7}{4}}
\|\triplet\|^2~.
\end{equs}
Choosing $\gamma_1=1-2(1+\epsilon)\olddelta$,
$\gamma_2=1-4(1+\epsilon)\olddelta$,
$\gamma_3=1-2\big(\frac{1+\epsilon}{1-\beta_0}\big)\olddelta$ and
$\sigma=\frac{1}{2}+\epsilon\olddelta$ with $\epsilon>0$ completes the proof.
\end{proof}

\begin{lemma}\label{lem:onD}
Let $0\leq p_1,q_2<2$, and $p_2,q_1\geq0$, then there
exist a constant $C$ such that
\begin{equs}
\D{p_1,q_1}{p_2,q_2}(x,\xzero)
&\equiv
\int_{\xzero}^x
\hspace{-3mm}{\rm d}\tilde{x}
\hspace{-1mm}
\int_{\tilde{x}}^x
\hspace{-3mm}{\rm d}z~
\min\Big(
\frac{\chinese{z}^{-{q_1}}}{(x-\tilde{x})^{p_1}}~,~
\frac{\chinese{z}^{-{q_2}}}{(x-\tilde{x})^{p_2}}
\Big)
\leq C\big(
\chinese{x}^{2-p_1-q_1}+
\chinese{x}^{2-p_2-q_2}
\big)~.
\end{equs}
for all $x\geq2\xzero\geq2$. 
\end{lemma}
\begin{proof}
The proof follows at once from
\begin{equs}
\D{p_1,q_1}{p_2,q_2}(x,\xzero)
&\leq
\frac{C}{(x-\xzero)^{p_2}}
\int_{\xzero}^{\frac{x+\xzero}{2}}
\hspace{-3mm}{\rm d}\tilde{x}
\hspace{-1mm}
\int_{\tilde{x}}^x
\hspace{-3mm}{\rm d}z~\chinese{z}^{-q_2}+
C\chinese{x}^{-q_1}
\int_{\frac{x+\xzero}{2}}^x
\hspace{-3mm}{\rm d}\tilde{x}~(x-\tilde{x})^{1-p_1}\\
&\leq C\big(
\chinese{x}^{2-p_1-q_1}+
\chinese{x}^{2-p_2-q_2}
\big)~,
\end{equs}
see also the proof of Lemma \ref{lem:blackbox} for related results.
\end{proof}

\section{Refined asymptotics}\label{sec:asymptoticsrefined}

To complete the asymptotic description of solution of
(\ref{eqn:NavierStokes}), we still have to prove the Corollary
\ref{cor:asymptoticsrefined}. Since the asymptotic description of $\omega$
is already proved in Theorem \ref{thm:asymptoticsrap}, it only remains to
prove the
\begin{theorem}\label{thm:refined}
Let $\olddelta<\olddelta_0<\frac{1}{8}$. Assume that
$\||y|^{\frac{1}{2}}{\bf v}(\xzero)\|_4<\infty$ and
$\||y|^{\frac{1}{2}-\olddelta_0}\S\uzero\|_1+
\||y|^{\frac{1}{2}-\olddelta_0}\S\huzero\|_1<\infty$, and let 
$a_1=-\M(\I\Q\omegazero)-\int_{\Omega_{+}}\Q\qun(x,y)\,{\rm d}x{\rm d}y$, 
$a_2=\M(\S\uzero)-\int_{\Omega_{+}}\Q\qun(x,y)\,{\rm d}x{\rm d}y$ and
$a_3=\M(\S\huzero)$. Let ${\bf a}=(a_1,a_2,a_3,a_4,a_1^2,a_1\Q a_3)$ and
$u_{\bf a},v_{\bf a}$ as in (\ref{eqn:defuaomegaa}), then there exist
a constant $a_4$ such that for all $\epsilon>0$, solutions to
(\ref{eqn:NavierStokes}) in $\Cu$ satisfy for all $x\geq\xzero$
\begin{equs}
\|u(x)-u_{{\bf a}}(x)\|_{\infty}&\leq C(\xzero,\|\triplet\|)
\;\chinese{x}^{-\frac{9}{8}+\olddelta_0}
\label{eqn:asymptrapcompu}
~,\\
\|v(x)-v_{{\bf a}}(x)\|_{\infty}&\leq C(\xzero,\|\triplet\|)
\;\chinese{x}^{-\frac{3}{2}+\olddelta_0}
\label{eqn:asymptrapcompv}
\end{equs}
for $\olddelta_0=(1+\epsilon)\olddelta$ and some constant
$C(\xzero,\|\triplet\|)$.
\end{theorem}
Here again, we note that we need only prove the estimates on $u$ and $v$ for
$x\geq2\xzero$, and, using Lemma \ref{lem:equalasympt}, we can choose to
compare $u(x)$ and $v(x)$ either to $u_{\bf a}(x)$ and $v_{\bf a}(x)$ or to
$u_{\bf a}(x-\xzero)$ and $v_{\bf a}(x-\xzero)$. The proof of Theorem
\ref{thm:refined} then stands on three pillars, the partial description of
Theorem \ref{thm:asymptoticsrap}, Lemma \ref{lem:univ} and its equivalent
on $\falpha$, $\galpha$ and $K_{\kzero}$, which we now state:
\begin{lemma}\label{lem:univpoisson}
Let $0\leq\gamma\leq1$ and $f$ satisfying
$\|\chinese{y}^{\gamma}f\|_{1}<\infty$. Then for all $m\geq0$, there exist
a constant $C_{\gamma}$
such that
\begin{equs}
\|\partial_y^m\falpha(x)(f-\M(f))\|_{{\infty}}
&\leq
C_{\gamma}x^{-1-m-\gamma}~
\||y|^{\gamma}f\|_{1}~,\\
\|\partial_y^m\galpha(x)(f-\M(f))\|_{{\infty}}
&\leq
C_{\gamma}x^{-1-m-\gamma}~
\||y|^{\gamma}f\|_{1}~,\\
\|\partial_y^mK_{\kzero}(x)(f-\M(f))\|_{{\infty}}
&\leq
C_{\gamma}x^{-1-m-\gamma}~
\||y|^{\gamma}f\|_{1}~,\\
\|\partial_y^m\H K_{\kzero}(x)(f-\M(f))\|_{{\infty}}
&\leq
C_{\gamma}x^{-1-m-\gamma}~
\||y|^{\gamma}f\|_{1}~,
\end{equs}
where $\M(f)=\int_{{\bf R}}f(y){\rm d}y$.
\end{lemma}
\begin{proof}   
The proof follows along the same lines as that of Lemma \ref{lem:univ},
e.g.
\begin{equs}
\|\partial_y^m\falpha(x)(f-\M(f))\|_{{\infty}}
&\leq
\sum_{n\in{\bf Z}}
\int_{-\infty}^{\infty}
\hspace{-3mm}{\rm d}k~
\left||k|^{m+\gamma}\ed^{-|k|x}\right|~
\int_{-\infty}^{\infty}
\hspace{-3mm}{\rm d}y~
\left|\frac{\ed^{iky}-1}{|ky|^{\gamma}}\right|~
|y|^{\gamma}|f_n(y)|~.
\end{equs}
The other estimates are similar.
\end{proof}
For convenience, the proof of Theorem \ref{thm:refined} will now be split
in the following subsections. We will first come back to the terms
proportional to $\omegazero$, $\uzero$ and $\huzero$. Then, using the first
order results on $\omega$ and $u$, we will prove (\ref{eqn:asymptrapcompv})
in a first round of estimates on the nonlinear terms. We will then use
(\ref{eqn:asymptrapcompv}) to prove (\ref{eqn:asymptrapcompu}) in a second
round of estimates on the nonlinear terms. In principle, this `ping-pong'
strategy could be systematically used to get higher order asymptotic
developments. 

\subsection{Back to the `linear' terms}

In this subsection, we consider the asymptotics of
\begin{equs}
U(x)&=K_{\kun}(x-\xzero)\myast\Lu\omegazero+
K_{\kzero}(x-\xzero)\uzero
~,~~~~
V(x)=K_{\kun}(x-\xzero)\myast\Lv\omegazero+
K_{\kzero}(x-\xzero)\huzero
~,\label{eqn:foruvzero}
\end{equs}
as $x\to\infty$. We first note that by Lemma \ref{lem:kernelun} and
\ref{lem:lunforLu}, for all $m\geq0$ and $x\geq2\xzero$, we have
\begin{equs}
\|\P K_{\kun}(x-\xzero)\myast\Lu\omegazero\|_{\infty}
+\|\P K_{\kun}(x-\xzero)\myast\Lv\omegazero\|_{\infty}\leq
C(\xzero,\|\triplet\|,m)\chinese{x}^{-m}
\end{equs}
since $\|\P K_{\kun}(x)\|_{1}$ decays exponentially as $x\to\infty$. Then
we note that for $x\geq2\xzero$, we have
\begin{equs}
\|\Q K_{\kun}(x-\xzero)(\Lu+\I)\omegazero\|_{\infty}
&\leq 
C\|\partial_y^2K_{\kun}(x-\xzero)\|_{\infty}
\|\I(\I(\Lu+\I)\omegazero)\|_{1}\leq
C\chinese{x}^{-\frac{3}{2}}\|\I\omegazero\|_{1}~.
\end{equs}
Thus, as in Section \ref{sub:linear}, the asymptotics of $U$ and $V$ are
the same as those of $U_1(x)=K_{\kun}(x-\xzero) f$ and $V_1(x)=-\partial_y
K_{\kun}(x-\xzero)f$, where $f=-\I\Q\omegazero$. Using Lemma \ref{lem:univ}
and that by Lemma \ref{lem:key} we have $\|y^{\gamma}f\|_{1}\leq
C(\xzero,\|(0,0,\omegazero)\|_{\xzero})$ for all
$\gamma\leq\frac{3}{2}-2(1+\epsilon)\olddelta$, we
conclude that for ${\bf a}_{1}=(-\M(\I\Q\omegazero)$,
$a_{4,1}=-\M(y\I\Q\omegazero)$ and
${\bf a}_1=(a_{1,1},0,0,a_{4,1},0,0)$, we have for $x\geq2\xzero$ that
\begin{equa}[1][eqn:asymptrapfrapclap]
\|U_1(x)-u_{{\bf a}_1}(x-\xzero)
\|_{\infty}&\leq C(\xzero,\|\triplet\|)
\;\chinese{x}^{-\frac{5}{4}+(1+\epsilon)\olddelta}
~,\\
\|V_1(x)-v_{{\bf a}_1}(x-\xzero)\|_{\infty}
&\leq C(\xzero,\|\triplet\|)\;
\chinese{x}^{-\frac{3}{2}+(1+\epsilon)\olddelta}~.
\end{equa}
We then note that since $\huzero=\H\uzero$ and $\uzero=-\H\huzero$, we have
\begin{equs}
K_{\kzero}(x-\xzero)\uzero&=
K_{\kzero}(x-\xzero)\S\uzero-K_{\kzero}(x-\xzero)\H\S\H\uzero
=K_{\kzero}(x-\xzero)\S\uzero-\H K_{\kzero}(x-\xzero)\S\huzero~,\\
K_{\kzero}(x-\xzero)\huzero&=
K_{\kzero}(x-\xzero)\S\huzero-K_{\kzero}(x-\xzero)\H\S\H\huzero
=K_{\kzero}(x-\xzero)\S\huzero+\H K_{\kzero}(x-\xzero)\S\uzero~.
\end{equs}
Defining $a_{2,1}=\M(\S\uzero)$, $a_{3,1}=\M(\S\huzero)$
and ${\bf a}_{2}=(0,a_{2,1},a_{3,1},0,0,0)$, we get
by Lemma \ref{lem:univpoisson} that for $x\geq2\xzero$,
\begin{equs}
\|K_{\kzero}(x-\xzero)\uzero
-u_{{\bf a}_2}(x-\xzero)\|_{\infty}
&\leq C c(\uzero,\huzero)\,\chinese{x}^{-\frac{3}{2}+(1+\epsilon)\olddelta}\\
\|K_{\kzero}(x-\xzero)\huzero
-v_{{\bf a}_2}(x-\xzero)\|_{\infty}
&\leq C c(\uzero,\huzero)\,\chinese{x}^{-\frac{3}{2}+(1+\epsilon)\olddelta}~,
\end{equs}
where $c(\uzero,\huzero)=(\||y|^{\frac{1}{2}-(1+\epsilon)\olddelta}\S\uzero\|_1+
\||y|^{\frac{1}{2}-(1+\epsilon)\olddelta}\S\huzero\|_1)$. Using Lemma
\ref{lem:equalasympt} and \ref{lem:alittlelemma}, we get
\begin{equs}
\|U(x)-u_{{\bf a}_3}(x)\|_{\infty}&\leq C(\xzero,\|\triplet\|,c(\uzero,\huzero))
\;\chinese{x}^{-\frac{5}{4}+(1+\epsilon)\olddelta}
\label{eqn:asymptrapcompulin}
~,\\
\|V(x)-v_{{\bf a}_3}(x)\|_{\infty}&\leq C(\xzero,\|\triplet\|,c(\uzero,\huzero))
\;\chinese{x}^{-\frac{3}{2}+(1+\epsilon)\olddelta}
\label{eqn:asymptrapcompvlin}
\end{equs}
for ${\bf a}_3=(a_{1,1},a_{2,1},a_{3,1},a_{4,1},0,0)$ and
some constant $C(\xzero,\|\triplet\|,c(\uzero,\huzero))$.

\subsection{Nonlinear terms, first round}

To complete the proof of Theorem \ref{thm:refined} we now have to give
the asymptotic development of
\begin{equs}
u_1(x)&={\cal F}_{1,u}(x)+{\cal F}_{2,u}(x)+\LA \qB(x)-\LB \qA(x)~,\\
v_1(x)&={\cal F}_{1,v}(x)+{\cal F}_{2,v}(x)-\LA \qA(x)-\LB \qB(x)~.
\end{equs}
We first tackle the terms
$u_2(x)=\LA \qB(x)-\LB \qA(x)$
and $v_2(x)=-\LA \qA(x)-\LB \qB(x)$.
Let $\qzero_{{\bf a}}(x)=u_{{\bf a}_1}(x)\omega_{{\bf a}_1}(x)$
where ${\bf a}_1=(a_1,0,0,0,0,0)$ and $\Delta\qB=\qB(x)-
\I\qzero_{{\bf a}}(x)$. We first note that by Theorem
\ref{thm:asymptoticsrap}
\begin{equs}
\|\Delta\qB(x)\|_{\infty}
&\leq
\|v(x)\|_{\infty}^2+
\|u(x)-u_{{\bf a}_1}(x)\|_{\infty}
\|u(x)+u_{{\bf a}_1}(x)\|_{\infty}
\\&
\leq C(\xzero,\|\triplet\|)\,
\chinese{x}^{-\frac{3}{2}+(1+\epsilon)\olddelta}~.
\end{equs}
Then, since $\P\qzero_{\bf a}(x)=0$ and $\Q\LB=0$ implies
$\LB\qB=\LB\Delta\qB$, we get
\begin{equs}
\|\LA\qA(x)\|_{\infty}+
\|\LB\qA(x)\|_{\infty}&\leq
C(\xzero,\|\triplet\|)\,\chinese{x}^{-\frac{3}{2}+\olddelta}~,\\
\|\LA\Delta\qB(x)\|_{\infty}+
\|\LB\qB(x)\|_{\infty}&\leq
C(\xzero,\|\triplet\|)\,\chinese{x}^{-\frac{3}{2}+(1+\epsilon)\olddelta}~,
\end{equs}
while $\Q\LA=\dirac$ implies
\begin{equa}[1][eqn:partial]
\|u_{2}(x)-\I \qzero_{{\bf a}}(x)\|_{\infty}&\leq
C(\xzero,\|\triplet\|)\,\chinese{x}^{-\frac{3}{2}+(1+\epsilon)\olddelta}
\\
\|v_{2}(x)\|_{\infty}&\leq
C(\xzero,\|\triplet\|)\,\chinese{x}^{-\frac{3}{2}+(1+\epsilon)\olddelta}~.
\end{equa}
It then follows from (\ref{eqn:partial}) and Propositions
\ref{prop:thecalfisomegaevolve} to \ref{prop:thecalfisu} that
\begin{equs}
\|u_1(x)-{\cal F}_{1,3,u}(x)
-{\cal F}_{1,5,u}(x)
-{\cal F}_{1,6,u}(x)
\|_{\infty}
&\leq C(\xzero,\|\triplet\|)\,
\chinese{x}^{-\frac{3}{2}+(1+\epsilon)\olddelta}\\
\|v_1(x)-{\cal F}_{1,1,\omega}(x)-{\cal F}_{1,3,v}(x)\|_{\infty}
&\leq C(\xzero,\|\triplet\|)\,
\chinese{x}^{-\frac{3}{2}+(1+\epsilon)\olddelta}~,
\end{equs}
where
\begin{equs}
{\cal F}_{1,5,u}(x)&=
\int_{\xzero}^{x}\hspace{-3mm}{\rm d}\tilde{x}~
K_{\kdeux}(\tilde{x}-x)\qzero(\tilde{x})
+2\I\qzero_{{\bf a}}(x)\\
{\cal F}_{1,6,u}(x)&=
-\int_{\xzero}^{x}\hspace{-3mm}{\rm d}\tilde{x}~
K_{\kdouze}(\tilde{x}-x)\qun(\tilde{x})
-\I\qzero_{{\bf a}}(x)
\end{equs}
The asymptotic development of ${\cal F}_{1,1,\omega}(x)$ is established
in Proposition \ref{prop:forkdouze} above, that of 
${\cal F}_{1,3,u}(x)$ and ${\cal F}_{1,3,v}(x)$ in Proposition
\ref{prop:forwithQ} below, followed by that of 
${\cal F}_{1,5,u}(x)$ in Proposition
\ref{prop:forwithPKdeux}. The proof of Theorem \ref{thm:refined} will be
completed by the study of ${\cal F}_{1,6,u}(x)$ in
Section \ref{sec:secondround}.

\begin{proposition}\label{prop:forwithQ}
Assume that $\qun$ satisfies (\ref{eqn:nonlinbounds}), and
define
${\bf a}_4=(0,-\int_{\Omega_{+}}\Q\qun(x,y)\,{\rm d}x{\rm d}y,0,0,0,0)$,
then for all $\epsilon>0$, there exist a constant $C$ such that 
\begin{equs}
\|{\cal F}_{1,3,u}(x)-u_{{\bf a}_4}(x)\|_{{\infty}}+
\|{\cal F}_{1,3,v}(x)-v_{{\bf a}_4}(x)\|_{{\infty}}\leq
C\chinese{x}^{-\frac{3}{2}+(1+\epsilon)\olddelta}\|\triplet\|^2
\end{equs}
for all $x\geq2\xzero$.
\end{proposition}

\begin{proof}   
The proof is very similar to the one of Proposition \ref{prop:forkdouze}.
We first note that $\|\falpha\|_{\frac{1}{2\epsilon\olddelta},
\{0,1-2\epsilon\olddelta\}}+\|\galpha\|_{\frac{1}{2\epsilon\olddelta},
\{0,1-2\epsilon\olddelta\}}\leq C$, 
$\|\qun\|_{\frac{1}{1-2\epsilon\olddelta},\frac{3}{2}-(1-\epsilon)\olddelta}\leq
C\|\triplet\|^2$. Then, we define $T(x)=\int_{\xzero}^{x}{\rm
d}\tilde{x}~\falpha(x-\tilde{x})\qun(\tilde{x})$. Since
$\partial_x\falpha(x)=\partial_y\galpha(x)$, after integration by parts, we
have
\begin{equs}
T(x)&=
\falpha(x-\xzero)\int_{\xzero}^{\infty}
\hspace{-3mm}{\rm d}z~\qun(z)-
\falpha(x-\xzero)\int_{x}^{\infty}
\hspace{-3mm}{\rm d}z~\qun(z)
-\int_{\xzero}^{x}
\hspace{-3mm}{\rm d}\tilde{x}
~\partial_{y}\galpha(x-\tilde{x})
\int_{\tilde{x}}^{x}
\hspace{-3mm}{\rm d}z~\qun(z)~.
\end{equs}
Let $T_1(x)=\falpha(x-\xzero)\int_{x}^{\infty}{\rm d}z~\qun(z)$ and
$T_2(x)=\int_{\xzero}^{x}{\rm d}\tilde{x}~
\partial_{y}\galpha(x-\tilde{x})\int_{\tilde{x}}^{x}{\rm d}z~\qun(z)$.
Since $x\geq2\xzero$, we
have
\begin{equs}
\|T_1(x)\|_{\infty}&\leq C\|\triplet\|^2(x-\xzero)^{-1}
\int_{x}^{\infty}
\hspace{-3mm}{\rm d}z~z^{-\frac{3}{2}+\olddelta}\leq
C\|\triplet\|^2\chinese{x}^{-\frac{3}{2}+\olddelta}\\
\|T_2(x)\|_{\infty}&\leq C\|\triplet\|^2
\D{2-2\epsilon\olddelta,\frac{3}{2}-\olddelta(1-\epsilon)}
{2-2\epsilon\olddelta,\frac{3}{2}-\olddelta(1-\epsilon)}
(x,\xzero)
\leq
C\|\triplet\|^2
\chinese{x}^{-\frac{3}{2}+(1+\epsilon)\olddelta}~.
\end{equs}
Then, we define $T_3(x)=\falpha(x-\xzero)\int_{\xzero}^{\infty}{\rm
d}z~\qun(z)$. Since
$|\strouhal|\leq\chinese{\xzero}^{\olddelta}\leq\chinese{x}^{\olddelta}$ and
$x\geq2\xzero$, (using Lemma \ref{lem:onD} in the second inequality)
\begin{equs}
\|\P T_3(x)\|_{\infty} 
\leq C\chinese{x}^{-2+\olddelta}
\|\triplet\|^2
\int_{\xzero}^{\infty}
\hspace{-3mm}{\rm d}z~\chinese{z}^{-\frac{3}{2}+\olddelta}
\leq C\chinese{x}^{-2+\olddelta}
\|\triplet\|^2~.
\end{equs}
Finally, by Proposition \ref{prop:forwithQ} and using
$\||y|^{\frac{1}{2}-(1+\epsilon)\olddelta}\qun\|_{1}\leq
C\||y|^{1-\olddelta}\qun\|_{2}\leq
C\chinese{z}^{-1-\frac{1}{4}(1-2\olddelta)}\|(u,v,\omega)\|^2$, we have
\begin{equs}
\|\Q T_3(x)
-u_{{\bf a}_4}(x-\xzero)\|_{\infty}
&\leq
C\chinese{x}^{-\frac{3}{2}+(1+\epsilon)\olddelta}
\|\triplet\|^2
\int_{\xzero}^{\infty}
\hspace{-3mm}
{\rm d}z~\chinese{z}^{-1-\frac{1}{4}(1-2\olddelta)}~,
\end{equs}
where we used $\Q\falpha=\Q K_{\kzero}$. Since $\olddelta<\frac{1}{2}$, the
proof of the estimate on $\|{\cal F}_{1,3,u}(x)-u_{{\bf a}_4}(x)\|_{\infty}$ is completed
using Lemma \ref{lem:equalasympt}. The proof of the estimate on
$\|{\cal F}_{1,3,v}(x)-v_{{\bf a}_4}(x)\|_{\infty}$ being very similar, we
omit the details.
\end{proof}
We now turn to the asymptotics of ${\cal F}_{1,5,u}$.
\begin{proposition}\label{prop:forwithPKdeux}
Assume that $\qzero$ satisfies (\ref{eqn:nonlinbounds}) and let
$\qzero_{{\bf a}}(x)=u_{{\bf a}_1}(x)\omega_{{\bf a}_1}(x)$
where ${\bf a}_1=(a_1,0,0,0,0,0)$. Assume that
\begin{equs}
\||y|(u(\xzero)^2+v(\xzero)^2)\|_{2}\leq C\|\triplet\|^2~.
\end{equs}
Let ${\bf a}_{5}=(0,0,0,
-\int_{{\bf R}}\Q u(\xzero,y)v(\xzero,y)\,{\rm d}y,a_1^2,0)$,
then for all $\epsilon>0$, there exist a constant
$C=C(\xzero,\|\triplet\|)$ such that 
\begin{equs}
\|{\cal F}_{1,5,u}(x)-u_{{\bf a}_5}(x)\|_{{\infty}}\leq
C(\xzero,\|\triplet\|)\chinese{x}^{-\frac{5}{4}+\epsilon\olddelta}
\end{equs}
for all $x\geq2\xzero$.
\end{proposition}

\begin{proof}   
We first note that Theorem \ref{thm:asymptoticsrap} implies that 
$\|\qzero-\qzero_{{\bf a}}\|_{p,2-
(1+\epsilon)\olddelta-\frac{1}{2p}}\leq C(\xzero,\|\triplet\|)$
for all $1\leq p\leq\infty$ and $\epsilon>0$, so that for
$p=\frac{1}{1-\epsilon\olddelta}$, we have
\begin{equs}
\Big\|
\int_{\frac{x+\xzero}{2}}^x
\hspace{-4mm}{\rm d}\tilde{x}~
K_{\kdeux}(x-\tilde{x})\big(
\qzero(\tilde{x})-\qzero_{\bf a}(\tilde{x})
\big)
\Big\|_{\infty}&\leq
C\|\triplet\|^2
\int_{\frac{x+\xzero}{2}}^x
\frac{{\rm d}\tilde{x}~
\chinese{\tilde{x}}^{
-\frac{3}{2}+(1+\frac{3\epsilon}{2})\olddelta}
}
{(x-\tilde{x})^{1-\frac{\epsilon\olddelta}{2}}}\\
&\leq
C\chinese{x}^{-\frac{3}{2}+(1+2\epsilon)\olddelta}
\|\triplet\|^2~.
\end{equs}
Then, we note that
\begin{equs}
\qzero(x)-\qzero_{\bf a}(x)=
\partial_x \qA+
{\textstyle\frac{1}{2}}\partial_y\big(v(x)^2\big)
-
{\textstyle\frac{1}{2}}\partial_y\big(
(u(x)-u_{{\bf a}_1}(x))
(u(x)+u_{{\bf a}_1}(x))~.
\end{equs}
Now, let $T_{1}(x)=\int_{\xzero}^{\frac{x+\xzero}{2}}
{\rm d}\tilde{x}~K_{\kdeux}(x-\tilde{x})\partial_{\tilde{x}}\qA(\tilde{x})
+K_{\kdeux}(x-\xzero)\qA(\xzero)$. By (\ref{eqn:nonlinbounds}) and
$\|\partial_xK_{\kdeux}\|_{\infty}\leq	  
\|\partial_yK_{\khuit}(x-\tilde{x})\|_{\infty}
+\|\partial_yK_{\kdix}(x-\tilde{x})\|_{\infty}
$, integrating by parts, we find
\begin{equs}
\|T_1(x)\|_{\infty}\leq
C\|\triplet\|^2
\chinese{x}^{-\frac{3}{2}+\olddelta}
\Big(1+\chinese{\xzero}^{\olddelta}\Big)~,
\end{equs}
while Lemma \ref{lem:univ} and \ref{lem:alittlelemma}, together with
$\||y|^{\frac{1}{2}-\epsilon\olddelta}\qA(\xzero)\|_1\leq
\||y|(u(\xzero)^2+v(\xzero)^2\|_2
\leq C\|\triplet\|^2$ show that
\begin{equs}
\|K_{\kdeux}(x-\xzero)\qA(\xzero)+
u_{{\bf a}_5}(x-\xzero)\|_{\infty}&\leq
C\|\triplet\|^2
\chinese{x}^{-\frac{5}{4}+\epsilon\olddelta}~.
\end{equs}
Then, let $\qB_1(x)=v(x)^2$ and
$\qB_2(x)=(u(x)-u_{{\bf a}_1}(x))(u(x)+u_{{\bf a}_1}(x))$.
By Theorem \ref{thm:asymptoticsrap}, we have
$\|\qB_2(x)\|_{2}\leq C\chinese{x}^{-1+\epsilon\olddelta}
\|\triplet\|^2$ and
$\|\qB_1(x)\|_{1}\leq C\chinese{x}^{-1+2\olddelta}\|\triplet\|^2$. 
Therefore, for $x\geq2\xzero$, we have
\begin{equs}
\Big\|
\int_{\xzero}^{\frac{x+\xzero}{2}}
\hspace{-4mm}
{\rm d}\tilde{x}~\partial_{y}K_{\kdeux}(x-\tilde{x})\qB_1(\tilde{x})
\Big\|&\leq
C\chinese{x}^{-\frac{3}{2}+2\olddelta}
\|\triplet\|^2\\
\Big\|
\int_{\xzero}^{\frac{x+\xzero}{2}}
\hspace{-4mm}
{\rm d}\tilde{x}~\partial_{y}K_{\kdeux}(x-\tilde{x})\qB_2(\tilde{x})
\Big\|&\leq
C\chinese{x}^{-\frac{5}{4}+\epsilon\olddelta}
\|\triplet\|^2~.
\end{equs}
We then define $P_{6}(x)=\Q
\int_{\xzero}^{x}{\rm d}\tilde{x}~
\big(K_{\kdeux}(x-\tilde{x})-\partial_yK_{\kc}(x-\tilde{x})\big)
\qzero_{\bf a}(\tilde{x})$ and we get
\begin{equs}
\|P_6(x)\|_{\infty}&\leq
C\|\triplet\|^2
\int_{\xzero}^{x}\hspace{-3mm}{\rm d}\tilde{x}
\int_{-\infty}^{\infty}\hspace{-3mm}{\rm d}k~
\Q\ed^{\Lambdam(x-\tilde{x})-\frac{k^2}{2}\tilde{x}}
\big(|k|^5(x-\tilde{x})+|k|^3\big)\tilde{x}^{-\frac{1}{2}}\\
&\leq C\|\triplet\|^2
\int_{\xzero}^{x}\hspace{-3mm}{\rm d}\tilde{x}
\min\Big(
\frac{\chinese{x-\tilde{x}}^3}{(x-\tilde{x})^{5}},
\frac{x-\tilde{x}}{\tilde{x}^3}+
\frac{1}{\tilde{x}^2}
\Big)\tilde{x}^{-\frac{1}{2}}
\leq
C\|\triplet\|^2\chinese{x}^{-\frac{3}{2}}~.
\end{equs}
It remains to establish the asymptotic comportment of
\begin{equs}
T(x)&=
\int_{\xzero}^{x}\hspace{-3mm}{\rm d}\tilde{x}~
\partial_yK_{\kc}(x-\tilde{x})\qzero_{\bf a}(\tilde{x})
+2\I\qzero_{\bf a}(x)~.
\end{equs}
We first note that $T(x)$ is conveniently computed in
terms of its Fourier transform, which reads
\begin{equs}
\hat{T}(x,k)&=
{\textstyle
\frac{ika_1^2}{4}{\rm erf}\big(
\frac{ik\sqrt{\xzero}}{\sqrt{2}}
\big)\ed^{-k^2x}-
\frac{ika_1^2}{4}{\rm erf}\big(
\frac{ik\sqrt{x}}{\sqrt{2}}
\big)\ed^{-k^2x}
-\frac{a_1^2\ed^{-\frac{k^2x}{2}}}{2\sqrt{2\pi x}}}~,\\
&=\hat{T}_2(x,\xzero,k)+x^{-\frac{1}{2}}H\big(k\sqrt{x}\big)~.
\end{equs}
For $T_2(x,\xzero,y)$, we note that for $x\geq\xzero$, we have
\begin{equs}
\|T_2(x)\|_{\infty}\leq
\int_{-\infty}^{\infty}
\hspace{-3mm}{\rm d}k~
{\textstyle
\big|\frac{ika_1^2}{4}{\rm erf}\big(
\frac{ik\sqrt{\xzero}}{\sqrt{2}}
\big)\big|}\ed^{-k^2x}\leq
a_1^2
\int_{-\infty}^{\infty}
\hspace{-3mm}{\rm d}k~k^2\ed^{-\frac{k^2x}{2}}
\leq C\|\triplet\|^2\chinese{x}^{-\frac{3}{2}}~.
\end{equs}
To complete the proof, we only have to prove that the inverse Fourier
transform of $\hat{H}$ is $-\frac{a_1^2h}{2}$. To do so, we note
that $\hat{H}(k)$ satisfies
\begin{equs}
k\hat{H}'(k)+(2k^2-1)\hat{H}(k)-{\textstyle
\frac{a_1^2\ed^{-\frac{k^2}{2}}}{2\sqrt{2}\pi}}=0~,~~~
\int_{-\infty}^{\infty}
\hspace{-3mm}{\rm d}k~\hat{H}(k)=-{\textstyle\frac{a_1^2}{4}}
~,~~~
\int_{-\infty}^{\infty}
\hspace{-3mm}{\rm d}k~k\hat{H}(k)=0~,
\end{equs}
which after inverse Fourier transform leads to
\begin{equs}
H''(y)+{\textstyle\frac{y}{2}}H'(y)+H(y)
+{\textstyle\frac{a_1^2\ed^{-\frac{y^2}{2}}}{8\pi}}=0~,~~~~
H(0)=-{\textstyle\frac{a_1^2}{8\pi}}~,~~~~
H'(0)=0~,
\end{equs}
whose unique solution is $H(y)=-\frac{a_1^2h(y)}{2}$.
\end{proof}

\subsection{Nonlinear terms, second round}\label{sec:secondround}

In view of Remark \ref{rem:heat} and the corresponding theory on nonlinear
heat equations, (see e.g.\ \cite{bricmont}), we may guess that the decay
rates of Proposition \ref{prop:forkdouze} on ${\cal F}_{1,2,u}$ would be
improved using higher moments of $\qun$, i.e.\ after substraction of
$u_{{\bf a}_0}$ with ${\bf a}_0=
(\Q\int_{\Omega_{+}}\qun(x,y)\,{\rm d}x{\rm d}y,0,0,
\Q\int_{\Omega_{+}}y\qun(x,y)\,{\rm d}x{\rm d}y,0,0)$. This is wrong since
the first moment $\int_{\Omega_{+}}y\qun(x,y)\,{\rm d}x{\rm d}y$ of $\qun$
is infinite in general\footnote{except for symmetric flows where $\int_{{\bf
R}}y\qun(x,y)\,{\rm d}y=0$}. However, with the estimates obtained so far on
$v-v_{{\bf a}}$ and $\omega-\omega_{{\bf a}}$, we can show that higher
moments are well defined for $\qun-\qun_{{\bf a}}$ as shows the
\begin{lemma}\label{lem:universalQ}
Let $\qun_{\bf a}=v_{\bf a}\omega_{\bf a}$ where
$v_{\bf a}$ and $\omega_{\bf a}$ are defined in (\ref{eqn:defuaomegaa})
and ${\bf a}=(a_1,a_2,a_3,0,0,0)$.
Then for all $\epsilon>0$, we have
\begin{equs}
\|\qun-\qun_{\bf a}\|_{\infty,\frac{5}{2}-(1+\epsilon)\olddelta}+
\|\qun-\qun_{\bf a}\|_{1,2-(1+\epsilon)\olddelta}&\leq
C\|\triplet\|^2
\label{eqn:untrouc}
\\
\||y|^{\gamma}(\qun-\qun_{\bf a})\|_{1,\frac{9}{4}-\gamma-2\olddelta(1+\frac{3\epsilon}{4})}
&\leq C\|\triplet\|^2
\end{equs}
for all $\frac{1}{2}\leq\gamma\leq\frac{5}{4}-2\olddelta(1+\epsilon)$.
\end{lemma}

\begin{proof}
The estimate (\ref{eqn:untrouc}) follows at once from the boundedness of
$\|(v-v_{\bf a})\|_{\infty,\frac{3}{2}-(1+\epsilon)\olddelta}$ and Theorem
\ref{thm:asymptoticsrap}. Now let
$\frac{1}{2}\leq\gamma\leq\frac{5}{4}-2\olddelta(1+\epsilon)$ and define
$\epsilon_1=1-\frac{1}{2\gamma}(1-(4+\epsilon)\olddelta)$ and
$\beta_0=(1-\epsilon_1)\gamma+\frac{1+\epsilon}{2}\olddelta$. By hypothesis on
$\gamma$, we have $\gamma\epsilon_1\leq1$ and
$0\leq\beta_0\leq1-2(1+\epsilon)\olddelta$, so that
\begin{equs}
\||y|^{\gamma}(\qun(x)-\qun_{\bf a}(x))\|_{1}&\leq
\|v(x)-v_{\bf a}(x)\|_{\infty}
\||y|^{\gamma}\omega(x)\|_{1}+
\||y|^{\epsilon_1\gamma}v_{\bf a}(x)\|_{\infty}
\||y|^{\beta_0}(\omega(x)-\omega_{\bf a}(x))\|_{2}\\
&\leq
C\|\triplet\|^2\Big(
\chinese{x}^{-2+\frac{\gamma}{2}+2\olddelta(1+\frac{3\epsilon}{4})}
+
\chinese{x}^{-\frac{9}{4}+\gamma+2\olddelta(1+\frac{3\epsilon}{4})}
\Big)~.
\end{equs}
This completes the proof since $\gamma\geq\frac{1}{2}$.
\end{proof}
We can now conclude the proof of Theorem \ref{thm:refined} by proving the
following
\begin{proposition}
Assume that $\qun$ satisfies (\ref{eqn:nonlinbounds}), let 
${\bf a}_{6}=(\int_{\Omega_{+}}\Q\qun(x,y)\,{\rm d}x{\rm d}y,0,0,
a_{4,2},0,a_1\Q a_3)$, $\qun_{\bf a}$ as in Lemma \ref{lem:universalQ}.
Then there exist $a_{4,2}\in{\bf R}$ such that
\begin{equs}
\|{\cal F}_{1,6,u}(x)
-u_{{\bf a}_6}(x)
\|_{\infty}
\leq C(\xzero,\|\triplet\|)\,\chinese{x}^{-\frac{9}{8}+(1+\epsilon)\olddelta}~.
\end{equs}
for some constant $C=C(\xzero,\|\triplet\|)$.
\end{proposition}
\begin{proof}
We first note that we can write
\begin{equs}
\Q Q_{\bf a}(x,y)&=
{\textstyle
\frac{a_1}{x}f_1\big(\frac{y}{\sqrt{x}}\big)
\Big(
\frac{a_1}{x}f_1\big(\frac{y}{\sqrt{x}}\big)+
\frac{b}{x}g_0\big(\frac{y}{x}\big)+
\frac{c}{x}g_1\big(\frac{y}{x}\big)
\Big)}~,\\
&=
\underbrace{\textstyle
\frac{a_1^2}{x^2}f_1\big(\frac{y}{\sqrt{x}}\big)^2+
\frac{a_1b}{x^2}f_1\big(\frac{y}{\sqrt{x}}\big)}_{\equiv \qun_{{\bf a},1}(x,y)}
+\underbrace{\textstyle\frac{a_1c}{x^{\frac{5}{2}}}f_2\big(\frac{y}{\sqrt{x}}\big)
g_0\big(\frac{y}{x}\big)
-
\frac{a_1b}{x^{3}}f_3\big(\frac{y}{\sqrt{x}}\big)
g_0\big(\frac{y}{x}\big)
}_{\equiv \qun_{{\bf a},2}(x,y)}
\end{equs}
where $f_m(z)=\frac{z^m\ed^{-\frac{z^2}{4}}}{4\sqrt{\pi}}$, 
$g_m(z)=\frac{z^m}{1+z^2}$, $b=\Q a_3$ and $c=\Q a_2$. Now, since
$|g_0(z)|\leq 1$, $\qun_{{\bf a},2}$ satisfies the same estimates as
$\qun(x)-\qun_{{\bf a}}(x)$ (with even better decay rates). To exploit
this, we define $\Delta\qun(x)=\Q(\qun(x)-\qun_{{\bf a}}(x))
+\qun_{{\bf a},2}(x)$ and
\begin{equs}
T_{3,1}(x)&=
-\int_{\frac{x+\xzero}{2}}^{x}\hspace{-3mm}{\rm d}\tilde{x}~
K_{\kdouze}(x-\tilde{x})\Delta\qun(\tilde{x})~,\\
T_{3,2}(x)&=
-\int_{\xzero}^{\frac{x+\xzero}{2}}\hspace{-3mm}{\rm d}\tilde{x}~\Big(
K_{\kdouze}(x-\tilde{x})\big(
\Delta\qun(\tilde{x})-\M(\Delta\qun(\tilde{x}))\big)
-\partial_yK_{\kdouze}(x-\tilde{x})
\M(y\Delta\qun(\tilde{x}))\Big)~.
\end{equs}
Using Lemma \ref{lem:univ} and \ref{lem:universalQ}, as well as
$x\geq2\xzero$, we get
\begin{equs}
\|T_{3,1}(x)\|_{\infty}&\leq C\,x\sup_{\xi\geq\frac{x+\xzero}{2}}
\|\Delta\qun(\xi)\|_{\infty}
\leq C\chinese{x}^{-\frac{3}{2}+(1+\epsilon)\olddelta}
\|\triplet\|^2~,\\
\|T_{3,2}(x)\|_{\infty}&\leq C
\chinese{x}^{-\frac{9}{8}+(1+\epsilon)\olddelta}
\int_{\xzero}^{\frac{x+\xzero}{2}}\hspace{-3mm}{\rm d}\tilde{x}~
\||y|^{\frac{5}{4}-2\olddelta(1+\epsilon)}\Delta\qun(\tilde{x})\|_1\\
&\leq C
\chinese{x}^{-\frac{9}{8}+(1+\epsilon)\olddelta}\|\triplet\|^2
\int_{\xzero}^{\infty}\hspace{-3mm}{\rm d}\tilde{x}~
\chinese{\tilde{x}}^{-1-\frac{\epsilon\olddelta}{2}}~.
\end{equs}
We then define
\begin{equs}
T_{3,3}(x)&=
K_{\kdouze}(x-\xzero)
\int_{\xzero}^{\frac{x+\xzero}{2}}\hspace{-3mm}{\rm d}\tilde{x}~
\M(\Delta\qun(\tilde{x}))
-\int_{\xzero}^{\frac{x+\xzero}{2}}\hspace{-3mm}{\rm d}\tilde{x}~
K_{\kdouze}(x-\tilde{x})\M(\Delta\qun(\tilde{x}))~,\\
T_{3,4}(x)&=
\partial_yK_{\kdouze}(x-\xzero)
\int_{\xzero}^{\frac{x+\xzero}{2}}\hspace{-3mm}{\rm d}\tilde{x}~
\M(y\Delta\qun(\tilde{x}))
-\int_{\xzero}^{\frac{x+\xzero}{2}}\hspace{-3mm}{\rm d}\tilde{x}~
\partial_yK_{\kdouze}(x-\tilde{x})\M(y\Delta\qun(\tilde{x}))~,
\end{equs}
and note that after integration by parts, using 
$\|\partial_xK_{\kdouze}(x)\|_{\infty}\leq
\|K_{\khuit}(x)\|_{\infty}+\|K_{\kdix}(x)\|_{\infty}\leq
C\chinese{x}^{-\frac{3}{2}}\chinese{\xzero}^{\olddelta}$ and
$\|\partial_x\partial_yK_{\kdouze}(x)\|_{\infty}\leq
\|\partial_yK_{\khuit}(x)\|_{\infty}+\|\partial_yK_{\kdix}(x)\|_{\infty}\leq
C\chinese{x}^{-2}\chinese{\xzero}^{\olddelta}$ if $x>0$, we get
\begin{equs}
\|T_{3,3}(x)\|_{\infty}&\leq
C\|\triplet\|^2
\chinese{x}^{-\frac{3}{2}+\olddelta}
\int_{\xzero}^{\frac{x+\xzero}{2}}\hspace{-4mm}{\rm d}\tilde{x}
\hspace{-1.5mm}
\int_{\tilde{x}}^{\frac{x+\xzero}{2}}\hspace{-4mm}{\rm d}z~
\chinese{z}^{-2+(1+\epsilon)\olddelta}\leq
C\|\triplet\|^2
\chinese{x}^{-\frac{3}{2}+(2+\epsilon)\olddelta}\\
\|T_{3,4}(x)\|_{\infty}&\leq
C\|\triplet\|^2
\chinese{x}^{-2+\olddelta}
\int_{\xzero}^{\frac{x+\xzero}{2}}\hspace{-4mm}{\rm d}\tilde{x}
\hspace{-1.5mm}
\int_{\tilde{x}}^{\frac{x+\xzero}{2}}\hspace{-4mm}{\rm d}z~
\chinese{z}^{-\frac{3}{2}+(1+\epsilon)\olddelta}\leq
C\|\triplet\|^2
\chinese{x}^{-\frac{3}{2}+(2+\epsilon)\olddelta}~,
\end{equs}
while also for $x\geq2\xzero$, we have
\begin{equs}
\Big\|K_{\kdouze}(x-\xzero)
\int_{\frac{x+\xzero}{2}}^{\infty}\hspace{-3mm}{\rm d}\tilde{x}~
\M(\Delta\qun(\tilde{x}))\Big\|_{\infty}&\leq
C\|\triplet\|^2\chinese{x}^{-\frac{1}{2}}
\int_{\frac{x+\xzero}{2}}^{\infty}\hspace{-3mm}{\rm d}\tilde{x}~
\chinese{\tilde{x}}^{-2+(1+\epsilon)\olddelta}
~,\\
&\leq C\|\triplet\|^2\chinese{x}^{-\frac{3}{2}+(1+\epsilon)\olddelta}\\
\Big\|\partial_yK_{\kdouze}(x-\xzero)
\int_{\frac{x+\xzero}{2}}^{\infty}\hspace{-3mm}{\rm d}\tilde{x}~
\M(y\Delta\qun(\tilde{x}))\Big\|_{\infty}&\leq
C\|\triplet\|^2\chinese{x}^{-1}
\int_{\frac{x+\xzero}{2}}^{\infty}\hspace{-3mm}{\rm d}\tilde{x}~
\chinese{\tilde{x}}^{-\frac{5}{4}+2(1+\frac{3\epsilon}{2})\olddelta}
~,\\
&\leq C\|\triplet\|^2
\chinese{x}^{-\frac{5}{4}+2(1+\frac{3\epsilon}{2})\olddelta}~.
\end{equs}
Now, let $a_{1,3}=\int_{\xzero}^{\infty}{\rm d}x~
\M(\Delta\qun(x))$ and $a_{4,3}=\int_{\xzero}^{\infty}{\rm d}x~
\M(y\Delta\qun(x))$. As is easily shown using Lemma
\ref{lem:universalQ}, $a_{1,3}$ and $a_{4,3}$ are bounded, and using
Lemma \ref{lem:alittlelemma}, we have for $x\geq2\xzero$
\begin{equs}
\|\big(K_{\kdouze}(x-\xzero)-K_{\kc}(x-\xzero)\big)a_{1,3}
\|_{\infty}&\leq
C\chinese{x}^{-\frac{3}{2}+\olddelta}\|\triplet\|^2~,\\
\|\partial_y\big(K_{\kdouze}(x-\xzero)-K_{\kc}(x-\xzero)\big)
a_{4,3}
\|_{\infty}&\leq
C\chinese{x}^{-\frac{3}{2}+\olddelta}\|\triplet\|^2~.
\end{equs}
for some constant $C$ possibly depending on $\xzero$. After collecting the
results obtained so far, and using
\begin{equs}
\|\big(K_{\kc}(x-\xzero)-K_{\kc}(x)\big)a_{1,3}
\|_{\infty}&\leq
C\chinese{x}^{-\frac{3}{2}+\olddelta}\chinese{\xzero}\|\triplet\|^2~,\\
\|\partial_y\big(K_{\kc}(x-\xzero)-K_{\kc}(x)\big)
a_{4,3}
\|_{\infty}&\leq
C\chinese{x}^{-\frac{3}{2}+\olddelta}\chinese{\xzero}\|\triplet\|^2~,
\end{equs}
we get for ${\bf a}_7=(a_{1,3},0,0,a_{4,3},0,0)$ that
\begin{equs}
\Big\|
\int_{\xzero}^{x}\hspace{-3mm}{\rm d}\tilde{x}~
K_{\kdouze}(x-\tilde{x})\Delta\qun(\tilde{x})
-u_{{\bf a}_7}(x)
\Big\|_{\infty}
&\leq C\|\triplet\|^2
\chinese{x}^{-\frac{5}{4}+2(1+\epsilon)\olddelta}~.
\end{equs}
In other words, since $\Q\qun=\Delta\qun+\Q\qun_{{\bf a},1}$, it only
remains to establish the asymptotic comportment of
\begin{equs}
T_{4}(x)=
-\int_{\xzero}^{x}\hspace{-3mm}{\rm d}\tilde{x}~
\Q K_{\kdouze}(x-\tilde{x})\qun_{{\bf a},1}(\tilde{x})~.
\end{equs}
To do so, we first define
\begin{equs}
T_5(x)=-
\int_{\xzero}^{x}\hspace{-3mm}{\rm d}\tilde{x}~
\Q\big(K_{\kdouze}(x-\tilde{x})+K_{\kc}(x-\tilde{x})\big)
\qun_{{\bf a},1}(\tilde{x})~,
\end{equs}
on which we get
\begin{equs}
\|T_5(x)\|_{\infty}&\leq
C(\xzero,\|\triplet\|)
\int_{\xzero}^{x}\hspace{-3mm}{\rm d}\tilde{x}
\int_{-\infty}^{\infty}
\hspace{-4mm}{\rm d}k~
\Q
\ed^{\Lambdam(x-\tilde{x})-\frac{k^2\tilde{x}}{4}}
\Big(
k^2+k^4(x-\tilde{x})
\Big)\tilde{x}^{-\frac{3}{2}}\\
&\leq
C(\xzero,\|\triplet\|)
\int_{\xzero}^{x}\hspace{-3mm}{\rm d}\tilde{x}
\min\Big(
\frac{\chinese{x-\tilde{x}}^{\frac{5}{2}}}{(x-\tilde{x})^4}~,~
\frac{1}{\tilde{x}^{\frac{3}{2}}}+
\frac{x-\tilde{x}}{\tilde{x}^{\frac{5}{2}}}
\Big)
\tilde{x}^{-\frac{3}{2}}
\leq
C\|\triplet\|^2\chinese{x}^{-\frac{3}{2}}~.
\end{equs}
We finally define
\begin{equs}
T_{6}(x)=
\int_{\xzero}^{x}\hspace{-3mm}{\rm d}\tilde{x}~
K_{\kc}(x-\tilde{x})\qun_{{\bf a},1}(\tilde{x})~.
\end{equs}
As in Proposition \ref{prop:forwithPKdeux}, $T_6$ is conveniently computed
in terms of its Fourier transform, which reads
\begin{equs}
\hat{D}_6(x,k)&=
a_1b(\ln(x)-\ln(\xzero))ik\ed^{-k^2x}+
\frac{a_1^2\ed^{-k^2(x-\frac{\xzero}{2})}}{4\sqrt{2\pi\xzero}}
-\frac{a_1^2\ed^{-\frac{k^2x}{2}}}{4\sqrt{2\pi x}}~,\\
&=
a_1b(\ln(x)-\ln(\xzero))ik\ed^{-k^2x}+
\frac{a_1^2\ed^{-k^2x}}{4\sqrt{2\pi\xzero}}
-\frac{a_1^2\ed^{-\frac{k^2x}{2}}}{4\sqrt{2\pi x}}
+\frac{a_1^2\ed^{-k^2(x-\frac{\xzero}{2})}
\big(
1-\ed^{-\frac{k^2\xzero}{2}}
\big)
}{4\sqrt{2\pi\xzero}}~,
\end{equs}
from which we get finally
\begin{equs}
T_6(x,y)={\textstyle
a_1b(\ln(\xzero)-\ln(x))\partial_yK_{\kc}(x,y)+
\frac{a_1^2K_{\kc}(x,y)}{4\sqrt{\pi\xzero}}
+\I\qzero_{\bf a}(x,y)}+
R(x,y)~,
\end{equs}
with
\begin{equs}
\|R(x)\|_{\infty}\leq Ca^2\sqrt{\xzero}
\int_{-\infty}^{\infty}
\hspace{-4mm}{\rm d}k~k^2\ed^{-\frac{k^2x}{2}}
\leq C\|\triplet\|^2\chinese{x}^{-\frac{3}{2}}~.
\end{equs}
We thus have proved that
\begin{equs}
\|{\cal F}_{1,6,u}(x)-u_{{\bf a}_8}(x)\|_{\infty}
\leq\chinese{x}^{-\frac{9}{8}+(1+\epsilon)\olddelta}
\|\triplet\|^2~,
\end{equs}
where ${\bf a}_8=(a_{1,3}+\frac{a_1^2}{4\sqrt{\pi\xzero}},0,0,
a_{4,3}+a_1b\ln(\xzero),0,a_1\Q a_3)$. It then follows by simple comparison
with the result of Proposition \ref{prop:forkdouze} that 
${\bf a}_8={\bf a}_6$ as claimed.
\end{proof}

\section{Estimates on the boundary data}\label{sec:discussion}

In this section, we complete the proof of Theorem \ref{thm:equivalence},
which is
\begin{theorem}\label{thm:equivalencehalf}
If $\xzero$ is sufficiently large and there exist a unique solution to
(\ref{eqn:NavierStokes}) in $\Cu$ with parameters satisfying 
(\ref{eqn:restrictions}), then $\uzero$ and $\omegazero$ are in the class
$\Ci$ with parameters satisfying (\ref{eqn:restrictions}). If furthermore
$\||y|^{\frac{1}{2}}{\bf v}(\xzero)\|_{4}
+\||y|^{\frac{1}{2}-(1+\epsilon)\olddelta}\S{\bf v}(\xzero)\|_{1}\leq
C\|\triplet\|$, then for all $\epsilon>0$, it holds
\begin{equs}
\||y|^{\frac{1}{2}-(1+\epsilon)\olddelta}\S\uzero\|_{1}+
\||y|^{\frac{1}{2}-(1+\epsilon)\olddelta}\S\huzero\|_{1}
&\leq C_1(\xzero,\|\triplet\|)~.
\label{eqn:additionalsymmetricrap}
\end{equs}
\end{theorem}
\begin{proof}
The functions $\uzero$ and $\omegazero$ are determined by the evaluation of
(\ref{eqn:foromega})-(\ref{eqn:forv}) at $x=\xzero$, which gives
\begin{equs}
\Lu\omegazero+\uzero&=u(\xzero)-\Fu(\xzero)
+\LA \qB(\xzero)-\LB \qA(\xzero)
\label{eqn:foromegazero}\\
\Lv\omegazero+\huzero&=v(\xzero)-\Fv(\xzero)
-\LA \qA(\xzero)-\LB \qB(\xzero)
\label{eqn:foruzero}\\
\omegazero&=\omega(\xzero)-\Fomega(\xzero)
~.
\label{eqn:forvzero}
\end{equs}
Denote by $(U,V,W)$ the r.h.s.\ of
(\ref{eqn:foromegazero})-(\ref{eqn:forvzero}). 
By Propositions \ref{prop:localterms}, \ref{prop:thecalfisomega},
\ref{prop:thecalfisv} and \ref{prop:thecalfisu}, $(U,V,W)$ are well defined
and
$\|(U,V,W)\|\leq\|\triplet\|_{\xzero}+C\chinese{\xzero}^{-\kappa}
\|\triplet\|^2$ for $\kappa=\min(\kappa_0,\kappa_2)$. Note that
unsurprisingly (the stationary Navier-Stokes system is elliptic), the
system (\ref{eqn:foromegazero})-(\ref{eqn:forvzero}) is overdetermined.
Nevertheless, since we know that the solution exists, the three relations
have to be satisfied. We now use this as an extra freedom to derive
properties on $\uzero$ and $\omegazero$. We first note that using
Propositions \ref{prop:localterms}, \ref{prop:thecalfisomega},
\ref{prop:thecalfisv} and \ref{prop:thecalfisu}, we get 
\begin{equs}
\|(\Lu\omegazero+\uzero,\Lv\omegazero+\huzero,\omegazero)\|_{\xzero}&\leq
\|\triplet\|_{\xzero}+C\chinese{\xzero}^{-\kappa}\|\triplet\|^2
\label{eqn:firstrhs}
\end{equs}
for some $\kappa>0$, since $(U,V,W)$ satisfies this estimate. In
particular, it implies at once that
\begin{equs}
\|(0,0,\omegazero)\|_{\xzero}\leq
\|\triplet\|_{\xzero}+C\chinese{\xzero}^{-\kappa}\|\triplet\|^2~.
\end{equs}
Then, by interpolation, we have 
\begin{equs}
\chinese{\xzero}^{\frac{1}{2}-\frac{1}{2p}}\|\tilde{\Lu}\omegazero\|_{\L^p}
&\leq
\|\tilde{\Lu}\omegazero\|_{\L^{1}}+
\chinese{\xzero}^{\frac{1}{2}}
\|\tilde{\Lu}\omegazero\|_{\L^{\infty}}\\
\chinese{\xzero}^{1-\frac{1}{2p}-\olddelta}\|\Lv\omegazero\|_{\L^p}
&\leq
\chinese{\xzero}^{\frac{1}{2}-\olddelta}
\|\Lv\omegazero\|_{\L^{1}}+
\chinese{\xzero}^{1-\olddelta}
\|\Lv\omegazero\|_{\L^{\infty}}~,
\end{equs}
where $\tilde{\Lu}=\Lu+\I\Q$. Using these inequalities,
$-\frac{1}{p}\leq-\frac{1}{2p}$ and Lemma \ref{lem:lunforLu}, we get
\begin{equs}
\|(\tilde{\Lu}\omegazero,\Lv\omegazero,0)\|_{\xzero}&
\leq
C_1\|(0,0,\omegazero)\|_{\xzero}~,
\end{equs}
so that from (\ref{eqn:firstrhs}), we get
\begin{equs}
\|(\uzero-\I\Q\omegazero,\huzero,\omegazero)\|_{\xzero}&\leq
(1+C_1)\big(\|\triplet\|_{\xzero}+C\chinese{\xzero}^{-\kappa}\|\triplet\|^2\big)~.
\label{eqn:secondrhs}
\end{equs}
In particular, this implies that $\huzero\in\L^{\pv}\cap\L^{\infty}$ and
$\partial_y\huzero\in\L^{\pdv}$, which gives $\uzero\in\L^{\pv}\cap\L^{\infty}$
using $\uzero=-\H\huzero$ (see Lemma \ref{lem:hilbert} below). Since
$\pu\geq\pv$, we get $\uzero\in\L^{\pu}$, and then (\ref{eqn:secondrhs})
also implies that $\I\Q\omegazero\in\L^{\pu}$ (because $\uzero\in\L^{\pu}$
and $\uzero-\I\Q\omegazero\in\L^{\pu}$). Thus $\I\Q\omegazero$ has to decay
as $|y|\to\infty$, though maybe only in a weak sense. On the other hand,
from the definition of $\I$ (see (\ref{eqn:defI})), we have
$\lim_{y\to\pm\infty}\I\Q\omegazero(y)=\pm\M(\Q\omegazero)$ (the limit
exists since $(1+|y|^{\beta})\omega\in\L^2$ implies $\omegazero\in\L^1$).
This is compatible with $\I\Q\omegazero\in\L^{\pu}$ only if
$\M(\Q\omegazero)$ vanishes. We can thus use Lemma \ref{lem:key} and get
that
\begin{equs}
\|\I\Q\omegazero\|_{\L^1}\leq C
\big(\chinese{\xzero}^{\frac{3}{4}}
\|\omegazero\|_{\L^2}\big)^{1-\frac{3}{2\beta}}
\big(
\big(\chinese{\xzero}^{\frac{3}{4}-\frac{\beta}{2}}
\||y|^{\beta}\omegazero\|_{\L^2}\big)^{\frac{3}{2\beta}}
\leq
C\|(0,0,\omegazero)\|_{\xzero}~.
\end{equs}
Using again Lemma \ref{lem:lunforLu}, we thus get
\begin{equs}
\|(\Lu\omegazero,\Lv\omegazero,0)\|_{\xzero}&
\leq
C_2\|(0,0,\omegazero)\|_{\xzero}~,
\end{equs}
so that again from (\ref{eqn:firstrhs}), we get
\begin{equs}
\|(\uzero,\huzero,\omegazero)\|_{\xzero}&\leq
(1+C_2)\big(\|\triplet\|_{\xzero}+C\chinese{\xzero}^{-\kappa}\|\triplet\|^2\big)~.
\label{eqn:thirdrhs}
\end{equs}
To complete the proof of the first part of Theorem \ref{thm:equivalence},
we still have to prove that (\ref{eqn:additionalsymmetricrap}) holds.
This is done in Proposition \ref{prop:onthesymmetric} below.
\end{proof}

\begin{lemma}\label{lem:hilbert}
Let $p,q>1$. There exist a constant $C_{p,q}$ such that for all
$f$ satisfying $(f,\partial_y f)\in\L^{p}\cap\L^{\infty}\times\L^{q}$, we
have $({\cal H}f,\partial_y\H f)\in\L^{p}\cap\L^{\infty}\times\L^{q}$ and
$\|{\cal H}f\|_{\L^{\infty}}\leq
C_{p,q}(\|f\|_{\L^{p}}+\|\partial_yf\|_{\L^{q}})$.
\end{lemma}
\begin{proof}
Note that ${\cal H}f\in\L^{p}$ and $\partial_y\H f\in\L^{q}$ for 
$1<p,q<\infty$ is a classical result which follows from
Lemma \ref{lem:hormander} (see page \pageref{lem:hormander}). Then, if
$q'\equiv\frac{q}{q-1}\geq p$, the $\L^{\infty}$ estimate follows from
$\|\H f\|_{\L^{\infty}}\leq(\|\H f\|_{q'}\|\partial_y\H
f\|_{q})^{\frac{1}{2}} \leq C(\|f\|_{q'}\|\partial_y
f\|_{q})^{\frac{1}{2}}$. However the $q'\geq p$ restriction is not
essential: using the Cauchy-Schwartz inequality and integration by parts,
we have
\begin{equs}
|{\cal H}f(y)|&=
\left|
\lim_{\epsilon\to0}
\int_{|z|\geq\epsilon}\frac{f(y-z)}{z}{\rm d}z\right|\leq
C\|f\|_{\L^{p}}+
\left|
\lim_{\epsilon\to0}
\int_{\epsilon\leq|z|\leq1}\frac{f(y-z)}{z}{\rm d}z\right|\\
&\leq C\|f\|_{\L^{p}}+
\lim_{\epsilon\to0}
\left|
\ln(\epsilon)
\int_{y-\epsilon}^{y+\epsilon}
\partial_zf(z){\rm d}z
\right|+
\left|
\int_{-1}^1\ln|z|\partial_yf(y-z){\rm d}z
\right|\\
&\leq
C_{p,q}(\|f\|_{\L^{p}}+\|\partial_yf\|_{\L^{q}})+
\|\partial_yf\|_{\L^q}\lim_{\epsilon\to0}(2\epsilon)^{1-\frac{1}{q}}|\ln(\epsilon)|~.
\end{equs}
This completes the proof.
\end{proof}

\begin{proposition}\label{prop:onthesymmetric}
Assume that $\||y|^{\frac{1}{2}}{\bf v}(\xzero)\|_{4}
+\||y|^{\frac{1}{2}-(1+\epsilon)\olddelta}\S{\bf v}(\xzero)\|_{1}\leq
C\|\triplet\|$, then for all $\epsilon>0$, it holds
\begin{equs}
\||y|^{\frac{1}{2}-(1+\epsilon)\olddelta}\S\uzero\|_{1}+
\||y|^{\frac{1}{2}-(1+\epsilon)\olddelta}\S\huzero\|_{1}
&\leq C(\xzero,\|\triplet\|)~.
\end{equs}
\end{proposition}
\begin{proof}   
In this proof, we will use repeatedly that
$\||y|^{a}f\|_{p}\leq\|f\|_p^{1-a}\||y|f\|_p^{a}$ for all $p\geq1$ and
$0\leq a\leq1$, as well as $\||y|^{\frac{1}{2}-(1+\epsilon)\olddelta}f\|_{1}
\leq\||y|^{1-(1+\frac{\epsilon}{2})\olddelta}f\|_{2}$ or
$\||y|^{\frac{1}{2}-(1+\epsilon)\olddelta}f\|_{1}\leq\||y|f\|_{2}$. We first
note that by Lemma \ref{lem:lunforLu} and \ref{lem:key}, (using also that
the symbols $\tilde{\Lu}$ and $\Lv$, together with their derivatives w.r.t.
the Fourier variable `$k$' are bounded), we have
\begin{equs}
\||y|^{\frac{1}{2}-(1+\epsilon)\olddelta}\Lu\omegazero\|_{1}
&\leq
\||y|^{\frac{1}{2}-(1+\epsilon)\olddelta}\I\Q\omegazero\|_{1}+
\||y|\tilde{\Lu}\omegazero\|_{2}\leq
C\|\triplet\|~,\\
\||y|^{\frac{1}{2}-(1+\epsilon)\olddelta}\Lv\omegazero\|_{1}
&\leq
\||y|\Lv\omegazero\|_{2}\leq
C\|\triplet\|~.
\end{equs}
Then we have
\begin{equs}
\||y|^{\frac{1}{2}-\epsilon}\LA \qB\|_{1}&\leq
\|y(\LA-\dirac)\|_2\|\qB\|_{1}
+\big(1+\|(\LA-\dirac)\|_1\big)\|y\qB\|_{2}\\
&\leq
C\chinese{\xzero}^{\olddelta}\|\triplet\|^2+
\||y|^{\frac{1}{2}}u(\xzero)\|_{4}^2
+\||y|^{\frac{1}{2}}v(\xzero)\|_{4}^2
\\
\||y|^{\frac{1}{2}-\epsilon}\LB \qB\|_{1}
&\leq
\|y\LB\|_2\|\qB\|_{1}+\|\LB\|_1\|y\qB\|_{2}
\\
&\leq
C\chinese{\xzero}^{\olddelta}\|\triplet\|^2+
\||y|^{\frac{1}{2}}u(\xzero)\|_{4}^2
+\||y|^{\frac{1}{2}}v(\xzero)\|_{4}^2
\end{equs}
where we used $|\strouhal|^{-1}\leq\chinese{\xzero}^{\olddelta}$.
This shows that $\||y|^{\frac{1}{2}-\epsilon}\LA \qB\|_{1}+
\||y|^{\frac{1}{2}-\epsilon}\LB \qB\|_{1}\leq C(\xzero)\|\triplet\|^2$.
The same holds for $\||y|^{\frac{1}{2}-\epsilon}\LA \qA\|_{1}$ and
$\||y|^{\frac{1}{2}-\epsilon}\LB \qA\|_{1}$.
We then note that 
\begin{equs}
{\cal F}_{2,v}(x)&=
{\cal F}_{2,\omega}(x)+
{\cal F}_{2,1,v}(x)
+{\cal F}_{2,2,v}(x)~,~~~~~
{\cal F}_{2,u}(x)=
{\cal F}_{2,1,u}(x)
+{\cal F}_{2,2,u}(x)~,
\end{equs}
see Propositions \ref{prop:thecalfisv} and \ref{prop:thecalfisu}, or
(\ref{eqn:f21v}), (\ref{eqn:f22v}) and (\ref{eqn:f21u}) and
(\ref{eqn:f22u}) for the definitions of the various terms appearing in this
decomposition. By Proposition \ref{prop:thecalfisomega}, the contribution of
${\cal F}_{2,\omega}$ is bounded by $C\|\triplet\|^2$. Then, there are
exponents $p\geq0$ and $q<1$ such that
\begin{equs}
\|K_{\kdeux}\|_{1,\{p,q\}}+
\|K_{\kcinq}\|_{1,\{p,q\}}+
\|K_{\ksix}\|_{1,\{p,q\}}+
\|K_{\ksept}\|_{1,\{p,q\}}&\leq C~,\\
\||y|K_{\kdeux}\|_{2,\{p,q\}}+
\||y|K_{\kcinq}\|_{2,\{p,q\}}+
\||y|K_{\ksix}\|_{2,\{p,q\}}+
\||y|K_{\ksept}\|_{2,\{p,q\}}&\leq C~.
\end{equs}
Using $\||y|^{\frac{1}{2}-(1+\epsilon)\olddelta}f\|_{1}\leq\||y|f\|_{2}$,
this shows that the contributions of ${\cal F}_{2,1,v}$ and ${\cal
F}_{2,1,u}$ is also bounded by $C\|\triplet\|^2$. For the contribution
of ${\cal F}_{2,2,v}$ and ${\cal F}_{2,2,u}$, we note that
\begin{equs}
\|\S\galpha^{*}(\tilde{x}-x)\qun(\tilde{x})\|_{2}&\leq
C|\tilde{x}-x|^{-\frac{1}{2}}\chinese{\tilde{x}}^{-\frac{3}{2}+\olddelta}
\|\triplet\|^2~,\\
\|y\big(\P\galpha^{*}(\tilde{x}-x)\qun(\tilde{x})\big)\|_{2}&\leq
C\big(
|\tilde{x}-x|^{\frac{1}{2}}\chinese{\tilde{x}}^{-\frac{3}{2}+\olddelta}
+
\chinese{\tilde{x}}^{-\frac{5}{4}+\olddelta}
\big)
\|\triplet\|^2~,
\end{equs}
while
\begin{equs}
\|y\S\Q\galpha^{*}(\tilde{x}-x)\qun(\tilde{x})\|_{2}&\leq
\Big(
\int_{-\infty}^{\infty}
\hspace{-4mm}
{\rm d}k~
\Big(
\partial_k
\ed^{-|k||\tilde{x}-x|}
\Big)^2
|\qun(\tilde{x},k)|^2\Big)^{\frac{1}{2}}\\
&\phantom{\leq~}
+\Big(\int_{-\infty}^{\infty}
\hspace{-4mm}
{\rm d}k~\ed^{-|k||\tilde{x}-x|}
|\partial_k\big(i\sigma(\qun(\tilde{x},k)-\qun(\tilde{x},-k)\big)|^2
\Big)^{\frac{1}{2}}\\
&\leq C\big(
|\tilde{x}-x|^{\frac{1}{2}}\chinese{\tilde{x}}^{-\frac{3}{2}+\olddelta}
+
\chinese{\tilde{x}}^{-\frac{5}{4}+\olddelta}
\big)
\|\triplet\|^2~,
\end{equs}
where we used that $|\qun(\tilde{x},k)-\qun(\tilde{x},-k)|\leq
|k|^{\frac{1}{2}-\epsilon}\||y|^{\frac{1}{2}-\epsilon}\qun(\tilde{x})\|_1\leq
|k|^{\frac{1}{2}-\epsilon}\||y|\qun(\tilde{x})\|_2$, so that the
coefficient of the Dirac measure appearing when differentiating $\sigma$
w.r.t.\ $k$ in the above expression vanishes. This implies finally that
\begin{equs}
\||y|^{\frac{1}{2}-(1+\epsilon)\olddelta}
\S\galpha^{*}(\tilde{x}-x)\qun(\tilde{x})\|_{1}&\leq
\|\S\galpha^{*}(\tilde{x}-x)\qun(\tilde{x})\|_{2}^{(1+\frac{\epsilon}{2})\olddelta}
\|y\S\galpha^{*}(\tilde{x}-x)\qun(\tilde{x})\|_{2}^{1-(1+\frac{\epsilon}{2})\olddelta}\\
&\leq
C
|\tilde{x}-x|^{\frac{1}{2}-(1+\frac{\epsilon}{2})\olddelta}
\chinese{\tilde{x}}^{-\frac{3}{2}+\olddelta}\|\triplet\|^2\\
&\phantom{\leq~}+
C|\tilde{x}-x|^{-\frac{1}{2}(1+\frac{\epsilon}{2})\olddelta}
\chinese{\tilde{x}}^{-\frac{5}{4}+\frac{6-\epsilon}{8}\olddelta}
\|\triplet\|^2~.
\label{eqn:unautretrouc}
\end{equs}
The same estimate holds for $\||y|^{\frac{1}{2}-(1+\epsilon)\olddelta}
\S\falpha^{*}(\tilde{x}-x)\qun(\tilde{x})\|_{1}$.
Since $\epsilon>0$, integrating (\ref{eqn:unautretrouc}) from
$\tilde{x}=\xzero$ to $\tilde{x}=\infty$ completes the proof.
\end{proof}

\section{Checking the applicability to the usual exterior
problem}\label{app:checkingwithgaldi}

In this section, we prove the Proposition \ref{prop:fromphysically}.
We will use the notation $r=\sqrt{x^2+y^2}$. From
\cite{babenkovortex,cla70,Gal94}, we get that any "Physically Reasonable"
(PR) solution solution satisfies the estimates
\begin{equs}
|u(x,y)|&\leq C
\left\{
\begin{array}{ll}
r^{-\frac{1}{2}}&\mbox{ if }r\geq C\\
r^{-\min(\frac{1+\sigma}{2},1-\epsilon)}&\mbox{ if }
1-\cos(\phi)\geq
r^{-1+\sigma}
\end{array}\right.\\
|v(x,y)|&\leq C
r^{-1}\ln(r)
~,~~~~~
|\partial_y u(x,y)|\leq C
r^{-1}\ln(r)^2
~,~~~~~
|\partial_y v(x,y)|\leq C
r^{-\frac{3}{2}}\ln(r)^2\\
\omega(x,y)&=c_1\partial_x
\big(\ed^{\frac{x}{2}}{\rm K}_0(r)\big)+
c_2\partial_y
\big(\ed^{\frac{x}{2}}{\rm K}_0(r)\big)+
{\cal O}\Big(
\ed^{\frac{x-r}{4}}
r^{-\frac{3}{2}}\ln(r)^2
\Big)~,\\
\partial_y\omega(x,y)&=c_1\partial_y\partial_x
\big(\ed^{\frac{x}{2}}{\rm K}_0(r)\big)+
c_2\partial_y^2
\big(\ed^{\frac{x}{2}}{\rm K}_0(r)\big)+
{\cal O}\Big(
\ed^{\frac{x-r}{4}}
r^{-2}\ln(r)^2
\Big)~,
\end{equs}
where $\epsilon$ is arbitrarily small, $\sigma\in[0,1]$,
$\tan(\phi)=\frac{y}{x}$, $c_1$ and $c_2$ are constants and $K_0$ is the
modified Bessel function of the second type of order zero. From this, we
get immediately $\|\triplet\|\leq C$ if $\xzero$ is sufficiently large
and $\pdv>(2\min(\au,\av))^{-1}$ (using also $\ln(x)\leq
C\chinese{x}^{\olddelta}$). Namely, for the estimates of the velocity fields
$u$ and $v$, the only difficulty is to prove that
$\|u\|_{\pu,\frac{1}{2}-\frac{1}{\pu}}\leq C$. This follows since for
$\sigma=\frac{1}{\pu}$, $\epsilon=\frac{1}{2}-\frac{1}{2\pu}$ and $\xzero$
sufficiently large, we have
\begin{equs}
|u(x,y)|&\leq C
\left\{
\begin{array}{ll}
r^{-\frac{1}{2}}&\mbox{ if }x\geq\xzero\mbox{ and }|y|< cx\\
r^{-\frac{1}{2}(1+\frac{1}{\pu})}
&\mbox{ if }x\geq\xzero\mbox{ and }|y|\geq cx
\end{array}\right.~,
\end{equs}
which gives
\begin{equs}
\|u\|_{\pu,\frac{1}{2}-\frac{1}{\pu}}\leq C
\Big(
\frac{\chinese{x}}{x}
\Big)^{\frac{1}{2}-\frac{1}{\pu}}
\left(
\Big(
\int_{-c}^{c}
\frac{{\rm d}y}{(1+y^2)^{\frac{q}{4}}}
\Big)^{\frac{1}{\pu}}
+
\frac{2}{x^{\frac{1}{2q}}}
\Big(
\int_{c}^{\infty}
\hspace{-3.5mm}
\frac{{\rm d}y}{(1+y^2)^{\frac{1+q}{4}}}
\Big)^{\frac{1}{\pu}}
\right)~.
\end{equs}
For the estimates on the vorticity, it follows, using that
$|z|^{p}\ed^{-z}\leq C_p$ for all $p\geq0$ and the asymptotic
development of $K_0$, that for $x\geq\xzero$ sufficiently large we have
\begin{equs}
|\omega(x,y)|&\leq C
\ed^{\frac{x}{4}-\frac{r}{4}}r^{-\frac{3}{2}}
\big(
|y|+\ln(x)^2
\big)~,
~~~~~~
|\partial_y\omega(x,y)|\leq C
\Big(\ed^{\frac{x}{4}-\frac{r}{4}}r^{-\frac{3}{2}}
\Big)~.
\end{equs}
This shows at once that $\|\partial_y\omega\|_{\infty,\frac{3}{2}}\leq C$.
Then, for all $\alpha\geq0$, after the change of
variable $y=\sqrt{2xz+z^2}$ and using again that $|z|^{p}\ed^{-z}\leq C_p$,
we get that
\begin{equs}
\||y|^{\alpha}\omega\|_{\L^2}
&\leq C\Big(
\int_{0}^{\infty}
\hspace{-3mm}
{\rm d}z~\frac{\ed^{-\frac{z}{2}}}{\sqrt{z}}
\frac{
\big(
\ln(x)^2+\sqrt{z}\sqrt{2x+z}
\big)^2
(z(2x+z))^{\alpha}
}{(x+z)^2\sqrt{2x+z}}
\Big)^{\frac{1}{2}}\\
&\leq
Cx^{-\frac{3}{4}+\frac{\alpha}{2}}\Big(
\int_{0}^{\infty}
\hspace{-3mm}
{\rm d}z~\frac{\ed^{-\frac{z}{4}}}{\sqrt{z}}
\Big)^{\frac{1}{2}}
\leq C \chinese{x}^{-\frac{3}{4}+\frac{\alpha}{2}}~,
\label{eqn:withalpha}
\\
\|\partial_y\omega\|_{\L^1}
&\leq C
\int_{0}^{\infty}
\hspace{-2mm}
\frac{{\rm d}z~\ed^{-\frac{z}{4}}}{\sqrt{z}\sqrt{x+z}\sqrt{2x+z}}
\leq
Cx^{-1}
\int_{0}^{\infty}
\hspace{-3mm}
{\rm d}z~\frac{\ed^{-\frac{z}{4}}}{\sqrt{z}}
\leq C \chinese{x}^{-1}~.
\end{equs}
Using the estimate (\ref{eqn:withalpha}) with $\alpha=0$ and
$\alpha=\beta$ achieves the proof of $\|(0,0,\omega)\|\leq C$. We then
note that for $|y|\geq cx\geq c\xzero$ with $\xzero$, we have for all $q>1$
\begin{equs}
|u(x,y)|+|v(x,y)|&\leq C r^{-\frac{1}{2}(1+\frac{1}{q})}
\label{eqn:foryldeux}~,
\end{equs}
from which we deduce that
$\||y|^{\frac{1}{2}}u(x)\|_{4}+\||y|^{\frac{1}{2}}v(x)\|_{4}\leq C$. Finally, it
follows from e.g.\ \cite{Gal94}, section X.6, that there exist constants
${\bf m}=(m_1,m_2)$ such that for all $|y|\geq c x\geq c\xzero$, we  have
\begin{equs}
|u(x,y)-u_{{\bf m}}(x,y)|+
|v(x,y)-v_{{\bf m}}(x,y)|&\leq C r^{-1}
\label{eqn:useuvadefgaldi}~,
\end{equs}
where $u_{{\bf m}}$ and $v_{{\bf m}}$ are defined in terms of Oseen's
tensor ${\bf E}$ by
\begin{equs}
\Big(\begin{matrix}
u_{{\bf m}}(x,y)\\
v_{{\bf m}}(x,y)
\end{matrix}\Big)
&={\bf m}\cdot{\bf E}(x,y)~.
\label{eqn:vadefgaldi}
\end{equs}
It then follows from (\ref{eqn:useuvadefgaldi}), (\ref{eqn:vadefgaldi})
and the explicit form of Oseen's tensor that
\begin{equs}
\||y|^{\frac{1}{2}-(1+\epsilon)\olddelta}\S u(x)\|_{1}
+\||y|^{\frac{1}{2}-(1+\epsilon)\olddelta}\S v(x)\|_{1}\leq C~,
\end{equs}
where $(\S f)(y)\equiv f(y)+f(-y)$.

\section*{Acknowledgements}
This work was completed during the stay of the author at the {\em Institut
Fourier} in Grenoble and at the {\em Fields Institute} in Toronto. The
author greatly benefited from the hospitality of these institute and wishes
in particular to express his gratitude to Jerry L. Bona, Walter Craig,
Thierry Gallay, Jean-Claude Saut and Peter Wittwer for helpful discussions.

\appendix

\newappendix{Kernels estimates}\label{app:kernelestimates}

\subsection{Definitions and preliminaries}
This section is devoted to estimates of all the kernels in
various $\L^p$ and Sobolev spaces of the `$y$' variable. Note that the
Kernels are most conveniently expressed in terms of their Fourier
transform, and though it is sometimes possible to calculate explicitly the
inverse Fourier transform of the kernels, we will estimate the norms in
Fourier space as often as possible. To do so, we will use the following
Lemma which relates the $\L^1$ norm in direct space to the ${\rm H}^1$ in
Fourier space, and the $\L^2$ norm with weight $|y|^{\beta}$ for
non-integer $\beta$ to integer ones.
\begin{lemma}\label{lem:Lun}
Let $\beta>\frac{1}{2}$. There exist a constant $C_{\beta}$ such that for all $f$
with $\|(1+|y|^{\beta})f\|_{\L^2}<\infty$, we have
\begin{equs}
\|f\|_{\L^1}&\leq
\Big\{
\begin{array}{l}
C_{\beta}\|f\|_{\L^2}^{1-\frac{1}{2\beta}}
\||y|^{\beta}f\|_{\L^2}^{\frac{1}{2\beta}}~,\\[1mm]
C\sqrt{\|f\|_{\L^2}\|yf\|_{\L^2}}
\leq C\sqrt{\|\hat{f}\|_{\L^{2}}\|\hat{f}'\|_{\L^{2}}}~,
\end{array}
\end{equs}
where $\hat{f}$ denote the (continuous) Fourier transform of $f$. Then, 
for all $s_1\in[0,3]$ and $s_2\in[0,2]$, we have
\begin{equs}
\||y|^{s_1}f\|_{\L^2}&\leq
\|f\|_{\L^2}^{1-\frac{s_1}{3}}
\||y|^3f\|_{\L^2}^{\frac{s_1}{3}}~,~~
\||y|^{1+s_2}f\|_{\L^2}\leq
\||y|f\|_{\L^2}^{1-\frac{s_2}{2}}
\||y|^3f\|_{\L^2}^{\frac{s_2}{2}}~.
\end{equs}
\end{lemma}

\begin{proof}
Let $a>0$, then
\begin{equs}
\|f\|_{\L^1}\leq \|(a+|y|^{\beta})f\|_{\L^2}\|(a+|y|^{\beta})^{-1}\|_{\L^2}
\leq C_{\beta}\Big(
a^{\frac{1}{2\beta}}\|f\|_{\L^2}+
a^{\frac{1}{2\beta}-1}\||y|^{\beta}f\|_{\L^2}
\Big)
\end{equs}
for some finite $C_{\beta}$. Setting $a=\||y|^{\beta}f\|_{\L^2}/\|f\|_{\L^2}$
completes the proof of the first inequality. The second one follows from
Plancherel's inequality, while the last two follow trivially from 
Young's inequality.
\end{proof}
We then introduce the functions
\begin{equs}
B_{\mu,\olddelta}(x,n\strouhal)=
\int_{-\infty}^{\infty}\hspace{-4mm}{\rm d}k~
|k|^{\olddelta}
{\textstyle\big|\frac{k}{\Lambdazero}\big|^{2\mu}}~
\ed^{2~\Re(\Lambdam)x}~,~~~~
B_{\olddelta}(x,n\strouhal)=
\int_{-\infty}^{\infty}\hspace{-4mm}{\rm d}k~
{\textstyle\big|\frac{k}{\Lambdazero}\big|^{2\olddelta}
\frac{1}{|\Lambdazero|^2}}~\ed^{2~\Re(\Lambdam)x}~.
\end{equs}
through which most estimates on the kernels can be easily obtained, and
which satisfy the
\begin{lemma}\label{lem:L2}
Let $\mu\geq\frac{1}{2}$. Then for all $\olddelta\geq0$, there exist a
constant $C_{\olddelta}$ such that for all $1\leq\xi_1\leq\mu+\frac{1}{2}$ 
we have
\begin{equs}
B_{0,\olddelta}(x,n\strouhal)
&\leq C_{\olddelta}
\frac{\ed^{b(n\strouhal)x}\chinese{x}^{\frac{\olddelta+1}{2}}}{x^{\olddelta+1}}
~,~~~~
B_{\mu,\olddelta}(x,n\strouhal)
\leq C_{\olddelta}\frac{\ed^{b(n\strouhal)x}\chinese{x}^{\frac{\olddelta}{2}}}{x^{\xi_1+\olddelta}}
~,~~~~
B_{\olddelta}(x,n\strouhal)\leq
C_{\olddelta}\frac{\ed^{b(n\strouhal)x}}{\chinese{x}^{\frac{1}{2}+\olddelta}}
\end{equs}
for all $x\geq0$ and $n\strouhal\in{\bf R}$.
\end{lemma}
\begin{proof}
We first have
\begin{equs}
B_{0,\olddelta}(x,n\strouhal)&\leq
C\ed^{2b(n\strouhal)x}\left(\int_{|k|>1}\hspace{-6mm}{\rm
d}k~|k|^{\olddelta}\ed^{-|k|x}+
\int_{|k|\leq1}\hspace{-6mm}{\rm d}k~
|k|^\olddelta\ed^{-2c(n\strouhal)xk^2}\right)
~,\\
&\leq
{\textstyle
C\frac{\ed^{b(n\strouhal)x}}{x^{\olddelta+1}}\left(
1+\big(\olddelta+1\big)^{\frac{\olddelta+1}{2}}\left(
\frac{c(n\strouhal)^{-1}x\ed^{\frac{b(n\strouhal)x}{\olddelta+1}}}{\olddelta+1}
\right)^{\frac{\olddelta+1}{2}}
\right)}\leq
C\frac{\ed^{b(n\strouhal)x}\chinese{x}^{\frac{\olddelta+1}{2}}}{x^{\olddelta+1}}~,
\end{equs}
because $c(n\strouhal)^{-1}\zeta\ed^{b(n\strouhal)\zeta}\leq C(1+\zeta)$
for all $\zeta\geq0$. Then, we note that since
$\big|\frac{k}{\Lambdazero}\big|$ is
uniformly bounded in $k$ and $n\strouhal$, we trivially have
$B_{\mu,\olddelta}(x,n\strouhal)\leq C_{\mu}B_{0,\olddelta}(x,n\strouhal)$ for
all $\mu\geq0$. To get the more precise bound of the Lemma in the case
$\mu\geq\frac{1}{2}$, we use that $\big|\frac{k}{\Lambdazero}\big|\leq C$
and that by hypothesis on $\xi_1$, we have 
$0\leq\xi_1-1\leq2\xi_1-1\leq2\mu$, hence
\begin{equs}
B_{\mu,\olddelta}(x,n\strouhal)&\leq C
\ed^{2b(n\strouhal)x}\left(
\int_{|k|>1}\hspace{-6mm}{\rm d}k~
{\textstyle
|k|^{\olddelta+\xi_1-1}\ed^{-|k|x}}+
\int_{|k|\leq1}\hspace{-6mm}{\rm d}k~
{\textstyle
\frac{|k|^{\olddelta+2\xi_1-1}\ed^{-2c(n\strouhal)xk^2}}{(1+(n\strouhal)^2)^{
\frac{\mu}{2}}}}\right)
\\
&\leq
\frac{C}{x^{\xi_1+\olddelta}}
\Big({\textstyle
\ed^{2b(n\strouhal)x}+
\frac{(c(n\strouhal)^{-1}x)^{\frac{\olddelta}{2}}\ed^{2b(n\strouhal)x}
}{
c(n\strouhal)^{\xi_1}(1+(n\strouhal)^2)^{\frac{\mu}{2}}
}}
\Big)~.
\end{equs}
Since $c(n\strouhal)^{-\mu-\frac{1}{2}}(1+(n\strouhal)^2)^{-\frac{\mu}{2}}\leq C$ by
hypothesis on $\mu$ and $\xi_1$, this completes the proof of the second
inequality if $\olddelta=0$. If $\olddelta>0$, we use
$c(n\strouhal)^{-1}\zeta\ed^{b(n\strouhal)\zeta}\leq C(1+\zeta)$, so that
\begin{equs}
B_{\mu,\olddelta}(x,n\strouhal)
&\leq C\frac{\ed^{b(n\strouhal)x}}{x^{\xi_1+\olddelta}}
\Big({\textstyle
1+\big(\frac{\olddelta}{2}\big)^{\frac{\olddelta}{2}}\big(
c(n\strouhal)^{-1}2x\olddelta^{-1}\ed^{b(n\strouhal)2x\olddelta^{-1}}
\big)^{\frac{\olddelta}{2}}}
\Big)
\leq C\frac{\ed^{b(n\strouhal)x}\chinese{x}^{\frac{\olddelta}{2}}}{x^{\xi_1+\olddelta}}
~.
\end{equs}
For the last inequality, we first note that 
$B_{\olddelta}(x,n\strouhal)\leq C_{\olddelta}B_{0}(0,n\strouhal)$
(this follows again because $\big|\frac{k}{\Lambdazero}\big|$ is uniformly bounded). Then we have
$B_{0}(0,n\strouhal)\leq C$, so
we only have to show that $B_{\olddelta}(x,n\strouhal)$ decays at least like
$\ed^{b(n\strouhal)x}x^{-\frac{1}{2}-\olddelta}$ as
$x\to\infty$, and this follows since
\begin{equs}
B_{\olddelta}(x,n\strouhal)&\leq
C\ed^{2b(n\strouhal)x}\Big(\int_{|k|\leq1}\hspace{-6mm}{\rm d}k~
{\textstyle
\frac{|k|^{2\olddelta}\ed^{-2c(n\strouhal)xk^2}}{
(1+(n\strouhal)^2)^{\frac{1+\olddelta}{2}}
}}
+
\int_{|k|>1}\hspace{-6mm}{\rm d}k~
{\textstyle
\frac{|k|^{2\olddelta}\ed^{-|k|x}}{1+k^2}}\Big)\\
&\leq
\frac{C\ed^{b(n\strouhal)x}}{x^{\frac{1}{2}+\olddelta}}\left(
\big(c(n\strouhal)^{\frac{1}{2}+\olddelta}(1+(n\strouhal)^2)^{\frac{1+\olddelta}{2}}\big)^
{-1}+
x^{-\frac{1}{2}-\olddelta}\right)~.
\end{equs}
This completes the proof.
\end{proof}
Note that in the bound on $B_{\mu,\olddelta}(x)$ in Lemma \ref{lem:L2}, the
best decay rate as $x\to\infty$ improves as $\mu$ grows. The `free'
parameter $\xi_1$ gives a way to limit the growth of the divergence rate as
$x\to0$.

\subsection{Actual estimates}

We begin this section by an easy estimate on $\LA$ and $\LB$:
\begin{lemma}\label{lem:onLALB}
Let $\hat{\LA}=\frac{k^2}{k^2+(n\strouhal)^2}$ and
$\hat{\LB}=\frac{|k|n\strouhal}{k^2+(n\strouhal)^2}$, then
\begin{equs}
\|\LA-\dirac\|_{1,\{0,0\}}+\|\LB\|_{1,\{0,0\}}\leq C~.
\end{equs}
In particular, $\LA$ and $\LB$ are $\L^p\to\L^p$ bounded operators for all
$p\in[1,\infty)$.
\end{lemma}

\begin{proof}
The proof follows immediately since using Fourier transform, we get that
for fixed $n$, it holds
$\|\hat{\LA}-1\|_{\L^2}+\|\hat{\LB}\|_{\L^2}\leq C|n\strouhal|$ and
$\|\partial_k(\hat{\LA}-1)\|_{\L^2}+\|\partial_k\hat{\LB}\|_{\L^2} \leq
C|n\strouhal|^{-1}$.
\end{proof}

\begin{lemma}\label{lem:sourcelikeesti}
For all $p>1$, $q\geq2$ and $m\in{\bf N}$, there exists a constant $C>0$
such that
\begin{equs}
\|\Q\falpha\|_{\K{p}{0}{1-\frac{1}{p}}}+
\|\Q\galpha\|_{\K{p}{0}{1-\frac{1}{p}}}&\leq C\\
\|\partial_y^m\falpha\|_{\K{q}{0}{1+m-\frac{1}{q}}}+
\|\partial_y^m\galpha\|_{\K{q}{0}{1+m-\frac{1}{q}}}
&\leq C\\
\|\chinese{\strouhal x}\P\partial_y^m\falpha\|_{\K{q}{0}{1+m-\frac{1}{q}}}+
\|\chinese{\strouhal x}\P\partial_y^m\galpha\|_{\K{q}{0}{1+m-\frac{1}{q}}}
&\leq C\\
\|\P\falpha\|_{\K{1}{0}{\frac{1}{4}}}+
\|\P\galpha\|_{\K{1}{0}{\frac{1}{4}}}&\leq
C|\strouhal|^{-\frac{1}{4}}
~.
\end{equs}
\end{lemma}
\begin{proof}
After the change of variables $k=\xi/x$, we get
\begin{equs}
\|\chinese{x\strouhal}\partial_y^m\falpha\|_{\K{q}{0}{1+m-\frac{1}{q}}}
&\leq
\sup_{x\geq0}
\sup_{n\in{\bf Z}}
\left(
\int_{-\infty}^{\infty}
\hspace{-4mm}{\rm d}\xi~
\Big(\frac{\xi^{2}(1+(\strouhal x)^2)}{\xi^2+(n\strouhal x)^2}\Big)^{\frac{q}{2(q-1)}}
|\xi|^{\frac{qm}{q-1}}
\ed^{-\frac{q|\xi|}{q-1}}
\right)^{\frac{q-1}{q}}\\
&\leq
\left[
\int_{|\xi|\leq1}\hspace{-5.5mm}{\rm d}\xi~|\xi|^{\frac{qm}{q-1}}
\ed^{-\frac{q|\xi|}{q-1}}+
\int_{|\xi|\geq1}\hspace{-5.5mm}{\rm d}\xi~|\xi|^{\frac{q(1+m)}{q-1}}
\ed^{-\frac{q|\xi|}{q-1}}
\right]^{1/q}\leq C
~,\\
\|\partial_y^m\falpha\|_{\K{q}{0}{1+m-\frac{1}{q}}}
&\leq
\sup_{x\geq0}
\sup_{n\in{\bf Z}}
\left(
\int_{-\infty}^{\infty}
\hspace{-4mm}{\rm d}\xi~
|\xi|^{\frac{qm}{q-1}}
\ed^{-\frac{q|\xi|}{q-1}}
\right)^{\frac{q-1}{q}}\leq C~,
\end{equs}
for any $m\in{\bf N}$ and $q\geq2$. The same holds for $\galpha$. We next
note that $\galpha=-i\sigma\falpha$, so that
\begin{equs}
\partial_k\galpha(x,k)&=-i\delta(k)\falpha(x,k)
-i\sigma\partial_k\falpha(x,k)\\
&=\frac{-i\delta(k)}{1-\frac{in\strouhal}{|k|}}
-i\sigma\partial_k\falpha(x,k)
=
-i\delta_{n,0}
-i\sigma\partial_k\falpha(x,k)~,
\label{eqn:deriv}
\end{equs}
where $\delta_{n,0}=1$ if $n=0$ and $\delta_{n,0}=0$ if 
$n\neq0$. We thus have
$\partial_k\Q\galpha(x,k)\notin\L^{2}$, so that we cannot use
Lemma \ref{lem:Lun} to bound $\|\Q\galpha(x)\|_{\L^{1}}$. In fact, 
$\Q\falpha$ and $\Q\galpha$ can be explicitly computed, giving
$\Q\falpha(x,y)=\frac{1}{\pi}\frac{x}{x^2+y^2}$ and
$\Q\galpha(x,y)=\frac{1}{\pi}\frac{y}{x^2+y^2}$. This
shows that $\Q\galpha(x,y)\notin\L^{1}$, and gives an easy way to prove the
estimate on $\|\Q\falpha\|_{\K{p}{0}{1-\frac{1}{p}}}+\|\Q\galpha\|_{\K{p}{0}{1-\frac{1}{p}}}$ for
$p>1$ in direct space. On the other hand, (\ref{eqn:deriv}) shows that 
$\|\P\partial_k\galpha(x)\|_{\L^{2}}=
 \|\P\partial_k\falpha(x)\|_{\L^{2}}$, and we have
\begin{equs}
\|\partial_k\P\falpha\|_{\K{2}{0}{0}}
+\|\partial_k\P\galpha\|_{\K{2}{0}{0}}
&=
\sup_{x\geq0}
\sqrt{x}
\sup_{n\in{\bf Z},n\neq0}
\left(
\int_{-\infty}^{\infty}
\hspace{-4mm}{\rm d}\xi~
{\textstyle
\frac{\ed^{-2|\xi|}
\left(
\xi^4+(n\strouhal x)^2(1-|\xi|)^2
\right)}{(\xi^2+(n\strouhal x)^2)^2}}\right)^{1/2}\\
&\leq
\sup_{x\geq0}
\frac{\sqrt{x}}{2}
\sup_{n\in{\bf Z},n\neq0}
\left(
\int_{|\xi|\leq1}
\hspace{-5.5mm}{\rm d}\xi~
{\textstyle\frac{1}{\xi^2+(n\strouhal x)^2}}
+
{\textstyle\frac{C}{\chinese{n\strouhal x}^2}}
\int_{|\xi|\geq1}
\hspace{-5.5mm}{\rm d}\xi~\ed^{-|\xi|}
\right)^{1/2}
\\&
\leq
\frac{C}{\sqrt{|\strouhal|}}~.
\end{equs}
Using Lemma \ref{lem:Lun}, this proves the estimates on
$\|\P\falpha\|_{\K{1}{0}{\frac{1}{4}}}+\|\P\galpha\|_{\K{1}{0}{\frac{1}{4}}
}$ and completes the proof.
\end{proof}

\begin{lemma}\label{lem:kernelun}
There exist a constant $C>0$ such that for all $1\leq\beta\leq3$, it holds
\begin{equs}
\|K_{\kun}\|_{\K{1}{0}{0}}+
\|K_{\kun}(x)\|_{\K{\infty}{\frac{1}{2}}{1}}+
\||y|^{\beta}K_{\kun}\|_{\K{2}{-\frac{1}{4}+\frac{\beta}{2}}{0}}
\leq C~,\\
\|\partial_yK_{\kun}\|_{\K{1}{\frac{1}{2}}{1}}+
\|\partial_yK_{\kun}\|_{\K{\infty}{1}{2}}+
\||y|^{\beta}\partial_yK_{\kun}(x)\|_{\K{2}{-\frac{3}{4}+\frac{\beta}{2}}{0}}
\leq C~,\\
\|\partial_y^2K_{\kun}\|_{\K{\infty}{\frac{3}{2}}{3}}+
\|\partial_y^2K_{\kun}\|_{\K{1}{1}{2}}
\leq C~,\\
\|K_{\kdeux}\|_{\K{1}{0}{\frac{1}{2}}}+
\|K_{\kdeux}\|_{\K{\infty}{0}{1}}+
\||y|^{\beta}K_{\kdeux}\|_{\K{2}{-\frac{3}{4}+\frac{\beta}{2}}{0}}
\leq C~,\\
\|\partial_yK_{\kdeux}\|_{\K{\infty}{\frac{1}{2}}{2}}+
\|\partial_yK_{\kdeux}\|_{\K{1}{\frac{1}{2}}{\frac{3}{2}}}
\leq
C~,\\
\|K_{\kcinq}\|_{\K{1}{0}{\frac{1}{2}}}+
\|K_{\ksix}\|_{\K{1}{0}{\frac{1}{2}}}+
\|K_{\ksept}\|_{\K{1}{\frac{1}{4}}{\frac{1}{4}}}
\leq C~,\\
\||y|K_{\kcinq}\|_{\K{2}{0}{\frac{1}{4}}}+
\||y|K_{\ksix}\|_{\K{2}{0}{\frac{1}{4}}}+
\||y|K_{\ksept}\|_{\K{2}{\frac{1}{2}}{\frac{1}{4}}}
\leq C~.
\end{equs}
The same estimates hold with $K_{n}$ replaced by 
$\ed^{-\frac{b(\strouhal)x}{4}}\P K_{n}$ for $n=1,2,5,6,7$.
\end{lemma}

\begin{proof}
We have $\left|\partial_k\ed^{\Lambdam x}\right|\leq
\frac{|k|x\ed^{{\rm Re}(\Lambdam)x}}{|\Lambdazero|}$ and
\begin{equs}
\left|
\partial_k^3
\ed^{\Lambdam x}
\right|
&\leq
\ed^{{\rm Re}(\Lambdam)x}
{\textstyle\left(
 \frac{x^3|k|^3}{|\Lambdazero|^3}
+\frac{x^2|k|}{|\Lambdazero|^2}
+\frac{x|k|}{|\Lambdazero|^3}
\right)}\\
\left|
\partial_k^3
(k\ed^{\Lambdam x})
\right|
&\leq
\ed^{{\rm Re}(\Lambdam)x}
{\textstyle\left(
 \frac{x^3|k|^4}{|\Lambdazero|^3}
+\frac{x^2|k|^2}{|\Lambdazero|^2}
+\frac{x}{|\Lambdazero|}
\right)}\\
\left|
\partial_k\big(
{\textstyle\frac{k}{\Lambdazero}}\ed^{\Lambdam x}
\big)
\right|
&\leq
\ed^{{\rm Re}(\Lambdam)x}
{\textstyle\left(
\frac{c_1}{|\Lambdazero|}
+\frac{c_{2}k^2x}{|\Lambdazero|^2}
\right)}~,\\
\left|
\partial_k^3\big(
{\textstyle\frac{k}{\Lambdazero}}\ed^{\Lambdam x}
\big)
\right|
&\leq
\ed^{{\rm Re}(\Lambdam)x}
{\textstyle\left(
 \frac{x^3|k|^4}{|\Lambdazero|^4}
+\frac{x^2|k|^2}{|\Lambdazero|^3}
+\frac{x}{|\Lambdazero|^2}
+\frac{1}{|\Lambdazero|^3}
\right)}~,
\\
\left|
\partial_k\big(
{\textstyle\frac{k^2}{\Lambdazero}}\ed^{\Lambdam x}
\big)
\right|
&\leq
|k|\ed^{{\rm Re}(\Lambdam)x}
{\textstyle\left(
\frac{c_1}{|\Lambdazero|}
+\frac{c_{2}k^2x}{|\Lambdazero|^2}
\right)}~.
\end{equs}
Similarly, we have
\begin{equs}
\left|K_{\kcinq}(x,k)\right|+\left|K_{\ksix}(x,k)\right|
&\leq\frac{|k|\ed^{\Re(\Lambdam)x}}{|\Lambdazero|}~,\\
 \left|\partial_kK_{\kcinq}(x,k)\right|
+\left|\partial_kK_{\ksix}(x,k)\right|
&\leq\ed^{\Re(\Lambdam)x}
{\textstyle\left(
\frac{c_3}{|\Lambdazero|}
+\frac{c_4k^2x}{|\Lambdazero|^2}
\right)}~,\\
\left|K_{\ksept}(x,k)\right|&\leq
C\ed^{\Re(\Lambdam)x}~,\\
\left|\partial_{k}K_{\ksept}(x,k)\right|&\leq
C\ed^{\Re(\Lambdam)x}
{\textstyle\left(
\frac{1}{|\Lambdazero|}
+\frac{|k|x}{|\Lambdazero|}
\right)}~.
\end{equs}
Finally, we note that for fixed $x$ and $n$, we have
\begin{equs}
\|\partial_y K_1\|_{\L^{1}}&\leq C\left(
\|kK_{\kun}\|_{\L^{2}}^2
\left(
\|K_{\kun}\|_{\L^{2}}^2+
\|k\partial_kK_{\kun}\|_{\L^{2}}^2
\right)
\right)^{\frac{1}{4}}\\
&\leq C\left(
\|\partial_y K_{\kun}\|_{\L^{2}}^2
\left(
\|K_{\kun}\|_{\L^{2}}^2+
x^2\|\partial_yK_{\kdeux}\|_{\L^{2}}^2
\right)
\right)^{\frac{1}{4}}~,\\
\|\partial^2_y K_1\|_{\L^{1}}&\leq C\left(
\|k^2K_{\kun}\|_{\L^{2}}^2
\left(
2\|kK_{\kun}\|_{\L^{2}}^2+
\|k^2\partial_kK_{\kun}\|_{\L^{2}}^2
\right)
\right)^{\frac{1}{4}}\\
&\leq C\left(
\|\partial_y^2 K_{\kun}\|_{\L^{2}}^2
\left(
2\|\partial_yK_{\kun}\|_{\L^{2}}^2+
x^2\|\partial^2_yK_{\kdeux}\|_{\L^{2}}^2
\right)
\right)^{\frac{1}{4}}~,
\end{equs}
where here, $\L^1\equiv\L^1({\bf R},{\rm d}y)$ and $\L^2\equiv\L^2({\bf R},{\rm
d}k)$. The proof is then easily completed using Lemma \ref{lem:L2}
and $\P\ed^{\frac{b(n\strouhal)x}{4}}\leq\ed^{\frac{b(\strouhal)x}{4}}$, we omit
the details.
\end{proof}

\begin{lemma}\label{lem:kerneldeux}
For all $1\leq\beta\leq3$, $\frac{1}{4}\leq\xi_{2}\leq1$ and
$1\leq\xi_3\leq\frac{5}{2}$, there exists a constant $C>0$ such that
\begin{equs}
\|K_{\khuit}\|_{\K{1}{0}{\xi_2}}+
\|K_{\khuit}\|_{\K{2}{0}{\frac{\xi_3}{2}}}+
\|K_{\khuit}\|_{\K{\infty}{\frac{1}{2}}{2}}+
\|\partial_yK_{\khuit}\|_{\K{\infty}{1}{3}}
&\leq C\\
\||y|^{\beta}K_{\khuit}\|_{\K{2}{-\frac{5}{4}+\frac{\beta}{2}}{0}}+
\||y|^{\beta}\partial_yK_{\khuit}\|_{\K{2}{-\frac{3}{4}+\frac{\beta}{2}}{1}}+
\|\partial_yK_{\khuit}\|_{\K{1}{\frac{1}{4}}{\frac{1+\xi_3}{2}}}
&\leq C
\label{eqn:lastdulemme}
\end{equs}
The same estimate holds with $K_{8}$ replaced
by $\ed^{-\frac{b(\strouhal)x}{4}}\P K_{8}$.
\end{lemma}

\begin{proof}
For any $0\leq\sigma\leq1$, we have
\begin{equs}
\left|K_{\khuit}(x,k)\right|&\leq
{\textstyle C\left|\frac{\Re(\Lambdam)}{\Lambdazero}\right|^{1-\sigma}
\frac{\ed^{{\rm Re}(\Lambdam)x/2}}
{(x|\Lambdazero|)^{\sigma}}
\leq
C\frac{\ed^{{\rm Re}(\Lambdam)\frac{x}{2}}}
{(x|\Lambdazero|)^{\sigma}}}~,\\
\left|\partial_kK_{\khuit}(x,k)\right|&\leq 
C{\textstyle\left(\frac{|k|}{|\Lambdazero|^2}
+\frac{x|k\Re(\Lambdam)|}{|\Lambdazero|^2}
\right)\ed^{{\rm Re}(\Lambdam)x}
\leq C\frac{|k|}{|\Lambdazero|^2}
\ed^{{\rm Re}(\Lambdam)\frac{x}{2}}}~,\\
\left|\partial_k^3K_{\khuit}(x,k)\right|&\leq C
{\textstyle\left|
\frac{\Re(\Lambdam)}{\Lambdazero}\right|\left|
\partial_k^3\ed^{\Lambdam x}
\right|
+\frac{|k\partial_k^2\ed^{\Lambdam x}|}{|\Lambdazero|^2}
+\frac{|\partial_k\ed^{\Lambdam x}|}{|\Lambdazero|^2}
+\frac{|k\ed^{\Lambdam x}|}{|\Lambdazero|^4}}~,\\
&\leq
C{\textstyle\left(
 \frac{x^2|k|^3}{|\Lambdazero|^4}
+\frac{x|k|}{|\Lambdazero|^3}
+\frac{|k|}{|\Lambdazero|^4}
\right)\ed^{\Re(\Lambdam)\frac{x}{2}}}~.
\end{equs}
Let $1\leq\xi_3\leq\frac{5}{2}$, $\sigma_3=\frac{\xi_3}{2}-\frac{1}{4}$ and
$\gamma_3=\frac{\xi_3}{2}-\frac{1}{2}$. Since
$0\leq\sigma_3,\gamma_3\leq1$, for any fixed $x$, we have
\begin{equs}
\|K_{\khuit}(x)\|_{\L^{2}}^2&\leq
C
\sup_{n\in{\bf Z}}\left(
x^{-2\sigma_3}\int_{|k|\leq1}\hspace{-6mm}{\rm
d}k~{\textstyle
\frac{
\ed^{b(n\strouhal)x-c(n\strouhal)xk^2}
}{
(1+(n\strouhal)^2)^{\frac{\sigma_3}{2}}}}
+							 
x^{-2\gamma_3}\int_{|k|>1}\hspace{-6mm}{\rm d}k~
\ed^{b(n\strouhal)x-\frac{|k|x}{2}}\right)
\\&\leq
C\sup_{n\in{\bf Z}}
\ed^{b(n\strouhal)x}{\textstyle\left(
\frac{x^{-\frac{1}{2}-2\sigma_3}}{(1+(n\strouhal)^2)^{\frac{4\sigma_3-1}{8}}}
+x^{-1-2\gamma_3}
\right)\leq 
C x^{-\xi_3}}
~.
\end{equs}
The bound on $\|K_{\khuit}\|_{\K{1}{0}{\xi_2}}+
\|K_{\khuit}\|_{\K{2}{0}{\frac{\xi_3}{2}}}+
\||y|^{\beta}K_{\khuit}\|_{\K{2}{-\frac{5}{4}+\frac{\beta}{2}}{0}}$ is
completed using Lemma \ref{lem:Lun}, \ref{lem:L2} and
$\P\ed^{\frac{b(n\strouhal)x}{4}}\leq\ed^{\frac{b(\strouhal)x}{4}}$. To
bound $\|\partial_yK_{\khuit}\|_{\K{1}{\frac{1}{2}}{1+\xi_2}}$, we note
that for fixed $x$
\begin{equs}
\|\partial_yK_{\khuit}(x)\|_{\L^{1}}
&\leq C
\sup_{n\in{\bf Z}}\left(
\|kK_{\khuit}(x)\|_{\L^{2}}^2
\left(
\|K_{\khuit}(x)\|_{\L^{2}}^2
+
\|k\partial_kK_{\khuit}(x)\|_{\L^{2}}^2
\right)
\right)^{\frac{1}{4}}
\leq
C\left({\textstyle
\frac{1+x}{x^{2\xi_3+2}}
}\right)^{\frac{1}{4}}~.
\end{equs}
This completes the proof.
\end{proof}

\begin{lemma}\label{lem:kerneldeuxx}
Let $x\geq0$, then there exists a constant $C>0$ such that for all
$1\leq\beta\leq3$ we have
\begin{equs}
\|\ed^{-\frac{b(\strouhal)x}{4}}
K_{\kdix}\|_{\K{\infty}{0}{1}}+
\|\ed^{-\frac{b(\strouhal)x}{4}}
K_{\kdix}\|_{\K{2}{0}{\frac{3}{4}}}+
\|\ed^{-\frac{b(\strouhal)x}{4}}
K_{\kdix}\|_{\K{1}{\frac{1}{8}}{\frac{5}{8}}}&\leq C\\
\|\ed^{-\frac{b(\strouhal)x}{4}}
|y|^{\beta}K_{\kdix}\|_{\K{2}{\frac{3}{8}+\frac{\beta}{8}}{-\frac{9}{8}
+\frac{3\beta}{8}}}
+
\|\ed^{-\frac{b(\strouhal)x}{4}}
\partial_yK_{\kdix}\|_{\K{\infty}{\frac{1}{2}}{2}}+
\|\ed^{-\frac{b(\strouhal)x}{4}}
\partial_yK_{\kdix}\|_{\K{1}{\frac{5}{8}}{\frac{13}{8}}}&\leq C~.
\end{equs}
\end{lemma}

\begin{proof}
We have $|K_{\kdix}(x,k)|\leq
C\frac{|n\strouhal|}{\chinese{n\strouhal}}\ed^{\Re(\Lambdam)x}
$, and
\begin{equs}
\left|\partial_kK_{\kdix}(x,k)\right|&\leq 
C{\textstyle\left(\frac{|n\strouhal k|}{|\Lambdazero|^4}
+\frac{x|kn\strouhal|}{\chinese{n\strouhal}|\Lambdazero|}
\right)}\ed^{{\rm Re}(\Lambdam)x}
\leq{\textstyle\frac{\chinese{x}|n\strouhal|}{\chinese{n\strouhal}}
\frac{|k|\ed^{\Re(\Lambdam x)}}{|\Lambdazero|}}~,\\
\left|\partial_k^3K_{\kdix}(x,k)\right|&\leq 
C{\textstyle\left|
\frac{\Im(\Lambdam)}{\Lambdazero}
\right|\left(
\left|
\partial_k^3\ed^{\Lambdam x}
\right|
+\left|
\frac{k\partial_k^2\ed^{\Lambdam x}}{|\Lambdazero|^2}
\right|
+\left|
\frac{\partial_k\ed^{\Lambdam x}}{|\Lambdazero|^2}
\right|
+\left|
\frac{k\ed^{\Lambdam x}}{|\Lambdazero|^4}
\right|
\right)}~,\\
&\leq
C{\textstyle\frac{|n\strouhal|}{\chinese{n\strouhal}}\ed^{\Re(\Lambdam)x}
\left(
\frac{x^3|k|^3}{|\Lambdazero|^3}+
\frac{x^2|k|}{|\Lambdazero|
}+
\frac{1+x}{|\Lambdazero|}
\right)}~.
\end{equs}
In particular, we have $\P K_{\kdix}=K_{\kdix}$. The proof is then
completed using Lemma \ref{lem:L2}, that
$\P\ed^{\frac{b(n\strouhal)x}{4}}\leq\ed^{\frac{b(\strouhal)x}{4}}$ and
$|n\strouhal|\chinese{n\strouhal}^{-1}\chinese{x}^{\frac{1}{2}}\ed^{\frac
{b(n\strouhal)x}{4}}\leq2$, and that for fixed $x$, we have
\begin{equs}
\|\partial_yK_{\kdix}(x)\|_{\L^{1}}
&\leq C
\left(
\|kK_{\kdix}(x)\|_{\L^{2}}^2
\left(
\|K_{\kdix}(x)\|_{\L^{2}}^2
+
\|k\partial_kK_{\kdix}(x)\|_{\L^{2}}^2
\right)
\right)^{\frac{1}{4}}~,
\end{equs}
where $\L^1\equiv\L^1({\bf R},{\rm d}y)$ and $\L^2\equiv\L^2({\bf R},{\rm
d}k)$. 
\end{proof}

\begin{lemma}\label{lem:withkr}
Let $p\geq2$. There exist a constant $C>0$ such that
\begin{equs}
\|K_{\kdouze}\|_{\K{\infty}{\frac{1}{2}}{1}}+
\|K_{\kdouze}\|_{\K{2}{\frac{1}{4}}{\frac{1}{2}}}+
\|\partial_yK_{\kdouze}\|_{\K{p}{1-\frac{1}{2p}}{2-\frac{1}{p}}}
&\leq C~,\\
\|\ed^{-\frac{b(\strouhal)x}{2}}K_{\ktreize}\|_{\K{\infty}{\frac{1}{2}}{1}}+
\|\ed^{-\frac{b(\strouhal)x}{2}}K_{\ktreize}\|_{\K{2}{\frac{1}{4}}{\frac{1}{2}}}+
\|\partial_y\ed^{-\frac{b(\strouhal)x}{2}}K_{\ktreize}
\|_{\K{p}{1-\frac{1}{2p}}{2-\frac{1}{p}}}
&\leq C~,
\end{equs}
while there exists a constant $C$ such that for all $x\geq0$, we have
\begin{equs}
\|K_{\kdouze}(x,n\strouhal)\|_{\L^{1}}
+\ed^{-\frac{b(\strouhal)x}{2}}
\|K_{\ktreize}(x,n\strouhal)\|_{\L^{1}}
&\leq
C{\textstyle\Big(1+
\frac{\chinese{\strouhal}}{|\strouhal|x^{\frac{1}{4}}}
\Big)}~.
\end{equs}
The estimates of this Lemma also hold with $K_{12}$ replaced by 
$\ed^{-\frac{b(\strouhal)x}{4}}\P K_{12}$.
\end{lemma}
\begin{proof}
We first note that $\Q K_{\ktreize}=0$ and $\P K_{\ktreize}=K_{\ktreize}$.
We then have $|K_{\kdouze}(x,k)|+|K_{\ktreize}(x,k)|\leq C\ed^{{\rm
Re}(\Lambdam)x}$ and
\begin{equs}
|\partial_kK_{\kdouze}(x,k)|+|\partial_kK_{\ktreize}(x,k)|&\leq 
C\ed^{{\rm Re}(\Lambdam)x}{\textstyle\left(
\frac{m_{n}}{|\Lambdazero|}
+\frac{|k|x}{|\Lambdazero|}
\right)}
~,
\end{equs}
with $m_{n}=1$ if $n=0$ and
$m_{n}=\frac{\chinese{n\strouhal}}{|n\strouhal|}$ if $n\neq0$.
We then get e.g.
\begin{equs}
\|\partial_kK_{\kdouze}(x)\|_{\L^{2}}^2
&\leq
\sup_{n\in{\bf Z}}\left(
\int_{|k|\leq1}\hspace{-6mm}{\rm d}k~
{\textstyle\frac{m_{n}^2+k^2x^2}{\chinese{n\strouhal}}}
\ed^{2b(n\strouhal)x-2c(n\strouhal)xk^2}
+
\int_{|k|>1}\hspace{-6mm}{\rm d}k~
{\textstyle\frac{m_{n}^2+k^2x^2}{1+k^2}}
\ed^{2b(n\strouhal)x-|k|x}\right)
\\&
\leq C
\sup_{n\in{\bf Z}}\left(
\ed^{b(n\strouhal)x}\left(
m_{n}^2
+
\sqrt{x}
\right)\right)~.
\end{equs}
The proof is completed using Lemmas \ref{lem:Lun} and \ref{lem:L2}, that
$|n\strouhal|\chinese{n\strouhal}^{-1}\chinese{x}^{\frac{1}{2}}\ed^{\frac
{b(n\strouhal)x}{4}}\leq2$ and
$\P\ed^{\frac{b(n\strouhal)x}{4}}\leq\ed^{\frac{b(\strouhal)x}{4}}$ (see
also the proof of Lemma \ref{lem:kerneldeuxx}), we omit the details.
\end{proof}

\begin{lemma}\label{lem:withki}
Let $p\geq2$. There exist a constant $C>0$ such that
\begin{equs}
\|\ed^{-\frac{b(\strouhal)x}{4}}
K_{\kr}\|_{\K{\infty}{\frac{1}{2}}{1}}+
\|\ed^{-\frac{b(\strouhal)x}{4}}
K_{\ki}\|_{\K{\infty}{\frac{1}{2}}{1}}+
&\leq C~,\\
\|\ed^{-\frac{b(\strouhal)x}{4}}
\partial_yK_{\kr}\|_{\K{p}{1-\frac{1}{2p}}{2-\frac{1}{p}}}+
\|\ed^{-\frac{b(\strouhal)x}{4}}
\partial_yK_{\ki}\|_{\K{p}{1-\frac{1}{2p}}{2-\frac{1}{p}}}
&\leq
C~,
\end{equs}
while there exists a constant $C$ such that for all $x\geq0$ and $p\geq1$,
we have
\begin{equs}
\|K_{\kr}(x)\|_{\L^{1}}+\|K_{\ki}(x)\|_{\L^{1}}
&\leq
C\ed^{\frac{b(\strouhal)x}{4}}{\textstyle\Big(
\frac{1}{x^{\frac{1}{2}}}+
\frac{\chinese{x}^{\frac{1}{8}}}{x^{\frac{1}{8}}}
\left(1+\frac{1}{|\strouhal|\sqrt{x}}\right)^{\frac{1}{4}}
\Big)}
~,\\
\|K_{\kr}(x)\|_{\L^{p}}+\|K_{\ki}(x)\|_{\L^{p}}
&\leq
C\ed^{\frac{b(\strouhal)x}{4}}{\textstyle\Big(
\frac{\chinese{x}^{\frac{1}{2}-\frac{1}{2p}}}{x^{1-\frac{1}{2p}}}+
\frac{\chinese{x}^{\frac{1}{2}-\frac{3}{8p}}}{x^{1-\frac{7}{8p}}}+
\frac{\chinese{x}^{\frac{1}{2}-\frac{3}{8p}}}{|\strouhal|^{\frac{1}{4p}}x^{1-\frac{3}{4p}}}
\Big)}~.
\end{equs}
\end{lemma}

\begin{proof}
We first note that $\Q K_{\ki}=0$, $\P K_{\ki}=K_{\ki}$.
We then have $|K_{\kr}(x,k)|\leq C
\frac{|n\strouhal|}{\chinese{n\strouhal}}
\ed^{{\rm Re}(\Lambdam)x}$, 
$|K_{\ki}(x,k)|\leq C
\frac{(n\strouhal)^2\ed^{{\rm Re}(\Lambdam)x}}{k^2+(n\strouhal)^2}
\leq
C\min\left(\frac{(n\strouhal)^2\ed^{b(n\strouhal)x}}{k^2+(n\strouhal)^2}~,~
\ed^{{\rm Re}(\Lambdam)x}
\right)$ and
\begin{equs}
|\partial_kK_{\kr}(x,k)|&\leq C\ed^{{\rm Re}(\Lambdam)x}
{\textstyle\frac{|n\strouhal|}{\chinese{n\strouhal}}
\left(
\frac{1}{|\Lambdazero|}+
\frac{|k|x}{|\Lambdazero|}
\right)}~,\\
|\partial_kK_{\ki}(x,k)|&\leq C
\ed^{{\rm Re}(\Lambdam)x}{\textstyle\left(
\frac{1}{|n\strouhal|}+\frac{|k|x}{|\Lambdazero|}
\right)}~.
\end{equs}
This shows that
\begin{equs}
\|K_{\ki}(x)\|_{\L^{\infty}}
&\leq
C\ed^{\frac{b(\strouhal)x}{2}}
\frac{\chinese{x}^{\frac{1}{2}}}{x}~,\\
\|K_{\ki}(x)\|_{\L^{1}}
&\leq
C
\sup_{n\in{\bf Z},n\neq0}
\ed^{\frac{b(n\strouhal)x}{2}}
{\textstyle
\min\Big(|n\strouhal|,\frac{\chinese{x}^{\frac{1}{2}}}{x}\Big)^{\frac{1}{4}}
\Big(\frac{\chinese{x}^{\frac{1}{2}}}{(n\strouhal)^2x}+\sqrt{x}\Big)
^{\frac{1}{4}}}~.
\end{equs}
The proof is completed using $|n\strouhal|^{-1}\leq C|\strouhal|^{-1}$ if
$|n|\geq1$,
$|n\strouhal|\chinese{n\strouhal}^{-1}\chinese{x}^{\frac{1}{2}}\ed^{\frac
{b(n\strouhal)x}{4}}\leq2$ and
$\P\ed^{\frac{b(n\strouhal)x}{4}}\leq\ed^{\frac{b(\strouhal)x}{4}}$ (see
also the proof of Lemma \ref{lem:kerneldeuxx}), we omit the details.
\end{proof}

\begin{lemma}\label{lem:alittlelemma}
Let $K_{\kc}(x,y)=\Q\frac{\ed^{-\frac{y^2}{4x}}}{\sqrt{4\pi x}}$. We have
\begin{equs}
\|\partial_y^m(K_{\kun}-K_{\kc})\|_{\infty,\{\frac{m+5}{2},m+4\}}+
\|\partial_y(K_{\kun}-K_{\kc})\|_{1,\{3,\frac{9}{2}\}}
&\leq C
\strouhal^{-2}\chinese{\strouhal}^{2}\\
\|\partial_y^m(K_{\kdouze}-K_{\kc})\|_{\infty,\{\frac{m+5}{2},m+4\}}+
\|K_{\kdeux}-\partial_yK_{\kc}\|_{\infty,\{3,5\}}
&\leq C
\strouhal^{-2}\chinese{\strouhal}^{2}
\end{equs}
for all $m\in{\bf N}$.
\end{lemma}

\begin{proof}
We first note that $\Q|\Lambdam+k^2|\leq Ck^4$, so that
\begin{equs}
\|\partial_y^m(\Q K_{\kun}(x)-K_{\kc}(x))\|_{\L^\infty}&\leq
\sup_{n\in{\bf Z}}
\int_{-\infty}^{\infty}
\hspace{-4mm}{\rm d}k~|k|^{m}\Q
\ed^{\Re(\Lambdam)x}\big|
1-\ed^{-(k^2+\Lambdam)x}
\big|\\
&\leq
Cx B_{0,4+m}(x/2,0)\leq
C\chinese{x}^{\frac{m+5}{2}}x^{-m-4}~,\\
\|\partial_y^m\P K_{\kun}(x)\|_{\L^\infty}&\leq
\sup_{n\in{\bf Z}}
\int_{-\infty}^{\infty}
\hspace{-4mm}{\rm d}k~|k|^m\P
\ed^{\Re(\Lambdam)x}
\leq C B_{0,m}(x/2,\strouhal)
\\
&\leq C
\chinese{x}^{\frac{m+5}{2}}x^{-m-4}
\sup_{x\geq0}\big(x^3\chinese{x}^{-2}\ed^{\frac{b(\strouhal)x}{4}}\big)
\leq
C\strouhal^{-2}\chinese{x}^{\frac{m+5}{2}}x^{-m-4}~,\\
\|\partial_y^m(\Q K_{\kun}(x)-K_{\kc}(x))\|_{\L^2}^2&\leq
\sup_{n\in{\bf Z}}
\int_{-\infty}^{\infty}
\hspace{-4mm}{\rm d}k~|k|^{2m}\Q
\ed^{2\Re(\Lambdam)x}\big|
1-\ed^{-(k^2+\Lambdam)x}
\big|^2\\
&\leq
C x^2B_{0,8+2m}(x,0)
\leq
C\chinese{x}^{\frac{9+2m}{2}}x^{-7-2m}~,\\
\|\partial_y^m\P K_{\kun}(x)\|_{\L^2}^2&\leq
\sup_{n\in{\bf Z}}
\int_{-\infty}^{\infty}
\hspace{-4mm}{\rm d}k~|k|^{2m}\P
\ed^{2\Re(\Lambdam)x}\leq
C B_{0,2m}(x,\strouhal)
\\
&\leq C
\chinese{x}^{\frac{9+2m}{2}}x^{-7-2m}
\sup_{x\geq0}\big(x^6\chinese{x}^{-4}\ed^{\frac{b(\strouhal)x}{4}}\big)
\leq
C\strouhal^{-4}\chinese{x}^{\frac{9+2m}{2}}x^{-7-2m}~,\\
\|\partial_y\big(y(\Q K_{\kun}(x)-K_{\kc}(x))\big)\|_{\L^2}^2&\leq
Cx^2
\sup_{n\in{\bf Z}}
\int_{-\infty}^{\infty}
\hspace{-4mm}{\rm d}k~|k|^{4}\Q
\ed^{2\Re(\Lambdam)x}\Big|
\frac{2\Lambdam}{\Lambdazero}+1-\ed^{-(k^2+\Lambdam)x}
\Big|^2\\
&\leq
C\big(
x^2B_{0,8}(x,0)+
x^4B_{0,12}(x,0)
\big)
\leq
C\chinese{x}^{\frac{13}{2}}x^{-9}~,\\
\|\partial_y\big(y\P K_{\kun}(x)\big)\|_{\L^2}^2&\leq
x^2\sup_{n\in{\bf Z}}
\int_{-\infty}^{\infty}
\hspace{-4mm}{\rm d}k~|k|^{4}\P
\ed^{2\Re(\Lambdam)x}\leq
C x^2B_{0,4}(x,\strouhal)
\\
&\leq C
\chinese{x}^{\frac{13}{2}}x^{-9}
\sup_{x\geq0}\big(x^6\chinese{x}^{-4}\ed^{\frac{b(\strouhal)x}{4}}\big)
\leq
C\strouhal^{-4}\chinese{x}^{\frac{13}{2}}x^{-9}~.
\end{equs}
The proof is completed using
$\|\partial_yf\|_{\L^1}\leq
\big(
\|\partial_yf\|_{\L^2}(
\|f\|_{\L^2}+
\|\partial_y(yf)\|_{\L^2})\big)^{\frac{1}{2}}
$, $K_{\kdeux}(x)=\partial_y(K_{\kun}(x)+K_{\khuit}(x)+K_{\kdix}(x))$
and $K_{\kdouze}(x)=K_{\kun}(x)+K_{\khuit}(x)+K_{\kdix}(x)$.
\end{proof}

\bibliographystyle{JPE}
\markboth{\sc \refname}{\sc \refname}
\bibliography{refs}

\def\Rom#1{\uppercase\expandafter{\romannumeral #1}}\def\u#1{{\accent"15
  #1}}\def\cprime{$'$} \def\cprime{$'$}
\begin{thebibliography}{10}

\bibitem{babenkovortex}
K.~I. Babenko.
\newblock The asymptotic behaviour of a vortex far away from a body in plane
  flow of a viscous fluid.
\newblock {\em J. App. Math. Mech.\/} {\bf 34} (1970), 869--881.

\bibitem{Batchelor}
G.~K. Batchelor.
\newblock {\em An introduction to fluid dynamics\/}.
\newblock Cambridge Mathematical Library (Cambridge: Cambridge University
  Press, 1999), paperback edition.

\bibitem{bricmont}
J.~Bricmont, A.~Kupiainen, and G.~Lin.
\newblock Renormalization group and asymptotics of solutions of nonlinear
  parabolic equations.
\newblock {\em Comm. Pure Appl. Math.\/} {\bf 47} (1994), 893--922.

\bibitem{cla70}
D.~C. Clark.
\newblock The vorticity at infinity for solutions of the stationary
  {N}avier-{S}tokes equations in exterior domains.
\newblock {\em Indiana Univ. Math. J.\/} {\bf 20} (1970/1971), 633--654.

\bibitem{Dusek94}
J.~Du{\v{s}}ek, P.~Le~Gal, and P.~Frauni{\'e}.
\newblock A numerical and theoretical study of the first {H}opf bifurcation in
  a cylinder wake.
\newblock {\em J. Fluid Mech.\/} {\bf 264} (1994), 59--80.

\bibitem{PR1}
R.~Finn and D.~R. Smith.
\newblock On the stationary solutions of the {N}avier-{S}tokes equations in two
  dimensions.
\newblock {\em Arch. Rational Mech. Anal.\/} {\bf 25} (1967), 26--39.

\bibitem{Gal94}
G.~P. Galdi.
\newblock {\em An introduction to the mathematical theory of the
  {N}avier-{S}tokes equations\/}, volume~38 of {\em Springer Tracts in Natural
  Philosophy\/} (New York: Springer-Verlag, 1994).

\bibitem{PRgaldi}
G.~P. Galdi.
\newblock {\em Mathematical questions relating to the plane steady motion of a
  {N}avier-{S}tokes fluid past a body\/}, volume~16 of {\em Lecture Notes
  Numer. Appl. Anal.\/} (Tokyo: Kinokuniya, 1998).

\bibitem{Peter2D2003}
F.~Haldi and P.~Wittwer.
\newblock Leading order down-stream asymptotics of non-symmetric stationary
  navier-stokes flows in two dimensions.
\newblock {\em Preprint\/}  (2003), 1--28.

\bibitem{Homescu02}
C.~Homescu, I.~Navon, and Z.~Li.
\newblock Suppression of vortex shedding for flow around a circular cylinder
  using optimal control.
\newblock {\em Int. J. Num. Meth. Fluids\/} {\bf 38} (2002), 43--69.

\bibitem{Hormander}
L.~H{\"o}rmander.
\newblock Estimates for translation invariant operators in {$L\sp{p}$}\ spaces.
\newblock {\em Acta Math.\/} {\bf 104} (1960), 93--140.

\bibitem{Posdziech01}
O.~Posdziech and R.~Grundmann.
\newblock Numerical simulation of the flow around an infinitely long circular
  cylinder in the transition regime.
\newblock {\em Theor. Comp. Fluid Dyn.\/} {\bf 15} (2001), 121--141.

\bibitem{Protas02}
B.~Protas and J.-E. Wesfreid.
\newblock Drag force in the open-loop control of the cylinder wake in the
  laminar regime.
\newblock {\em Phys. Fluids\/} {\bf 14} (2002), 810--826.

\bibitem{Provansal87}
M.~Provansal, C.~Mathis, and L.~Boyer.
\newblock B\'enard-von karman instability: transient and forced regimes.
\newblock {\em J. Fluid Mech.\/} {\bf 182} (1987), 1--22.

\bibitem{Sazonov03}
L.~I. Sazonov.
\newblock On the stability of periodic solutions of the {N}avier-{S}tokes
  system in a three-dimensional exterior domain.
\newblock {\em Izv. Ross. Akad. Nauk Ser. Mat.\/} {\bf 67} (2003), 155--170.

\bibitem{Schumm94}
M.~Schumm, E.~Berger, and P.~Monkewitz.
\newblock Self-excited oscillations in the wake of two dimensional bluff bodies
  and their control.
\newblock {\em J. Fluid Mech.\/} {\bf 271} (1994), 17--53.

\bibitem{Sreenivasan86}
K.~Sreenivasan, P.~Strykowski, and D.~Olinger.
\newblock Hopf bifurcation, landau equation and vortex shedding behind circular
  cylinders.
\newblock In: K.~Ghia, ed., {\em Proc. Forum on Unsteady Separation\/} (ASME
  FED 52, 1986).

\bibitem{wake}
G.~van Baalen.
\newblock Stationary solutions of the {N}avier-{S}tokes equations in a
  half-plane downstream of an obstacle: ``universality'' of the wake.
\newblock {\em Nonlinearity\/} {\bf 15} (2002), 315--366.

\bibitem{Wesfreid96}
J.-E. Wesfreid, S.~Goujon-Durand, and B.~Zielinska.
\newblock Global mode behavior of the streamwise velocity in wakes.
\newblock {\em J. Phys. II\/} {\bf 6} (1996), 1343.

\bibitem{Peter03}
P.~Wittwer, S.~B\"onisch, and V.~Heuveline.
\newblock Adaptative boundary conditions for exterior flow problems.
\newblock To appear in J. Math. Fluid Mech. (2003).

\bibitem{Peter04}
P.~Wittwer, S.~B\"onisch, and V.~Heuveline.
\newblock Second order adaptative boundary conditions for exterior flow
  problems: non-symmetric stationary flows in two dimensions.
\newblock Preprint, Heidelberg/Geneva (2004).

\bibitem{Zielinska95}
B.~Zielinska and J.-E. Wesfreid.
\newblock On the spatial structure of global modes in wake flow.
\newblock {\em Phys. Fluids\/} {\bf 7} (1995), 1418.

\end{thebibliography}

\end{document}